\documentclass{article}
\usepackage[a4paper, total={6in, 8in}]{geometry}
\usepackage{svg}
\usepackage{longtable}
\usepackage{rotating}
\usepackage{graphicx}
\usepackage{subcaption}
\usepackage{parskip} 
\usepackage{multirow}
\usepackage{array}
\usepackage{booktabs}
\usepackage{hyperref}

\AtBeginDocument{%
  }

\usepackage{float}

\begin{document}

\title{Repeatability, Reproducibility, Replicability, Reusability (4R) in Journals' Policies and Software/Data Management in Scientific
Publications: A Survey, Discussion, and Perspectives}

\author{
  José Armando Hernández González\\
  \texttt{jose.hernandez\_gonzalez@ens-paris-saclay.fr}\\
  ORCID 0000-0002-6692-8640
  \and
  Miguel Colom\\
  \texttt{mcolomba@ens-paris-saclay.fr}\\
  ORCID 0000-0003-2636-0656
  \and
  Université Paris Saclay, Université Paris Cité, ENS Paris Saclay, CNRS, SSA, INSERM,\\Centre Borelli. France.
}

\maketitle

\begin{abstract}
With the recognized crisis of credibility in scientific  research, there is a growth of reproducibility studies in computer science, and although existing surveys have reviewed reproducibility from various perspectives, especially very specific technological issues, they do not address the author-publisher relationship in the publication of reproducible computational scientific articles. This aspect requires significant attention because it is the basis for reliable research.
We have found a large gap between the reproducibility-oriented practices, journal policies, recommendations, publisher artifact Description/Evaluation guidelines, submission guides, technological reproducibility evolution, and its effective adoption to contribute to tackling the crisis. 

We conducted a narrative survey, a comprehensive
overview and discussion identifying the mutual efforts required from Authors, Journals, and Technological actors to achieve reproducibility research.
The relationship between authors and scientific journals in their mutual efforts to jointly improve the reproducibility of scientific results is analyzed. Eventually, we propose recommendations for the journal policies, as well as a unified and standardized Reproducibility Guide for the submission of scientific articles for authors.

The main objective of this work is to analyze the implementation and experiences of reproducibility policies, techniques and technologies, standards, methodologies, software, and data management tools required for scientific reproducible publications. Also, the benefits and drawbacks of such an adoption, as well as open challenges and promising trends, to propose possible strategies and efforts to mitigate the identified gaps.
To this purpose, we analyzed 200 scientific articles, surveyed 16 Computer Science journals, and systematically classified them according to reproducibility strategies, technologies, policies, code citation, and editorial business.

We conclude there is still a reproducibility gap in scientific publications, although at the same time also the opportunity to reduce this gap with the joint effort of authors, publishers, and technological providers.

\textbf{Keywords}: Repeatability, Reproducibility, Replicability, Reusability, Data Science AI/ML, RaaS, Scientific journal, Trustworthy, Data Citation, Rewarding Research, Reproducible Research.
\end{abstract}

\section{Introduction}
\label{sec:introduction}
Reproducibility is a broad and complex topic strongly related to the history of science and knowledge~\cite{ivie_reproducibility_2018} reflected in the cumulative technological and scientific development of humanity~\cite{hughes_history_2001}. Such development has been based on the evolutionary capacity of human beings to build new knowledge from previous discoveries and achievements, transmitting this knowledge to new generations in a continuous cycle of improvement. The evolution of Science through the reproducibility of knowledge could be metaphorically compared to the natural mechanisms of DNA replication~\cite{hu_evolution_2020} transmitted from generation to generation in a continuous cycle of refinement. Within these reproducible mechanisms, scientific journals play a significant role in the communication, divulgation, corroboration, validation, and acceptance of reliable and trustworthy knowledge.

The reproducibility of knowledge has recently become relevant to the scientific community given that there is a growing concern for ethics and transparency in the research results in scientific publications in the so-called \textit{reproducibility crisis}~\cite{plaven-sigray_readability_2017, gundersen_reproducibility_2020}. In addition, with the boom in artificial intelligence/machine learning (AI/ML), publications have evolved towards data-centric and model-centric developments that force journals to adapt their publishing business models to new dynamics accelerated by technological changes~\cite{hutson_artificial_2018}. 

In response to these developments, several recent articles have hypothesized what the future of academic publishing will be like~\cite{dodds_future_2019, baillieul_reflections_2018} ~\cite{ahmed_future_2023}, analyzing important changes, proposing technological tools~\cite{kitchenham_meta-analysis_2020, anchundia_resources_2020} and identifying significant gaps in publishing policies~\cite{stoddart_is_2016,kapoor_leakage_2022, lucic_reproducibility_2022, ahmed_measuring_2022, moradi_reproducible_2021, haibe-kains_transparency_2020, mckinney_reply_2020, kitamura_reproducible_2020, gibney_this_2020, ghanta_interpretability_2018, aviyente_explainability_2022, kerautret_reproducibility_2021, raff_research_2020}. These articles account for policies implemented by publishers and their evolution, which are crucial for understanding their evaluation processes oriented to the Reproducibility of knowledge. 

This paper analyzes the journal policies concerning the reproducibility of knowledge addressed to trustworthiness and transparency through a survey of computer science journals indexed in SCOPUS\& WoS. This survey identifies recent advancements in computational reproducibility and policies, develop strategies that engage with the reproducibility crisis, and analyze future perspectives in computer base scientific publications. In particular, the role of stakeholders~\cite{diaba-nuhoho_reproducibility_2021} in the reproducibility of computational scientific articles, especially ML and AI-related projects, is explored to understand better the relationship between authors and scientific journals driven by advanced cloud-computing technologies. By reviewing the most promising trends, this paper broadens the landscape~\cite{baillieul_reflections_2018} of possibilities for advances in reproducibility from both technological and methodological points of view, addressing the gap in effective journal policies that guarantee the reproducibility of computational works.

A wide variety of scientific journals deal with different topics or specialize in various disciplines and fields of knowledge. In several cases, these are subjects based completely on theory or physical experiments. Therefore, this article delimits its scope to specifically analyze scientific journals of computer science, especially the management and reproducibility of scientific articles based on software and data, where the objects to be replicated are information (\textit{bits}),  highlighting that many scientific disciplines converge in the computational field to solve their problems and build up their knowledge base.

\begin{figure*}
    \centering
    \includegraphics[width=0.50\textwidth]{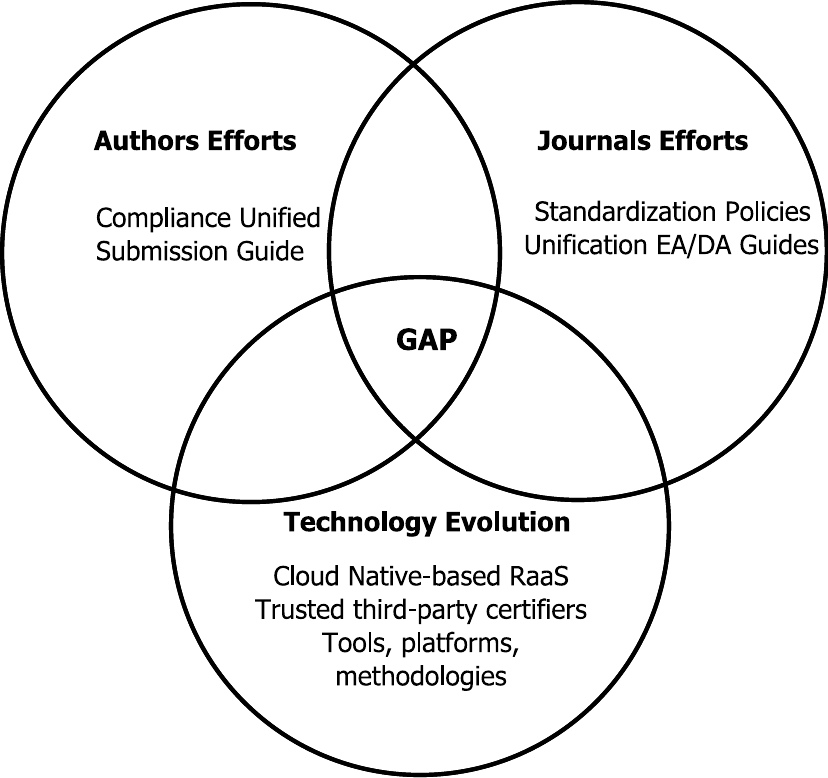}
    \caption{This diagram summarizes the structure of this paper, where three aspects are analyzed: efforts required by authors and publishers and the technological evolution required to reproduce results. The combined efforts of these three actors are required to close the reproducibility gap.}
    \label{fig:gap}
\end{figure*}

Figure~\ref{fig:gap} summarizes the structure of this paper, where three aspects are analyzed: efforts required by authors (Section~\ref{sec:authors_efforts}) and by publishers (Section \ref{sec:journal_efforts}), and the technological evolution required to reproduce results (Section \ref{sec:repro_priciples}).

The combined efforts of these three actors are required to close the reproducibility gap. Our systematic literature is a PRISMA-based review (see Appendix~\ref{sec:literature_review}). We discuss the corresponding definitions, difficulties, and measures related to Reproducibility, which allows us to define the technological evolution as necessary to reproduce results and the fundamental strategies of reproducibility in Section~\ref{sec:repro_priciples}. Section~\ref{sec:Tools} presents a landscape of existing tools, data management platforms, and techniques that are helpful in reproducible research. Section~\ref{sec:scientific_publishers} introduces the role of publishers in reproducible research, including new types of publications with code, and discusses the problem of evaluating research artifacts. Section~\ref{sec:survey} surveys 16 computer science journals and provides insights about experiences implementing data-code sharing policies based on the reviewed reproducibility platforms and technologies. The methodology that we followed is described in Appendix ~\ref{sec:survey_methodology}.

Section~\ref{sec:discussion} includes
our technological discussion of the topics presented before. Section~\ref{sec:shared_responsibility} focuses on and highlights the shared responsibility between authors and publishers supported by technological evolution. Section~\ref{sec:authors_publisher_efforts} asses and analyzes the combined efforts of these three actors required to close the reproducibility gap. The efforts required by authors in Section~\ref{sec:authors_efforts} and by publishers in section\ref{sec:journal_efforts}. Important dilemmas that emerge are addressed in sections ~\ref{sec:dilemma_docker}~\ref{sec:dilemma_policies}. Under the possibility of regarding reproducibility as a service provided by trusted third party, considering software as valuable research artifacts, and how to reward authors.

Section~\ref{sec:conclusion} concludes the paper.

\subsection{Definitions}
\label{sec:definitions} 

Several works~\cite{NIPS2015_86df7dcf, kitzes_practice_2018, baker_workshop_2019, parashar_research_2022, article_Thompson,raghupathi_reproducibility_2022} have addressed reproducibility from different points of view, as ~\cite{gundersen_fundamental_2021} reproducibility is considered a fundamental part of the scientific method. However, to our knowledge, no works have holistically reviewed the different dimensions and strategies of reproducibility in computer science, i.e., to consider their essential participation within an end-to-end data science project/experiment life cycle. This life cycle begins from scientific research and ends in mass industrial production for final customers. The life cycle also incorporates the responsibilities of the main stakeholders ~\cite{diaba-nuhoho_reproducibility_2021, feger_reproducibility_2022, macleod_improving_2022} in this process (e.g., journals, authors, industry, and the scientific community).

The report ~\cite{committee_on_reproducibility_and_replicability_in_science_reproducibility_2019} from the National Academies of Sciences, Engineering, and Medicine (NASEM) is a reference reproducibility study that gathers contributions from relevant specialized researchers. It focuses on strategies for obtaining consistent computational results using the same input data, computational steps, methods, code, analysis conditions, and replicability to get consistent results across studies. In NASEM’s definitions, \textit{reproducibility} involves the original data and code, whereas \textit{replicability} is related to the collection of new data and similar methods used in previous studies.

\begin{figure*}
    \centering
    \includegraphics[width=0.85\textwidth]{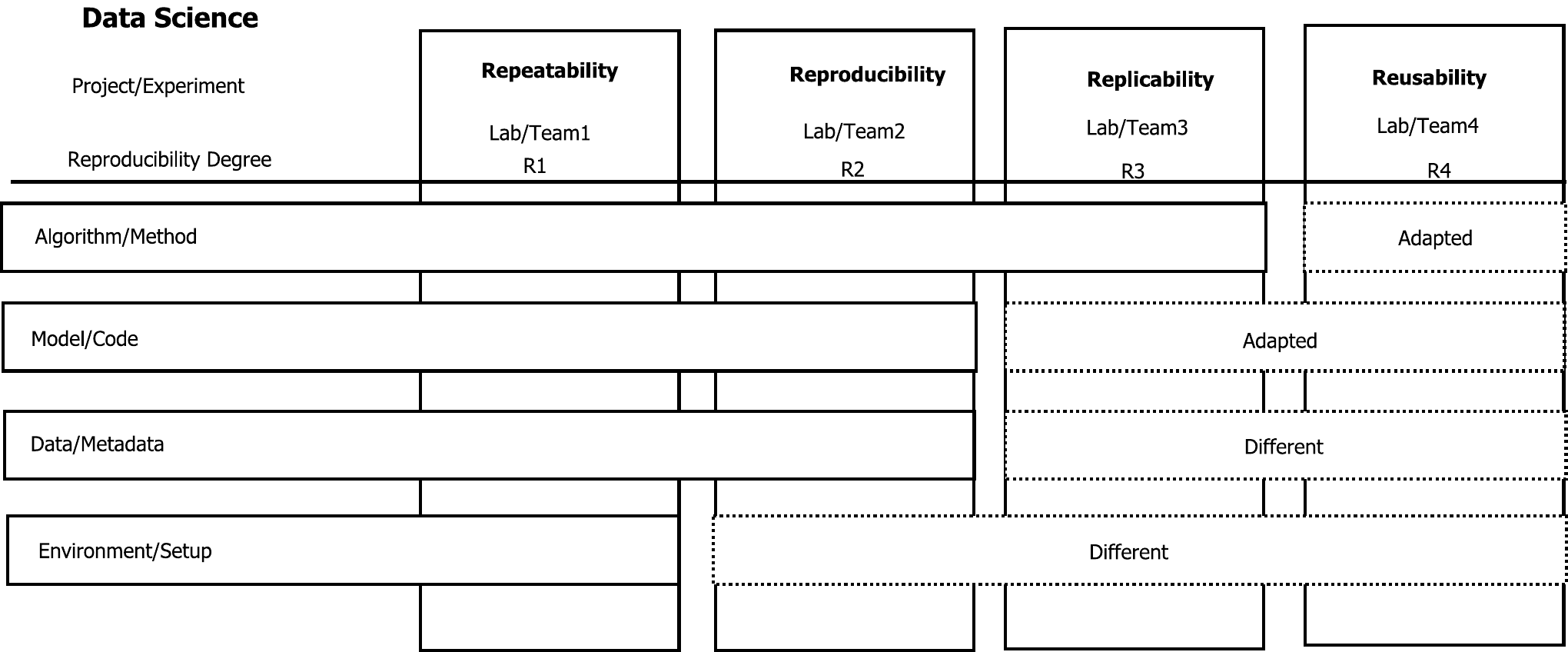}
    \caption{Definition of Reproducibility, Replicability, Repeatability, and Reusability (4R)~\cite{benureau_re-run_2018}. Different degrees of reproducibility can be considered according to the characteristics of the particular experiment or project.}
    \label{fig:mesh3}
\end{figure*}

The simplest definition of reproducibility extended and used in the different works is the one proposed by ACM in version 1.1 of their Artifact Review and Badging report\footnote{\url{https://www.acm.org/publications/policies/artifact-review-and-badging-current}}, as shown in Figure~\ref{fig:mesh3}. 

\begin{figure}
    \centering
    \includegraphics[width=0.50\textwidth]{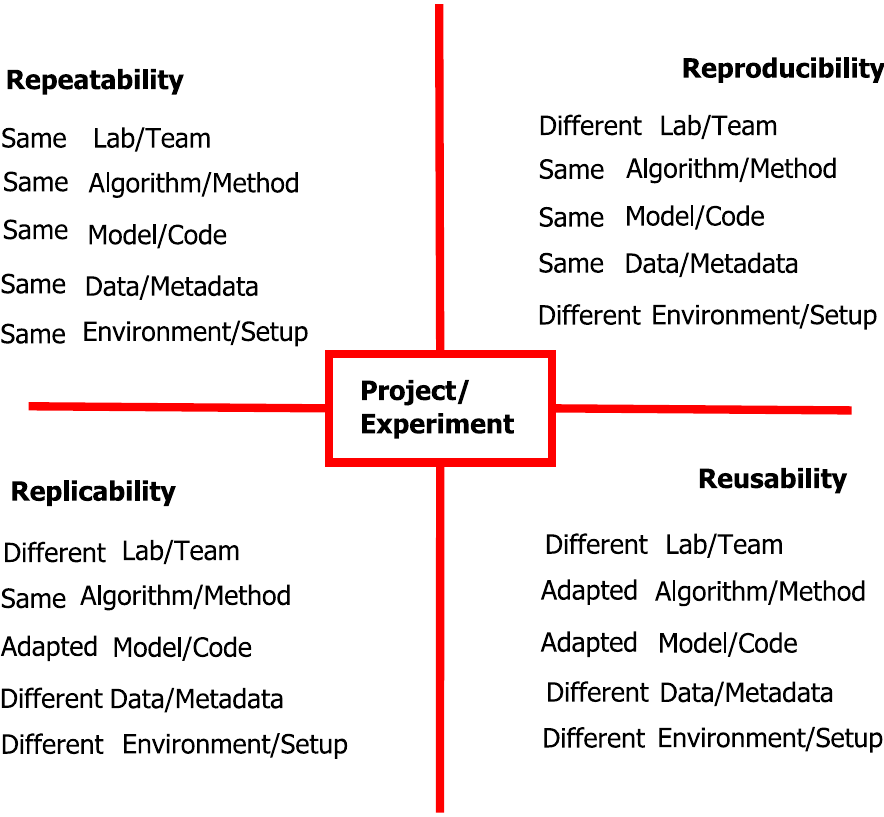} 
    \caption{A classification of Repeatability, Reproducibility, Replicability, and Reusability (4R)~\cite{benureau_re-run_2018} according to the characteristics a project or experiment.}
    \label{fig:mesh1}
\end{figure}

Figure~\ref{fig:mesh1} shows the four definitions concerning the team, method, code, metadata, and setup elements. 

\textbf{Reproducibility} (different team, same experimental setup): the experiment is done with different equipment, different environment, same code/algorithm.
\textbf{Repeatability} (same team, same experimental setup): the experiment is done by the same team, same environment (software/hardware), same code/algorithm.
\textbf{Replicability} (different team, different experimental setup): the experiment is done with different equipment, different environment, different code, same algorithm.
\textbf{Reusability} (different equipment, different and partial experimental configuration): the experiment is carried out with different equipment, different environments, different codes, and the algorithm partially implemented.

There is still some discussion, in some cases even confusion, about the definitions~\cite{plesser_reproducibility_2018} even from a taxonomic point of view~\cite{essawy_taxonomy_2020, heroux_toward_2018}. A very different interpretetion of reproducibility is presented in ~\cite{jose_reproducibility_2020} where it is a continuous improvement process rather than an achievable objective. However, following the discussions with the National Information Standards Organization (NISO), ACM accepted the recommendation to harmonize its terminology and definitions with those widely used in the community of scientific research. In this way, it interchanged the terms \textit{reproducibility} and \textit{replicability} with the existing definitions proposed by ACM to ensure consistency. 

In this article, we specifically discuss the reproducibility of complex ML/AI data science projects. A scientific publication in ML/AI can range from effectively a model developed in an experiment by a single researcher for a tiny device to large implementations of Distributed Big Data supercomputing developed by large consortiums of universities, governmental, or research institutions.

\subsection{Types of Reproducibility}

Defining what reproducibility is as important as determining the types of reproducibility, considering the nuances that conceptually appear when studying the various cases and possibilities. The term reproducibility is acceptable in the case where the same input can lead to statically equivalent same results. It is also important to note that reproducibility in data science does not necessarily imply obtaining the same numerical result from the same numerical input.

Previous works~\cite{gundersen_state_2018, raghupathi_reproducibility_2022} have defined three degrees of reproducibility: R1 (Experiment, Data, Method), R2 (Data, Method), and R3 (Method). It is only sometimes possible to obtain the same numerical result from different realizations of an experiment. In that case, one can consider the following definitions ~\cite{impagliazzo_reproducibility_2022} : 

\begin{itemize}
    \item \textbf{Experimental} reproducibility: similar input (data) + similar experimental protocol $\rightarrow$ similar results
    \item \textbf{Statistical} reproducibility: same input (data) + same analysis $\rightarrow$ same conclusions (independently from (random) sampling variability)
    \item \textbf{Computational} reproducibility: similar input (data) + same code/software + same software environment $\rightarrow$ exact same bit-wise results
\end{itemize}

\subsection{Side difficulties to achieve reproducibility}
\label{sec:side_difficulties}

Several difficulties identified in our review are not related to the actual shared source code used to reproduce an experiment. Rather, they involve external considerations such as dependency on 3rd-party libraries, complex and uncontrolled software dependencies, the quality of the writing in scientific articles, the documentation of the software, sustainability for the long term ~\cite{akhlaghi_toward_2021}, reproducibility, and the reward for the career advancement of the researchers~\cite{baker_why_2016}. 

\textbf{Complex and uncontrolled software dependencies} With the increasing demand for functionalities, the code has become more complex, relying on several 3rd party software dependencies. These include libraries, packages, or complete frameworks. Changes from different releases (or even from version to version) and irregular maintenance or support lead to what has been described as \textit{dependency hell}. This irregular maintenance represents one significant contemporary challenge for obtaining reproducible results from software made from different components that are sometimes obsolete. Using package managers, virtual environments, and container tools, keep an up-to-date list of dependencies, and perform continuous testing~\cite{fan_escaping_2020}, are recommended to mitigate this problem. With these strategies, the management of dependencies can be simplified, and projects can then be run in a reproducible manner across different configurations and systems.

\textbf{Low writing quality} The impact of the writing quality in scientific articles and the associated documentation of the software has already been studied ~\cite{mack_how_2018}. In practice, however, these aspects are sometimes overlooked. One can easily identify articles with confusing writing that is unnecessarily overloaded with complex academic jargon. Such writing is very difficult to interpret and, consequently, very hard or even impossible to reproduce.

\textbf{Compilation and infrastructure setup}. In applied computer science, significant time is generally spent compiling source code, debugging it, and configuring the running platform. The time spent configuring is comparable to the effort to solve the scientific problem. If building and running the program is too time-consuming, the software could be \textit{de facto} considered as nonreproducible. 

\textbf{Float point operations}. Given their deterministic nature, computer systems should theoretically reproduce any numerical result \textit{per se}. However, in practice, point floating operations can give slightly different numerical results in different systems  ~\cite{bailey_reproducibility_2020, jalal_apostal_improving_2020}. This aspect needs to be taken into account in the reproducibility assessment. Float point operations errors are also analyzed by Jezequel ~\cite{jezequel_first_2014} on numerical reproducibility with the IEEE 754 encoding. 

\textbf{Adapted operating systems}. Replicated environments require installing the exact version and its respective dependencies to obtain the equivalent result, and doing so is very difficult with no isolated systems.  Conda resolves this difficulty at the Python level by creating isolated environments (env); Equally, the same mechanism is applied at the operating system level in "NixOS/Guix:\textit{a Purely Functional Linux Distribution}~\cite{dolstra_nixos_2010}. One benefit to reproducibility is that Nix creates packages that are isolated from each other. These packages ensure that the environments are reproducible and have no undeclared dependencies. For example, BioNix~\cite{vallet_toward_2022} is based on the characteristics of Guix.

 \textbf{Lack of academic reward}. One major problem in reproducibility concerns the career advancement ~\cite{liu_promoting_2022} of researchers in academia and how their work is acknowledged and eventually rewarded ~\cite{ghimpau_incentives_2020, article5}. Traditionally, recognition and prestige have been associated with the number of publications and citations ~\cite{aksnes_citations_2019} in high-impact factor publications and have been decided according to metrics such as the h-index-h, Altmetrics, CiteScore, the Clarivate Analytic Journal Impact Factor (JIF), the Source Normalized Impact per Paper (SNIP), the SCImago Journal Rank (SJR), and the proposed Scientific Impact Factor (SIF) ~\cite{ATM15375}. The abuse of these metrics to decide career advancement promotes behaviors in scientists and journals that are detrimental to reproducibility, such as reticence or reluctance to publish code and data in articles or to split a single research into several non-significant articles because of the high pressure to publish. Much of the effort required to perform quality and reproducible research is not usually reflected in traditional metrics ~\cite{raff_does_2022}. In particular, the rewarding of computer science publications is limited to granting reproducibility badges ~\cite{frery_badging_2020,mauerer_beyond_2022}. Currently, at least 138 computer science journals award reproducibility badges~\footnote{\url{https://cos.io/our-services/open-science-badges/}}.

Moreover, some of the current incentives produce perverse behaviors in a hyper-competitive environment ~\cite{edwards_academic_2017} which certainly goes against ethics and scientific transparency. They have also promoted the rise of \textit{predatory journals} ~\cite{cukier_checklists_2020} based solely on Article Processing Charges (APC), articles of low-quality review standards, uncorroborated, or even false claims.

\begin{figure}
\centering
\begin{subfigure}{.5\textwidth}
    \centering
    \includegraphics[width=\textwidth]{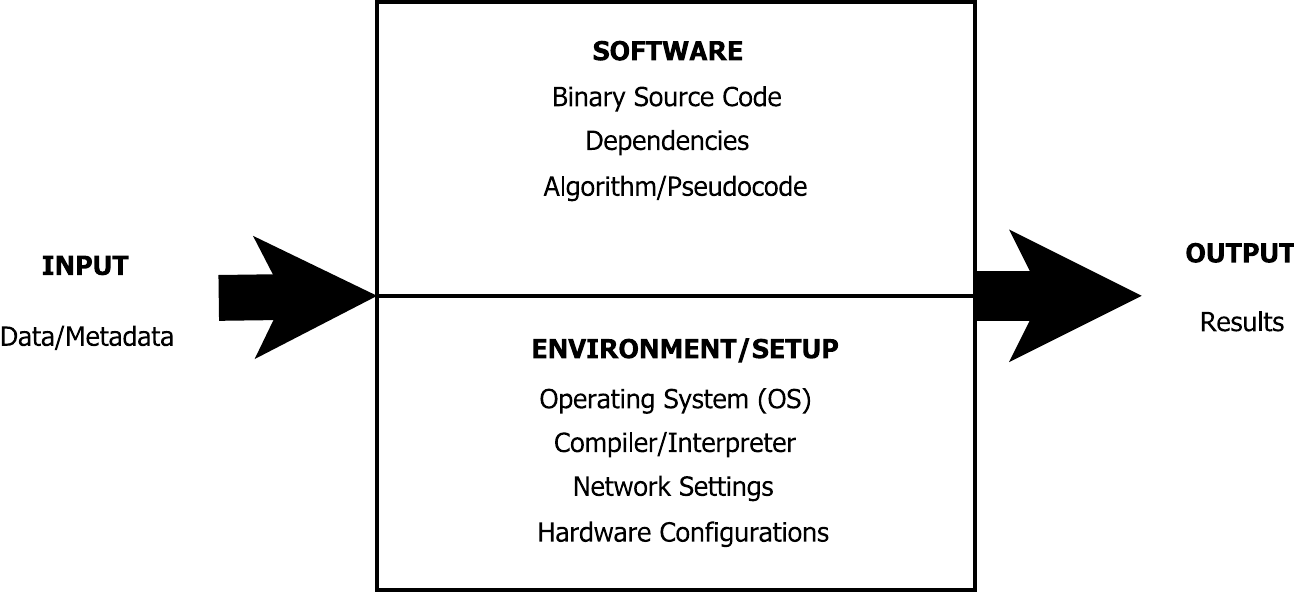} 
    \caption{Basic block Software/Data  and Hardware/Environment Reproducibility.}
    \label{fig:mesh4}
\end{subfigure}%
\begin{subfigure}{.5\textwidth}
    \centering
     \includegraphics[width=\textwidth]{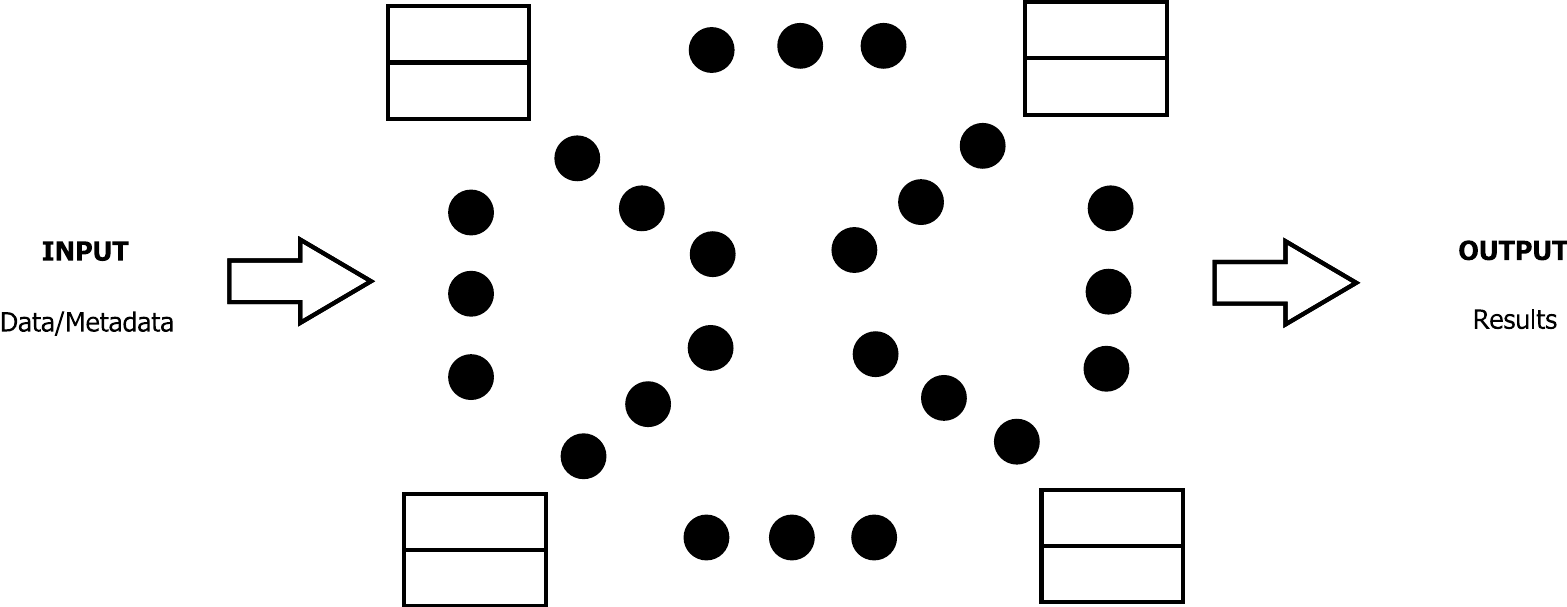} 
    \caption{System/Flow Reproducibility.}
    \label{fig:mesh2}
\end{subfigure}
\caption{Generalization of an architecture allowing for reproducible projects or experiments. It is made of basic blocks, interconnected to build complex systems, applications, and workflows.}
\label{fig:3d_eval}
\end{figure}

\subsection{Measure and Evaluate Reproducibility}
Reproducibility can be measured from different points of view, including the type of reproducibility which is evaluated (bitwise or statistical), the type of data~\cite{rosenblatt_epistemic_2023} and field of the research~\cite{ahmed_measuring_2022}. It has already been shown in multiple works~\cite{raff_research_2020} that executing the same code on a different machine might not necessarily produce the same numerical results, but one can establish that a result is statistically equivalent to other~\cite{raff_step_2019, nordling_literature_2022}.

The survival analysis proposed in \cite{raff_research_2020} permits to extract new insights that better explain past longitudinal data and extend a recent data set with \textit{reproduction times}, taking into account the number of days it took to reproduce an article~\cite{Collberg2014MeasuringRI}.

This point is certainly important because it is imperative to measure reproducibility to evaluate the degree and percentage of reproducibility of an article. As will be seen in the artifact evaluation section \ref{sec:Peer_Code_Reviews}, there is a wide disparity among journals/conferences in the criteria and policies for describing and evaluating artifacts partly as a result of the difficulty in measuring reproducibility.

\begin{figure*}
    \centering
    \includegraphics[width=0.90\textwidth]{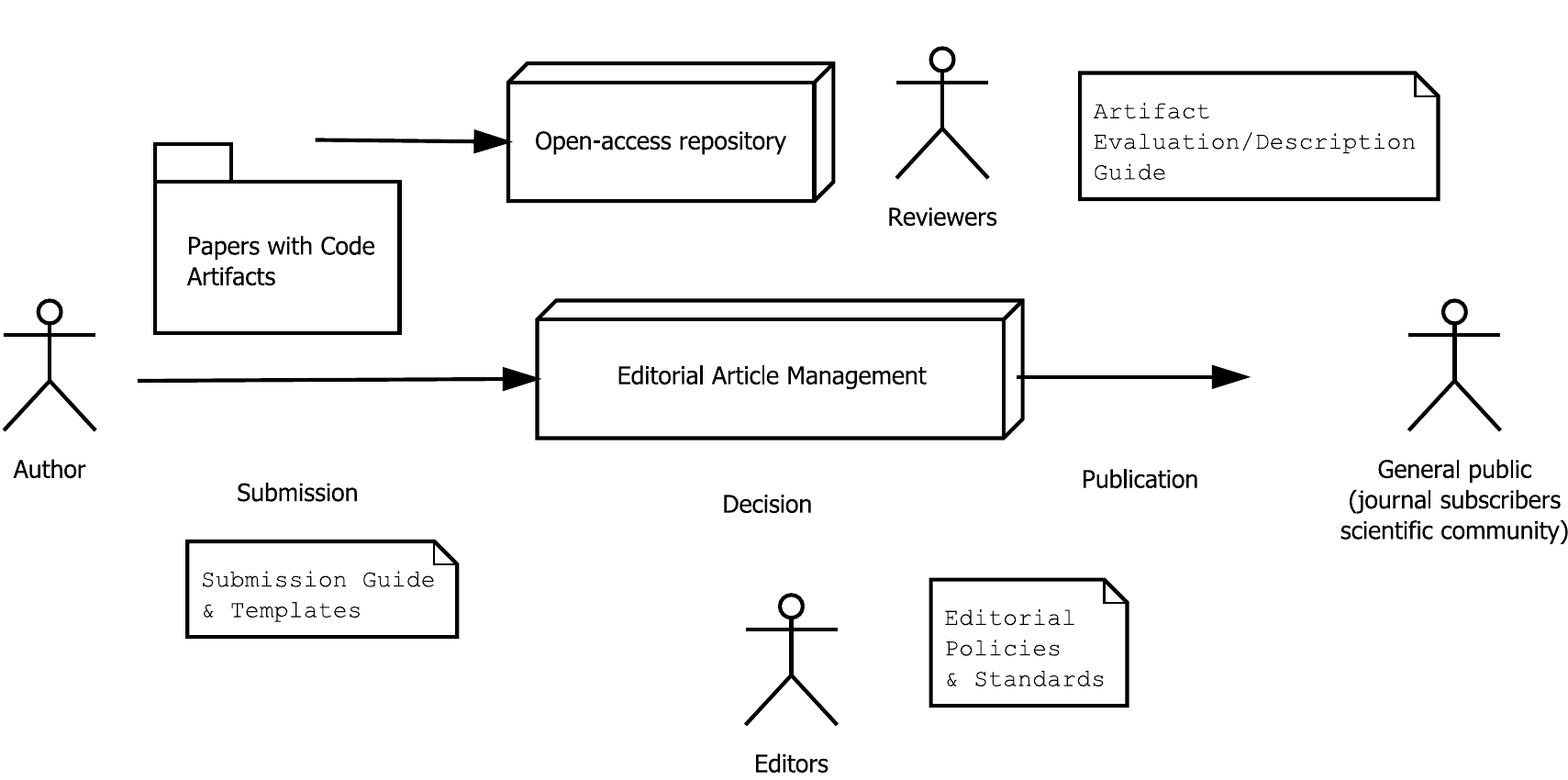}
    \caption{General description of an editorial process for the publication of a scientific article with code and the involved actors: authors, editors, reviewers, and readers.}
    \label{fig:process}
\end{figure*}

\section{Reproducibility Strategies and Technological evolution}
\label{sec:repro_priciples}
Motivated by the great reproducibility challenges~\cite{Schelter2015, freire_computational_2012}, there is extensive literature on data science projects, including current approaches for executing big data science projects~\cite{saltz_current_2022}, and the best coding practices to ensure reproducibility~\cite{bibliograph5}. Different strategies have been proposed~\cite{6171147} to tackle the problem of reproducibility of scientific works, specifically in ML/AI.

Given that the size and scope of data science projects that can range from small projects of the Internet of Things (IoT)~\cite{ray_review_2022} to very large high-end distributed HPC~\cite{pouchard_computational_2019} as complex infrastructure required for example for the recently popular Large Language Models (LLM), different strategies are required to address the complexity of each project/experiment specifically. 

In this section, we survey the most relevant characteristics that can be considered as general reproducibility strategies~\cite{gundersen_fundamental_2021}, such as the use of open source software, open repositories, and open data formats, the use of well-established methodologies and following good practices, or system architectures which are typically used in systems dedicated to run ML/AI applications.

We classify the strategies identified in the literature into four main classes: software and data, environment, system data management and workflows, and methods. Each strategy can be part of a more complex one. For example Workflows can be themselves made of Containers, and Code publications strategies require the Open code/data repositories strategy.

\begin{enumerate}
\item \textbf{Software and data reproducibility} 
    \begin{itemize}
        \item Adoption of free and open source software
        \item Tools with the potential to be used as reproducibility tools. For example, notebooks.
        \item Standardized automation benchmarks, open dataset formats, state of the art model baselines.
    \end{itemize}
\item \textbf{Environment reproducibility}
    \begin{itemize}
        \item Software reproducibility: containers and virtualization
        \item System architecture: monolithic, microservices, server-less functions
        \item Hardware reproducibility, including ambient configuration. For example, Infrastructure as Code.
    \end{itemize}
\item \textbf{System and workflow reproducibility}
    \begin{itemize}
        \item Data science project-life cycle management tools
        \item Metadata and provenance (lineage and traceability)
        \item Reproducibility as a Service. This includes 3rd-party specialized and trusted entities that certify reproducibility. They typically also offer services for the execution of algorithms on the infrastructure they provide.
    \end{itemize}
\item \textbf{Methodological reproducibility}
    \begin{itemize}
        \item Adaption of good practices and methodologies
        \item Teaching and reproducibility culture
        \item Performing evaluation specifically for research code and data artifacts
        \item Publications with code (journals and conferences)
    \end{itemize}
\end{enumerate}

\subsection{Open Source software and open repositories}
\label{sec:Open_repositories}
One could think that uploading code and data to a public repository and labeling it as open-source software could be a sufficient guarantee of reproducibility and transparency in research~\cite{macleod_improving_2022,barba_defining_2022}. There have been objections to this approach~\cite{9354557}, as well as proposals for evaluating the reproducibility level~\cite{gonzalez-barahona_revisiting_2023}. Others have proposed concrete solutions to the problem is reproducibility and transparency of scientific software~\cite{wang_how_2023, haim_how_2023, stodden_beyond_2020}. It cannot be ensured that code will not be modified after publication\footnote{Indeed, the history of a repository in Github can be altered with a \textit{hard push} command or using the corresponding tools provided by GitHub.}, or that the code is executed in exactly the same environment, dependencies, and parameters. In many cases, the full reproduction of the work cannot be achieved and often requires contacting the authors to obtain detailed information. It might happen that even the authors themselves cannot replicate the experiment due to changes in their own research infrastructure, lack of documentation, or being outdated as the project evolves~\cite{stodden_empirical_2018}.

As significant examples of these repositories and open source communities, we could cite, for example Github, Bitbucket, Gitlab, Zenodo, the Open Science Framework~\footnote{\url{https://osf.io}}, OpenAIRE (Open Access Infrastructure for Research in Europe), COAR (Confederation of Open Access Repositories), the French open document repository HAL, EOSC (European Open Science Cloud), HuggingFace, the Harvard DLhub~\footnote{\url{https://www.dlhub.org/}}, Dataverse~\footnote{\url{https://dataverse.org/}}, among many others.

\subsubsection{FAIR data}
\label{sec:FAIR_data}

Data in reproducible research should be Findable, Accessible, Interoperable, and Reusable (FAIR). Indeed, an open-source code can very well end up being non-reproducible without proper access to the data.

\textbf{Findable}: Metadata is assigned a globally unique and persistent identifier. For example, the Minimal Viable Identifiers (minids), or the Software Heritage SWHIDs;
\textbf{Accessible}: The metadata is retrievable by its identifier using a standardized communication protocol;
\textbf{Interoperable}: Metadata uses a formal, accessible, shared, and widely applicable specification for knowledge representation;
\textbf{Reusable}: metadata is described in detail with a plurality of precise and relevant attributes. 

The FAIR principles~\cite{parland-von_essen_supporting_2018} applied to data allow promoting the reproducibility of the scientific publications focusing specifically on its R (Reuse) aspect. However, each research work is a particular case. Compliance with the FAIR principles is a real challenge that is usually only partially achieved~\cite{albertoni_reproducibility_2023}. These principles aim at categorizing the data in a more extensive and systematic way~\cite{parland-von_essen_supporting_2018}, as a mean to improve research data services. Also, they promote a convenient tripartite categorization of research data artifacts.

Many data science projects and research labs have started to adopt the FAIR principles, but there are discrepancies in how to implement it, from considering how to handle big data and using cloud-native repositories~\cite{9354557}, as well as smaller scale data science projects that require an affordable means of sharing~\cite{article_Vanschoren}. Each team or laboratory ends up establishing its own means to comply with the guiding principles. Therefore, it becomes a challenging task to determine the degree of FAIR compliance and, to a certain extent, audit it.

\subsection{Open data formats and benchmarking}
\label{sec:open_data_benchmark}
There are cases in which it is not viable to publish datasets and codes because they contain sensitive data (for example, medical data corresponding to individuals) or simply because it is under industrial secret. In these cases, the assessment of the reproducibility of the methods is compromised ~\cite{rosenblatt_epistemic_2023}. However, is in developing the concept of Federated Learning ~\cite{baracaldo_towards_2022, zirpins_towards_2022} as a novel paradigm based on decentralized and private data for the shared training of models.

However, these exceptions are scarce and, in general, it is possible to approach open science by the use of open datasets, standardized formats, baselines, and benchmarks~\cite{vitek_repeatability_2011}, allowing the scientific community to check the results published methods reliably.

Even when data can not be made available because of confidentiality reasons, one can rely on benchmarking and comparing results without necessarily accessing the data itself. There are several recent tools for this purpose~\cite{vitek_repeatability_2011}, such as DataPerf, Mlperf, Collective Mind~\cite{Fursin2014CollectiveMT}, ReQuEST~\cite{fursin_invited_2018} or MLCommons \footnote{\url{https://mlcommons.org}} with MLcube\footnote{\url{https://mlcommons.org/en/mlcube/}} among others.  They attempt to determine the state of the art in certain disciplines by comparing the performances. Competitions such as Kaggle or BRATS (brain tumor segmentation)~\cite{kazerooni_brain_2023} challenge and others publish open datasets and have become a reference for the industry to evaluate and compare models.

\subsection{System architecture}
\label{sec:system_architecture}
In terms of architecture for ML/AI systems, one can find two major trends: the deployment of microservices and the use of serverless functions.

Microservices allow the building of scalable and flexible software systems for which each component works independently and can be reused in different contexts. Because many applications of ML/AI require large resources in terms of computations and storage, they are usually deployed as a distributed system. The fact that different modules can work autonomously contributes to the reproducibility and understanding of the system, compared to monolithic ones~\cite{fritzsch_adopting_2023}.

In particular, to reproduce scientific experiments, microservices can help improve the portability and reusability of the software. By dividing the components of an experiment into microservices, one can increase the flexibility and modularity of the software, making it easier to adapt the code for new tests or experiments and lessening the dependency on software specific to a development environment.

Also related to isolation, Serverless Computing is a popular cloud-based computing model~\cite{jonas_cloud_2019} where the cloud provider manages the server infrastructure and platform resources, allowing developers to focus on application logic. Depending on the provider, they can also be referred to as lambda-functions\footnote{Note that, despite their name, they are totally unrelated to lambda calculus!}.

The use of these serverless functions is beneficial for reproducibility in computer science as it reduces the complexity and variability of the underlying infrastructure, and it enables greater modularity and automation when developing applications and services.

\subsection{Tools and platforms}
\label{sec:Tools}
This section surveys tools and platforms that are commonly used in ML/AI applications and how they contribute to reproducibility. Specifically, we focus on containers and cloud computing and the Infrastructure as Code (IaC) technique.

\subsubsection{Notebooks}
\label{sec:notebooks}


In data science, the use of notebooks has been popularized because of the opportunity to incorporate executable code, rich visualization, and documentation in the same document. It has become a common practice to publish and share work and a step forward for reproducibility. However, it has been shown~\cite{pimentel_large-scale_2019} that this approach has some deficiencies, such as the lack of version control. Very recent studies~\cite{samuel2023computational} have also studied the low degree of reproducibility of Jupyter notebooks in biomedical publications.

Several solutions have been proposed to address these challenges ~\cite{pimentel_understanding_2021}, including the use of Python scripts and the adoption of best practices for documentation, version control or additional packages as ReproduceMeGit tool to analyze the reproducibility of ML pipelines in Notebooks~\cite{glavic_reproducemegit_2021} and Osiris~\cite{wang_restoring_2020}.

\subsubsection{Containers and cloud computing}
\label{sec:Containers_and_cloud}
Advances in cloud computing and containerization have undoubtedly contributed to the reproducibility of large distributed systems.

These systems are complex, made of several interacting components~\cite{wolke_reproducible_2016, congo_building_2015} along a pipeline. It is required to have control over the execution environment in order to not only reproduce the experiments but even trust them at all.

Given a code, the associated data, and the execution pipeline, one should be able to obtain the same results over and over again. To achieve this, the pipelines, dependencies of the software, and the environment need to be perfectly defined.
Virtual machines and lightweight containers such as Docker help by defining and fixing the execution environment~\cite{6193081}.
One could summarize these two concepts as
\begin{itemize}
    \item Virtualization = data + code + environment,
    \item Cloudcomputing = data + code + environment + resources + services.
\end{itemize}

We address the topics on lightweight containers such as Docker, the MLOps methodology, the management of scientific workflows, and techniques such as IaC in the following.

\subsubsection{Docker containers}
\label{sec:SW_repro_containers}

Since its appearance in 2007, Docker containers have quickly become popular in computer systems and have become a fundamental tool for reproducibility. Its lightweight nature allows having several containers dedicated to small microservices on the same machine, with limited consumption and sharing of resources. This is a major advantage with respect to full virtual machines such as VMWare or Hyper-V. The light containers eventually allow for better reuse, and many infrastructures are nowadays migrating to containers, e.g. RE3~\cite{bahaidarah_toward_2021}.

Docker is one of the most efficient and widely used tools with applications for reproducibility nowadays, but, at the same time, we identify some of its limitations~\cite{canon_role_2020} to this purpose compared to container alternatives such as for example, Singularity Containers for HPC.

The emergence of containerization technologies such as Docker and orchestrators such as Kubernetes~\cite{krzhizhanovskaya_reproducibility_2020} has allowed the rapid development and automation~\cite{bahaidarah_toward_2021, vasyukov_using_2018} of pipelines of experiments, thus making the reproduction of complex and computationally intensive experiments possible. Indeed, they can be divided into different functional blocks.

We can cite Repo2Docker~\cite{forde_reproducible_2018}, which with Binder, can fetch a notebook for a given repository, create a proper execution environment, and run it inside a container. This makes the code publicly available for anyone to reproduce the results.

Specifically for HPC, there are initiatives such as The Extreme-Scale Scientific Software Stack (E4S), a community effort to provide open-source software packages for developing, deploying, and running scientific applications on High-Performance Computing (HPC) platforms. As an important contribution to the reproducibility of such a complex, E4S builds from source and provides containers of a wide collection of HPC software packages.

Many scientific experiments are made up of pipelines that concatenate several processes~\cite{sugimura_building_2018}. In terms of reproducibility over time, highly specialized platforms have been developed to manage these complex workflow management systems~\cite{steidl_pipeline_2023} (e.g. watchdog ~\cite{kluge_watchdog_2020}), tools~\cite{glavic_machine_2021}, roadmaps~\cite{da_silva_community_2021} and general frameworks are proposed~\cite{franch_model-driven_2022}. They allow researchers can focus on solving their specific scientific problems rather than the underlying infrastructure, networking, or other technical specifics ~\cite{francoise_marcelle_2021}. Despite the great step forward, many interoperability and reproducibility difficulties still persist~\cite{prabhu_reproducible_2020,ghoshal_experiences_2020} considering the immense possibility of languages, open or private infrastructures that are currently available or under development in the ecosystem of ML/AI data science technologies.

\subsubsection{Workflow management systems}
\label{sec:scientific_workflows}
Scientific workflow management systems are useful for managing complex, cloud-distributed workflows \cite{rosendo_kheops_2023}  and automating repetitive processes~\cite{cohen-boulakia_scientific_2017}. They also enable detailed documentation and workflow sharing with other researchers, thus helping improve the reproducibility of results and speeding up scientific research~\cite{plale_reproducibility_2021}.

Formally, the workflows are represented as Direct Acyclic Graphs (DAG)~\cite{santana-perez_towards_2015}, where a task starts in a particular node to be processed, and then transferred to the next one in the chain until the final result is available in the last node. As pointed out in Section~\ref{sec:repro_priciples}, it is required that the pipeline of the workshop and the node themselves follow well-established reproducibility principles to obtain reliable results, including access to the source code running the computations, along with an accurate description of the environment, the use of FAIR data, and the use of open data formats for interoperability, among others.

Each scientific community has developed its own workflow managers. We can cite some well-known ones, such as Taverna (bioinformatics, cheminformatics, and ever social sciences) or the Galaxy project (bioinformatics), OpenAlea (Botanics), Chimera (cheminformatics), or Pegasus (physics and bioinformatics),  Knime(semantic workflow), Wings(graphical workbench). Pegasus was the workflow management system used by LIGO for the first detection of gravitational waves, certainly a paramount hit in physics.

The criteria to establish the reproducibility of a given pipeline can vary much from community to community. Although the basic principles remain (see Section~\ref{sec:repro_priciples}), there are specificities depending on the field. We invite the reader to check the work of Cohen-Boulakia and co-authors, who conducted a study~\cite{cohen-boulakia_scientific_2017} analyzing three cases of use of in-silico experiments in the domain of biological sciences with Taverna, Galaxy, OpenAlea, VisTrails, and Nextflow, proposing different criteria and discussing about these reproducible environments based on docker, Vagrant, Conda, and ReproZip.

The significant increase in articles on AI/ML inevitably forces an adaptation towards the management of both data and software because both are a source and contribution to knowledge. Therefore it is necessary to analyze tools, infrastructures, and technologies that have evolved to support these requirements. In this sense, AIOps/MLops evolves from the DevOps/DevSecOps (Development - Operations) concept to cover several of these aspects of reproducibility management infrastructures for computer-based scientific articles.

Transferring knowledge and prototypes from the academy to the industry is, most of the time, challenging~\cite{8258038}.
There are very well-specified methodologies for the development of software in the industry, such as DevOps, which include CI/CD (Continuous Integration/Continuous Delivery). However, in the academic environment, these practices are not necessarily followed. In part, this is explained by the lack of career reward pointed out in Section~\ref{sec:side_difficulties}.

MLOps~\cite{gift_practical_2021} can be considered the natural evolution of the DevOps best practices components adapted to the particular needs of ML-based software development ~\cite{8804457}. 
In general terms, within data science projects MLOps tries to harmonize the practices of two environments with very different characteristics, such as academic/research environments with ML production environments for a final client, where reproducibility plays a very important role. It is an end-to-end process from the research model to the final model, exploited by the end customer or reproducibility reviewer.

There are few works that deal with MLops from the point of view of reproducibility. Among these,~\cite{gundersen_machine_2022} does an excellent analysis of the reproducibility of various MLops tools.

Other articles made a benchmark for different MLops features~\cite{schlegel_management_2022} and products available in the open source such as private code~\cite{bibliograph3}, which is equally important when data and software management is required by a journal. Let us mention here the most relevant ones, from our review of the literature:

\begin{small}
\begin{itemize}
    \item Neptune. A metadata store for any MLOps workflow. It was built for both research and production teams that run a lot of experiments~\footnote{\url{https://neptune.ai/}}.

    \item Weights\&Biases. A machine learning platform built for experiment tracking, dataset versioning, and model management~\footnote{\url{https://wandb.ai/}}.

    \item Comet. An ML platform that helps data scientists track, compare, explain and optimize experiments and models across the model’s entire lifecycle~\footnote{\url{https://www.comet.com/}}.

    \item Sacred + Omniboard. Open-source software that allows machine learning researchers to configure, organize, log, and reproduce experiments~\footnote{\url{https://github.com/IDSIA/sacred}}.

    \item Tensorboard. A visualization toolkit for TensorFlow~\footnote{\url{https://www.tensorflow.org/tensorboard}}.

    \item Polyaxon. A platform for reproducible and scalable machine learning and deep learning applications~\footnote{\url{https://polyaxon.com/}}.

    \item ClearML. An open-source platform, a suite of tools to streamline your ML workflow~\footnote{\url{https://clear.ml/}}.

    \item Pachyderm. An  enterprise-grade, open-source data science platform that makes it possible for its users to control an end-to-end machine learning cycle~\footnote{\url{https://www.pachyderm.com/}}.

    \item MLflow. An open-source platform that helps manage the whole machine learning lifecycle. This includes experimentation but also model storage, reproducibility, and deployment~\footnote{\url{https://mlflow.org/}}.

    \item DVC (Data Version Control). It is a very popular tool in MLops and in the data science environment because it allows the versioning of training, testing and validation datasets in a very simple format~\footnote{\url{https://dvc.org/}}.

    \item NextFlow. In terms of reproducibility, it allows Docker and Singularity containers technology for the creation of workflows.~\footnote{\url{https://www.nextflow.io/}}

    \item Collective Knowledge~\cite{fursin_collective_2020} is an initiative that, based on its experience trying to reproduce hundreds of experiments, came to identify common patterns that are repeated from project to project. In a certain sense, it is a unifying proposal within the wide variety of existing MLOps solutions and seeks to resolve persistent integration issues.
\end{itemize}
\end{small}

With the emerging Internet of Things (IoT) technology and the advances in smaller devices with significant computing power, simplified ML models at the edge are possible with TinyMLOps~\cite{ray_review_2022}. Significant reproducibility challenges appear considering the strong restrictions of energy consumption, limited computing capacity, and heterogeneity between different devices and technologies. Also considering that you can no longer containerize/virtualize with Docker.

\subsubsection{Workflow languages}
Despite the efforts to unify existing workflows, each community has kept its own particularities, including the language to define the pipelines~\cite{cohen-boulakia_scientific_2017}.
This fragmentation~\cite{adams_yawl_2020} makes it harder for integration and interoperability between different academic groups. Indeed, some of the groups use a very particular language for their workflows.

There are initiatives such as SHIWA (SHaring Interoperable Workflows for Large-Scale Scientific Simulations on Available DCIs)~\cite{article_Korkhov} which try to provide a solution to this problem of interoperability. Multiple organizations and providers of workflow systems have also jointly worked to propose the Common Workflow Language (CWL) \cite{demchenko_experimental_2023} with the aim of standardizing the pipelines around a common language.

Those specifications propose a conceptual workflow language to describe high-level scientific tasks, with the aim of promoting workflow specification portability and reusability and addressing the heterogeneity of workflow languages.

\subsubsection{Infrastructure as code (IaC)}
\label{sec:IaC}
Much attention is paid to source code and containerization in order to address reproducibility, but unfortunately, not that much to hardware~\cite{bowman_improving_2023}. With the rise of cloud computing technologies, the possibility of replicating the exact execution environment for an experiment is viable. Indeed, for reproducibility purposes, it is a requirement to define the characteristics of the hardware, such as the type of CPU, TPU, GPU, memory amount, or network architecture. This is especially important for a large distributed system as, for example, HPC applications.

In this respect, IaC provides several advantages towards reproducibility in computer science.
One of the main benefits is that IaC allows researchers to accurately define and control their infrastructure in a format that can be easily stored, versioned, and shared, making it easy to reproduce experiments and obtain the same results at each execution. Defining infrastructure as code discharges from manually configuring infrastructure resources and allows researchers to easily version and share the infrastructure configuration with other colleagues.
According to the Octave 2022 report~\cite{bibliograph7}, the Hashicorp Configuration Language (HCL) programming Terraform languages were widely used by developers in 2022, indicating that IaC practices are becoming quite popular for Github projects.

Additionally, IaC can help improve consistency and accuracy by ensuring that all instances of the infrastructure are created and configured identically. This helps ensure that the test conditions are the same each time an experiment is performed. IaC in the academic environment can significantly help in many aspects, such as the quality of the software developed, and is a step forward in the reproducibility of scientific research. As a recent example, Adorno-Gomes and Serodio~\cite{bibliograph1} managed to define a complete experiment with IaC from a unique high-level code with Pulumi~\cite{bibliograph2}.

\subsubsection{Provenance and Metadata Traceability of Artifacts}


Provenance refers to the way in which the origin~\cite{silva_junior_provenance-and_2021} of the artifacts of an experiment is documented in metadata.
Provenance documentation is a commonly used technique to improve the reproducibility of scientific workflows and research artifacts. There are numerous articles proposing tools such as ProvStore~\cite{ludascher_provstore_2015}, ReproZip~\cite{inproceedings}, MERIT~\cite{wonsil_integrated_2023}, CAESAR~\cite{samuel_collaborative_2022}, Provbook~\cite{glavic_machine_2021} in several different disciplines and research areas~\cite{samuel_end--end_2022}, demonstrating how it can help improve traceability, linage and transparency of results.

The PROV standards allow the task to be carried out (see Openprovenance\footnote{\url{https://openprovenance.org/store/})}, for example. However, it is not yet complete and does not allow it to be generalized to multiple cases and languages. The foregoing requires the use of permanent, Unique Identifiers and tools that manage this aspect in order to have correct traceability of data sources and artifacts, even using new technologies such as blockchain~\cite{wittek_blockchain-based_2021} InterPlanetary File System (IPFS)~\cite{kawamoto_ai_2020} to achieve traceability and lineage of Software or code snippets.

\subsubsection{Reproducibility as a Service (RaaS)}
\begin{figure*}
    \centering
    \includegraphics[width=0.90\textwidth]{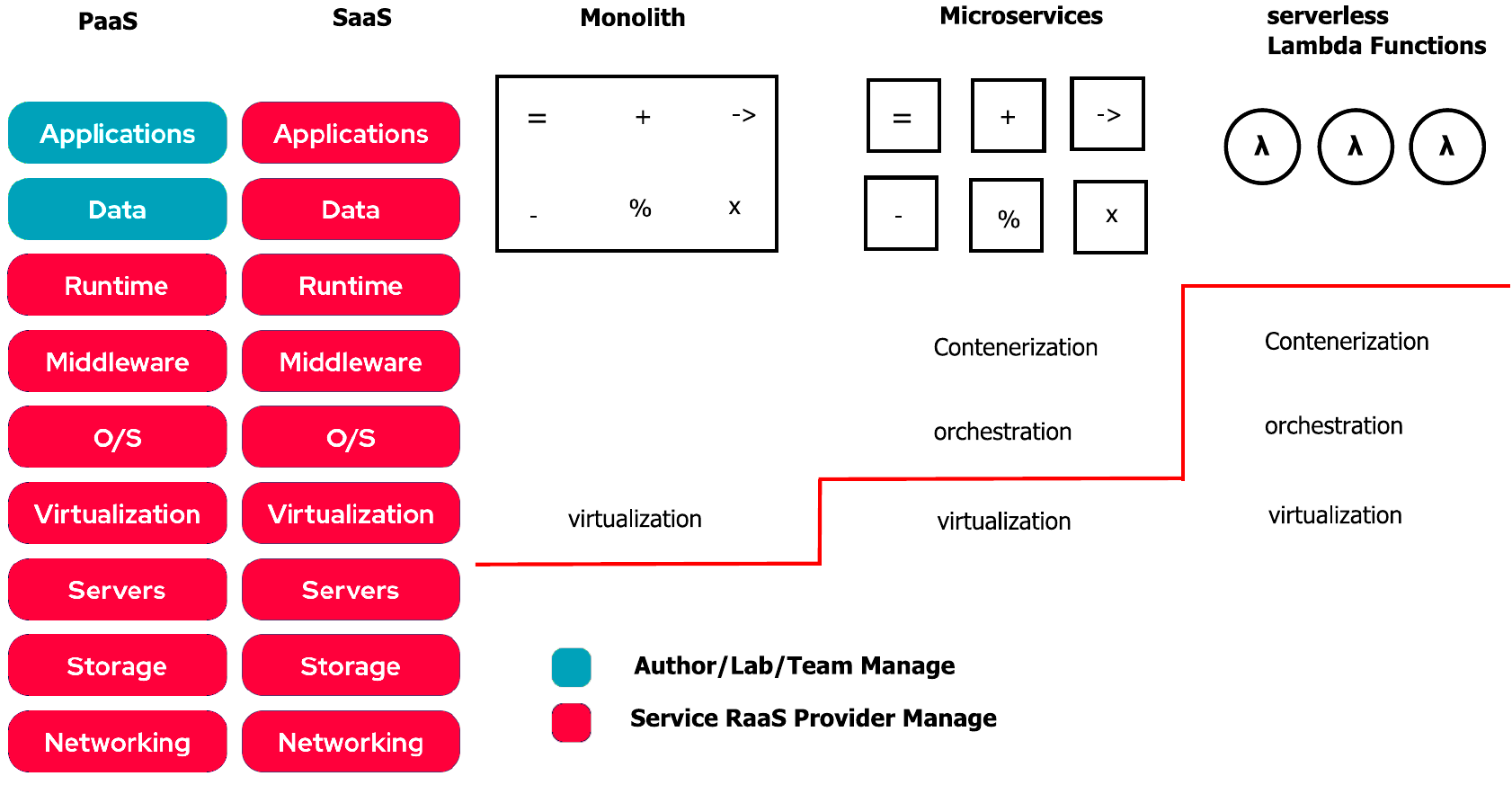}
    \caption{Representation of a RaaS-managed cloud infrastructure. The description layers, microservices, the serverless approach, and taking care of the granularity of the software help the reproducibility of a complex system.}
    \label{fig:serv1}
\end{figure*}

The \textit{Reproducibility as a Service} (RaaS) concept was proposed in 2021 by Wolsin~\cite{wonsil_reproducibility_2021}. An strategy based on RaaS takes advantage of the availability of cloud computing technology to offer reproducibility services. This include the reproduction and research artifacts after the execution of the software in the controlled environment and its evaluation, validation, and certification (related to this, see Section~\ref{sec:Peer_Code_Reviews} about code review). Also, granting reproducibility badges, tracking the provenance of software, or assigning persistent identifiers to the software at different granularity levels. Another responsibility of RaaS is to manage the underlying architecture if a way that makes it easier for authors to share and execute their code depending on the chosen complexity, from baremetal infrastructure to fully managed services. Figure~\ref{fig:serv1} shows how a SaaS architecture is organized in a complex system.

Crick and co-authors proposed~\cite{crick_reproducibility_2015} to make a first approach to the offer of reproducibility services for journals/conferences from an empirical and quantitative point of view. They presented a \textit{cyber-infrastructure} and the associated workflow for a reproducibility service as a high-level technical specification without delving into technical details. On the other hand, the work of Demchenko~\cite{demchenko_experimental_2023} addresses the topic of provisioning on demand of research environments and introduces the concept of Platform Research Infrastructure as a Service (PRIaaS) with the aim to ensure data quality and support effective data sharing.

For example, the IPOL journal~\cite{colom_ipol_2015} also partially meets the attributes of what can be considered as a RaaS tool, together with the article, makes available a technological platform for the creation and execution of online demos (simplified demonstrations of algorithms).

Equally, among other existing reference platforms, we could mention CodeOcean, Chameleon, and Whole Tale. They allow code to be executed in a wide range of languages, but they are still maintained at the demo level with certain technical restrictions to offer the mentioned features. They start to be actively taken into account by publishers.

\subsection{Good practices and data management methodologies}
\label{sec:data_management}
Agility and security are among the many quality attributes of software~\cite{milewicz_towards_2023}, even though the majority of them are not specifically designed for reproducibility in computer science.
However, data project management methodologies and well-known best practice guides are applied widely across the AI/ML industry to improve reproducibility.

Several studies~\cite{Schelter2015} and best practices guides ~\cite{stodden_best_2014} have proposed different tools for the management of data science project artifacts~\cite{schlegel_management_2022} as well as methodologies. For example, Goodman et al. propose ten simple rules to achieve reproducibility, and \textit{The Turing Way} handbook also provides a relevant compilation of good practices~\cite{community_illustrations_2022} 
 to reproducible, ethical, and collaborative data science projects.

There is a consensus that one of the main factors limiting the success of data science projects is the lack of reproducibility in the management platforms~\cite{baillieul_reflections_2018, alnoamany_towards_2018, martinez_survey_2021}.
From the many methodologies available, the most popular are CRISP-DM~\cite{article2}, KDD, SEMMA, Microsoft TDSP, Agile DS Lifecycle, Domino, DS Lifecycle, IBM FMDS, RAMSYS, and MIDST, among others. These are widely used in the industry, especially CRISP-DM. 
  
Certainly, as observed in the scarce literature, there is not a standard or unified methodology that is focused on reproducibility for data management itself. So far, only good practices, recommendations~\cite{turkyilmaz-van_der_velden_reproducibility_2020, merz_editorial_2020, samuel_understanding_2021, nichols_better_2021}, and guides from different fields of computer science of different needs are available.

\subsection{Scientific publishers and reproducible research}
\label{sec:scientific_publishers}
Historically one of the main forms of communication, recognition, socialization, and validation before the scientific community are the articles published in journals and conferences~\cite{stodden_empirical_2018}.
Publishers \textit{de facto} become auditors of the scientific activity, and indeed, the metrics (Impact Factor and others) that they have established are the typical indicators that are used to evaluate researchers in their career and their advancement. Publishers have, therefore, a responsibility to assure the scientific integrity of the work they make public, along with their own interest in maintaining their own reputation. This includes not only avoiding fraud but also establishing clear quality criteria. In scientific publications, reproducibility is fundamental since it allows others to verify if the same or equivalent results are obtained when repeating the experiment, thus allowing them to potentially refuse a paper containing wrong or inaccurate claims.
As pointed out by Heesen~\cite{Heesen2017-HEECAT}, \textit{the work that is not widely shared is not really scientific work}.

In the following, we discuss two significant initiatives which have been put into practice by publishers: the possibility of associating code with the publications and the proper evaluation of software artifacts.

\subsection{Publications with code }
\label{sec:Publications_with_code}
Associating source code with a particular publication is gaining great popularity in the scientific and technical community~\cite{bonsignorio_new_2017}. It allows for greater transparency and reproducibility, essential to guarantee the quality and reliability of the results~\cite{de_sterck_enhancing_2023}. However, the reproducibility aspects of this practice are evaluated in the Dataverse repository ~\cite{trisovic_advancing_2020}, Figure~\ref{fig:process} describes a typical editorial structure for publishing articles with code, which, as will be seen later, is not so easy to implement by the Journals considering technical and economic aspects.

Many conferences have started to request that the source code be given and made public. Others go one step further and perform an exhaustive evaluation of the artifacts. For example, Checklist NeuroIPS\cite{pineau_improving_2020}: It is a widely recognized checklist for the reproducibility assessment of conference papers.

From many examples, one might include here \textbf{Code Ocean}, used by IEEE's publishers after the integration of the CodeOcean's platform as a Computational Research Platform, \textbf{Whole Tale}~\cite{foster_toward_2020} allowing researchers to create and share scientific narratives that include data, code, and runtime environments~\cite{brinckman_computing_2019, chard_implementing_2019}, \textbf{Binder} as a platform that allows users to create and share code execution environments online, making it easy to reproduce and distribute results, \textbf{PapersWithCode} with open resources on ML, 
 \textbf{ReproducedPapers} with open teaching and structuring machine learning reproducibility~\cite{yildiz_reproducedpapersorg_2021}, or the \textbf{ReScience Journal}~\cite{kerautret_rescience_2019} which replicates computations from independent open-source implementations of original research and the advanced Chameleon\footnote{\url{https://www.chameleoncloud.org/}}  large-scale edge to cloud tool~\cite{inbook}.

Unfortunately in many cases, this is limited to providing a non-persistent link~\cite{salsabil_study_2022,corcho_documenting_2022} to the source code repository in public platforms (see Section~\ref{sec:Open_repositories}). Moreover, each journal sets its own strict criteria, formats, and procedures for authors. Aspects such as consistency, reproducibility, and reusability cannot be properly tracked or audited by other teams and research over time, thus limiting their impact~\cite{raff_siren_2022}.

\subsection{Review of Research Artifacts}
\label{sec:Peer_Code_Reviews}
To begin with, it must be understood that for different reasons~\cite{gomes_why_2022} an article is not 100\% reproducible, but rather certain elements (e.g. computational artifacts, pseudocode, algorithms, demos) that the author decides to share and considers sufficient grounds to legitimize his results.

The evaluation criteria to accept articles for publication is traditionally well defined for scientific journals. They are based typically on originality, novelty, or overall scientific interest. However, when considering a publication not only as the article but also all major research artifacts, including source code, the criteria is relaxed, if considered at all.
When the evaluation takes into account the associated source code, it is required to establish the proper evaluation criteria for peer review ~\cite{noauthor_supporting_2021}. 

Conferences have started to publish guides containing checklists for the evaluation of artifacts and to grant the so-called  \textit{reproducibility badges} ~\cite{frery_badging_2020,athanassoulis_artifacts_2022} if the conditions are met. Among the most important conferences, we can cite the checklist of NeurIPS 2019 \footnote{\url{https://nips.cc/Conferences/2019/CallForPapers}}, the ACM reproducibility badges
\footnote{\url{https://www.acm.org/publications/policies/artifact-review-badging}}, as well as other initiatives such as the Unified Artifact Appendix and the Reproducibility Checklist \footnote{\url{https://ctuning.org/ae/checklist.html}}, the CTuning artifact evaluate \footnote{\url{https://ctuning.org/ae/reviewing.html}} or the Empirical Evaluation Guidelines SIGPLAN NISO RP-31-2021
\footnote{\url{https://www.sigplan.org/Resources/EmpiricalEvaluation/}}, among others.

Following several of the published guides, recently, the SC23 supercomputing conference (one of the most important conferences in HPC) ~\cite{plale_reproducibility_2021} adopted the Reproducibility Initiative where \textit{accepted papers with available artifacts} were acknowledged with the corresponding ACM badges. The use of blockchain technology for artifact traceability has also been proposed ~\cite{radha_verifiable_2021,kawamoto_ai_2020}.

CTuning has participated in the artifact evaluation task for different ACM conferences~\cite{fursin_enabling_2020} and has defined a more detailed Unified Artifact Appendix and the Reproducibility Checklist based on the previous evaluation experience in ACM ASPLOS, MLSys, MICRO, and SCC'23 conferences.

Other specialized scientific journals have already implemented specific criteria to a greater or lesser degree. For example, Table~\ref{demo-table1} in the Appendix summarizes the checklist for Artifacts Description/Artifacts Evaluation (AD/AE) reproducibility~\cite{fostiropoulos_reproducibility_2023, malik_artifact_2020} for a data science experiments and projects of different publishers.

Finally, reproducibility-certifying agencies are starting to offer their evaluation as a service in different disciplines working with sensitive or confidential data, outsourcing this function as a trusted third party. Recently Cascad~\cite{perignon_certify_2019} has been proposed in the field of Economics and Management~\cite{radha_verifiable_2021}.

From our review of the data above, we observe that the existing criteria are still quite varied, not standardized, complex for the authors to fulfill, and time-consuming on the reviewer's side.

Table~\ref{demo-table} is an extensive summary of the reproducibility strategies and technologies that have been reviewed in this work. However, it needs to be analyzed how they are implemented according to the reproducibility policies of the different scientific journals. Our survey tries, from an empirical point of view, to provide insights on the application of these strategies directly from participating journals.

\section{Survey in Computer Science Journals}
\label{sec:survey}
The increasing number of published articles in computer sciences, as well as and the fast development of new and innovative AI/ML methods, pose several challenges to publishers. The reproducibility, legitimacy of the works, and adapting the policy of the journal and the procedures of evaluation of the works is challenging~\cite{cousijn_data_2018}.

Several papers~\cite{seibold_computational_2021} have explored the effectiveness of journal policies regarding open source code and data sharing to validate the research procedures in an attempt to mitigate the \textit{reproducibility crisis}.
Likewise, other studies address solutions, platforms, technologies, mechanisms, and procedures for the reproducibility of scientific articles that have been proposed to deal with the problem from the perspectives of the different actors involved: authors, publishers, industry, and scientific community. For example, the work of Gomes et al.~\cite{gomes_why_2022}, as well as Baker et al.~\cite{baker_why_2016}, focus on the barriers why authors might reluctant to share code and data in their publications and why that would be pertinent.

From another point of view, the \textit{reproducibility culture} has also been analyzed in previous works~\cite{karathanasis_reproducibility_2022,mauerer_beyond_2022, hofman_expanding_2020, fund_we_2023, lin_building_2022}, and along with the culture, it has been discussed how to teach reproducibility in academic environments to young students and as seen in previous sections incentives ~\cite{edwards_academic_2017}, Massive Open Online Courses (MOOC)\footnote{\url{https://www.fun-mooc.fr/en/courses/reproducible-research-methodological-principles-transparent-scie/}}\footnote{\url{https://www.coursera.org/learn/reproducible-research}}, good practices in the way of measuring and rewarding reproducibility, such as novel badging mechanisms, new measurement indices of valuation and code/data citation \footnote{\url{https://datacite.org/}}  ~\cite{parsons_history_2019}, the Scientist Impact Factor (SIF) for reputation and impact of researchers~\cite{ATM15375}, or applying statistical methods a mean to measure impact~\cite{dozmorov_github_2018}. However, all these elements have not been analyzed as a whole as part of an articulated, agreed-upon \textit{Reproducibility policy} in publishers.

Moreover, the experience, opinions, and results of journals implementing and adapting their policies with a strong focus on reproducibility have not yet been surveyed. Therefore, we surveyed SCOPUS-indexed journals specialized in computer science to know from them as a primary source of information about their experience in the application of reproducibility policies, insights, and the difficulties and successes derived from their policies.

In subsection~\ref{sec:survey_related_work} we explore which aspects were previously surveyed by other authors, including especially relevant questions. Subsection~ \ref{sec:survey_results} presents our survey, along with the answers, which eventually leverages the discussion at Section~\ref{sec:discussion}.

\subsection{Previous work}
\label{sec:survey_related_work}
From the existing literature one can conclude that there is still incipient and timid progress toward implementing sharing and open science policies in scientific works~\cite{stodden_toward_2013, stodden_empirical_2018}. The traditional peer review scheme is maintained, with slight variations, and it is, in general, limited to encouraging the publication of the source code and data in software repositories~\cite{lewis_policy_nodate, noauthor_digital_nodate}

For example, The Diamond OA Journals Study~\cite{bosman_oa_2021} makes a general survey; in our case, the results of its question 41 are highlighted. To the question ``\textit{Do you have any policy or practice to stimulate open sharing of research data?}" 42\% of the respondents declared to have policy or practice to stimulate open sharing of research data. The study finds that an equal number of respondents who did not have an established policy and an additional 15\% answered ``\textit{Unknown}". However, the factors that explain the adoption of
open-data policies are not analyzed and it only focuses on other aspects of the publishing business.

Question 54 asked ``\textit{Does the journal require linking to data, code, and other
research results?}". Although there is not much information available from journals about requiring links to data, code, and other research outputs in DOAJ, from the survey data the study found that nearly half of respondents reported not requiring this, against 24.8\% who do. For more than 25\% the answer was ``\textit{No}" or ``\textit{Unknown}".

The above questions are certainly limited to code-sharing policies in journals, but do not delve into actual reproducibility policies through article automation, evaluation, and preservation of reproducibility technologies. This represents a dilemma that is discussed in Section~\ref{sec:dilemma_policies}.

In the article~\cite{vasilevsky_reproducible_2017}, 318 biomedical journals were manually reviewed to analyze the journal’s data sharing requirements and characteristics
A total of 11.9\% of journals analyzed explicitly stated that data share a total of 11.9\% of journals analyzed explicitly stated that data sharing was required as a condition of publication. A total of 9.1\% of journals required data sharing but did not state that it would affect publication decisions. 23.3\% of journals had a statement encouraging authors to share their data but did not require it. A total of 9.1\% of journals mentioned data sharing indirectly, and only 14.8\% addressed protein, proteomic, or genomic data sharing. There was no mention of data sharing in 31.8\% of journals. Impact factors were significantly higher for journals with the strongest data-sharing policies compared to all other data-sharing criteria. Open-access journals were not more likely to require data sharing than subscription journals.

Another contribution by Konkol et al. from the point of view of the analysis of reproducibility technologies for Publishing computational research~\cite{konkol_publishing_2020} concludes that still, publishing reproducible articles is a demanding task and not achieved simply by providing access to code scripts and data files. Several platforms were analyzed, including Whole Tale, ReproZip, REANA, o2r, Manuscripts, Gigantum, Galaxy, eLife RDS, Code Ocean, Binder and its limitations as well, the facilities it offers for authors. The previous article is complemented by the work of Willis~\cite{willis_trust_2020}, who made an analysis of technical aspects and use of some of these technological infrastructures and repositories around seven reproducibility initiatives designed by journals to improve computational reproducibility.

In the work of Malik~\cite{malik_artifact_2020}, the technical difficulties are discussed, but also the benefits of implementing Artifact Description and evaluation policies for presenting scientific articles to journals and conferences.

The percentages of implementation of concrete reproducibility policies remain low. However, there is ongoing open discussion on the efforts and contributions that can be made by each of the actors in the reproducibility research ecosystem. In our case, this work analyzes the problem from the point of view of practical implementation of policies by publishers, based on your opinion and experiences with the following research questions: \textit{What is the best way reproducibility policy mandatory, or instead an incentive policy for authors and reviewers, to allow publishers improve the quality and impact of their publications?} \textit{What type of technological infrastructure best supports these types of reproducibility policies?}

\subsection{Survey and Results}
\label{sec:survey_results}
Following the methodology described in section~\ref{sec:survey_methodology} and considering the literature mentioned above and the technologies discussed in Section~\ref{sec:Tools}, a series of reproducibility-oriented questions were carefully designed for the evaluation of reproducibility policies and their implementation. This is the base of the brief gap analysis we do in Section~\ref{sec:gap_analysis}, where the answers are analyzed and discussed.

In the following we present the questions of the survey and the results.

\textbf{Question 1. Do you want to be mentioned in the acknowledgment section as a Survey participant?
}
\begin{figure}
 \centering
  \includegraphics[width=0.8\linewidth]{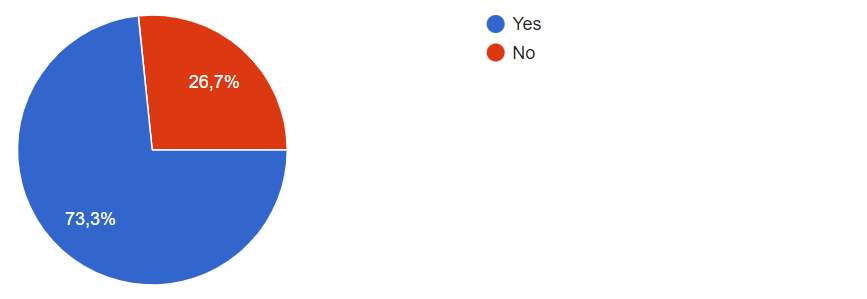}
  \caption{Do you want to be mentioned in the acknowledgment section as a Survey participant? }
  \label{fig:question}
\end{figure}

Despite having a policy of sharing and publishing code and data implemented at some level, some publishers refrained from being mentioned, probably due to not being able to match several items. Indeed, the survey asked for very specific questions about the implementation of infrastructures and technical details. Some publishers requested to be considered as anonymous in this question. Figure~\ref{fig:question} shows the results.

\textbf{Question 2. Respondent's role in the scientific journal}

\begin{figure}
 \centering
  \includegraphics[width=0.8\linewidth]{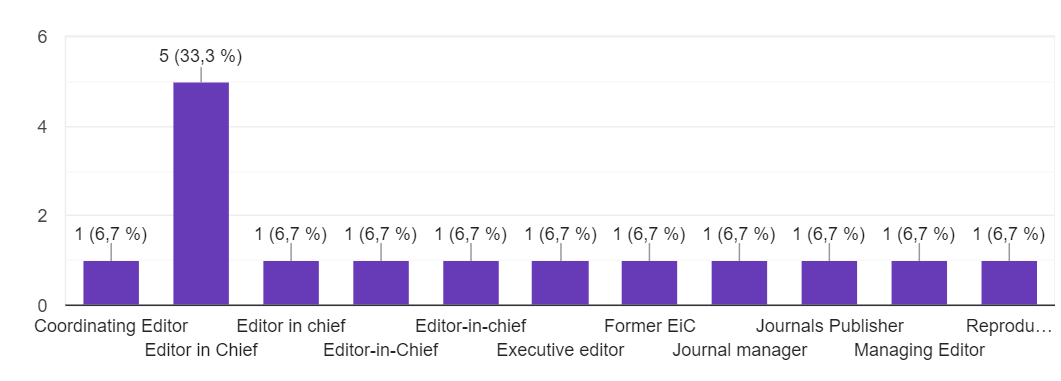}
  \caption{Respondent's role in the scientific journal.}
  \label{fig:question2}
\end{figure}

The answers came from a variety of different roles, with a slight predominance of Editors in Chief, Figure~\ref{fig:question2}.

\textbf{Question 3. Do you have a Reproducibility policy or similar in your guidelines for authors?}

\begin{figure}
 \centering
  \includegraphics[width=0.8\linewidth]{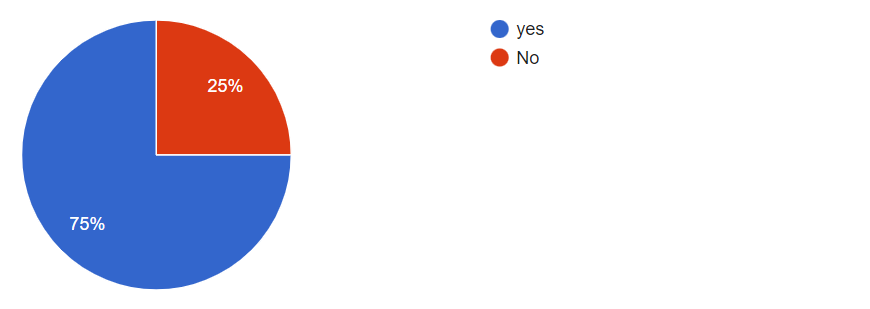}
  \caption{Do you have a Reproducibility policy or similar in your guidelines for authors? }
  \label{fig:question3}
\end{figure}

A large majority (75 \%) of the respondents indicate that they do have a reproducibility policy, as shows in Figure~\ref{fig:question3}.

\begin{figure}
 \centering
  \includegraphics[width=0.8\linewidth]{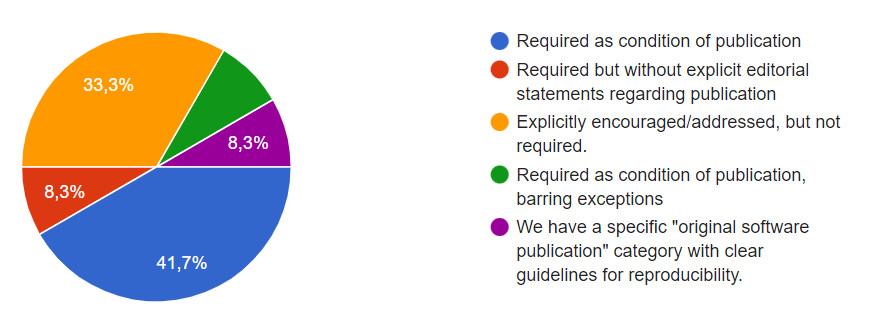}
  \caption{ How do you think the reproducibility policy requirements should be?}
  \label{fig:question3_1}
\end{figure}

Related to this, in Figure~\ref{fig:question3_1} we can observe that there is a significant percentage (41.7\%) of journals which request reproducibility as an essential condition for publication and thus make it mandatory. This decision has important consequences and, in general, it is counterproductive for almost all journals to add extra requirements for the publication because it reduced the publication rate\footnote{See \url{https://scholarlykitchen.sspnet.org/2018/09/25/does-adopting-a-strict-data-sharing-policy-affect-submissions/}}). On the other hand, it improves the overall quality of the publications. 

\textbf{Question 4. If you wish, you can indicate the link to the policy of the scientific journal or guides for authors.}

\begin{table}[htbp]
\centering
\begin{tabular}{p{4cm}|p{8cm}}
\toprule
\textbf{Journal} & \textbf{Policy Link} \\
\midrule
IPOL & \url{https://tools.ipol.im/wiki/ref/software_guidelines/} \\
\midrule
ACM Transactions on Graphics & \url{https://www.replicabilitystamp.org} \\
\midrule
GigaScience & \url{https://academic.oup.com/gigascience/pages/editorial_policies_and_reporting_standards?login=false#Reporting%20Standards} \\
\midrule
Anonymous & \url{https://www.springer.com/journal/12532/submission-guidelines#Instructions%20for%20Authors_MPC%20Reviewing%20Guidelines} \\
\midrule
Optical Memory and Neural Networks (Information Optics) & \url{https://www.pleiades.online/en/journal/optmem/authors-instructions/} \\
\midrule
Information Systems (Elsevier) & \url{https://www.elsevier.com/journals/information-systems/0306-4379/guide-for-authors} \\
& \url{http://doi.org/10.13140/RG.2.2.34277.22243/1} \\
\midrule
Science of Computer Programming & \url{https://www.journals.elsevier.com/science-of-computer-programming/call-for-software/a-new-software-track-on-original-software-publications-science-of-computer-programming} \\
\midrule
INFORMS Journal on Computing & \url{https://pubsonline.informs.org/page/ijoc/datapolicy}; \\
& \url{https://pubsonline.informs.org/page/ijoc/softwarepolicy} \\
\midrule
JETAI & \url{https://authorservices.taylorandfrancis.com/data-sharing-policies/open-data/} \\
\bottomrule
\end{tabular}
\caption{The nine journals who answered question \#4 about their policy, and the links they provided.}
\label{tbl:polLinks}
\end{table}

Nine journals provided a link to their reproducibility policy, Table~\ref{tbl:polLinks}.

\textbf{Question 5. How do you think the reproducibility policy requirements should be?}

\begin{figure}
 \centering
  \includegraphics[width=0.9\linewidth]{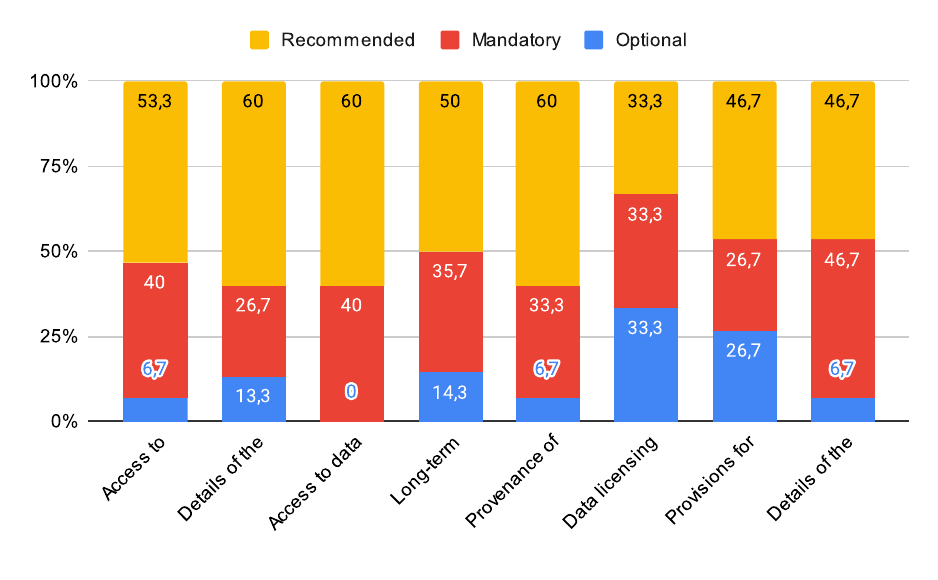}
  \caption{How do you think the reproducibility policy requirements should be?}
  \label{fig:question5}
\end{figure}

In this question we asked about what should be the most significant requirements for a reproducibility policy, regardless whether the journal actually implemented them or not. The results are given in Figure~\ref{fig:question5}, with a variety of different preferences and showing, in any case, gradual interest towards making them mandatory.

\textbf{Question 6. Do you follow any guide or checklist for the evaluation of research artifacts? If so, which one?}

\begin{figure}
 \centering
  \includegraphics[width=0.8\linewidth]{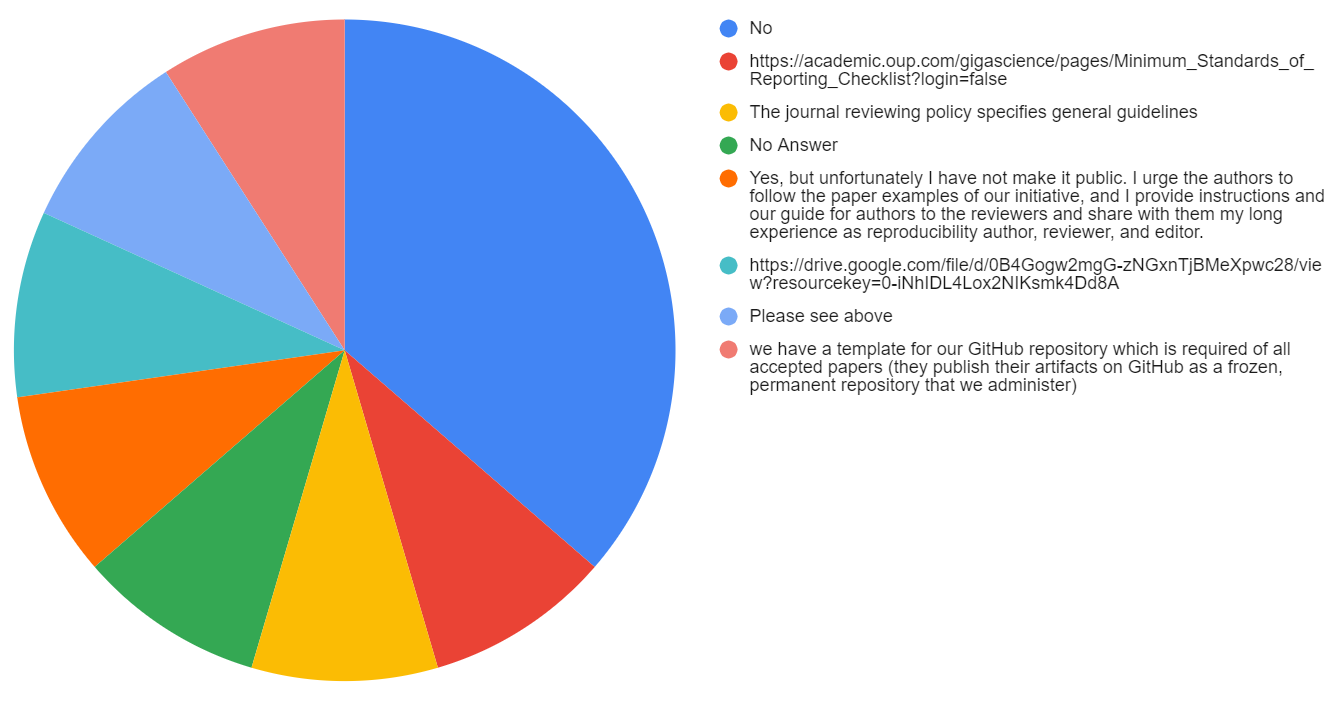}
  \caption{Do you follow any guide or checklist for the evaluation of research artifacts? If so, which one?}
  \label{fig:question6}
\end{figure}

The responses were very varied, which shows the lack of standardization in this matter. The problem of the evaluation fo the research artifacts has been extensively studied, yet without much agreement or formalization. See Figure~\ref{fig:question6}.

\textbf{Question 7. Journal access modality}

\begin{figure}
 \centering
  \includegraphics[width=0.8\linewidth]{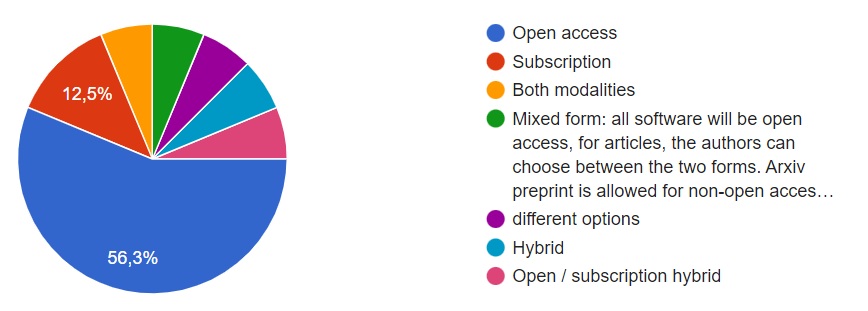}
  \caption{Journal access modality.}
  \label{fig:question7}
\end{figure}

Most of the journals answered that their publication modality was open-access, Figure~\ref{fig:question7}.

\textbf{Question 8. What is the range of your APC (Article Publication Charges)?}

The ACP are very relevant for the discussion about how the reproducibility costs are shared between authors, publishers, and technology providers (see Section~\ref{sec:authors_efforts}). Free publication costs predominate in the responses. In addition to question \#7, it is an indicator that the business model of these journals is based on open platforms and repositories.

\begin{figure}
 \centering
  \includegraphics[width=0.8\linewidth]{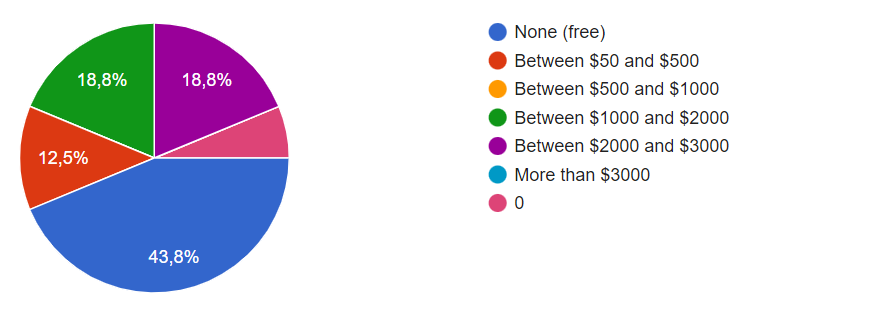}
  \caption{Which is the range of your APC (Article Publication Charges)?}
  \label{fig:question8}
\end{figure}

\textbf{Question 9. Preferred sharing method}

\begin{figure}
 \centering
  \includegraphics[width=0.8\linewidth]{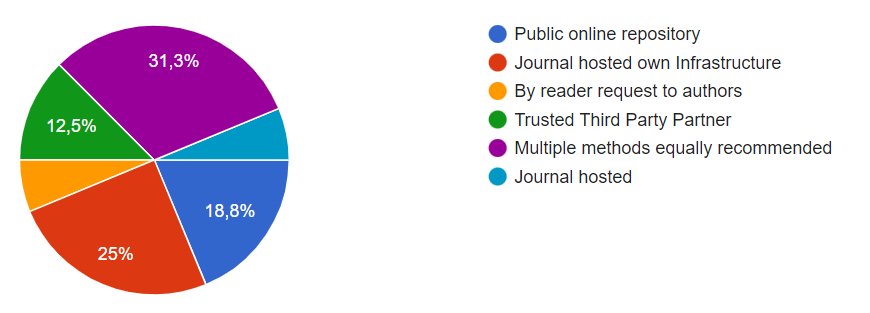}
  \caption{Preferred method to share research artifacts.}
  \label{fig:question9}
\end{figure}

This question confirms that free open platforms are used to share the code, and the number of journals that owe third parties, or have their own technological storage infrastructure, is very low. See Figure~\ref{fig:question9} for the results.

\textbf{Question 10. How compliant is your publication of software and data policy with FAIR-TLC?}

\begin{figure}
 \centering
  \includegraphics[width=0.8\linewidth]{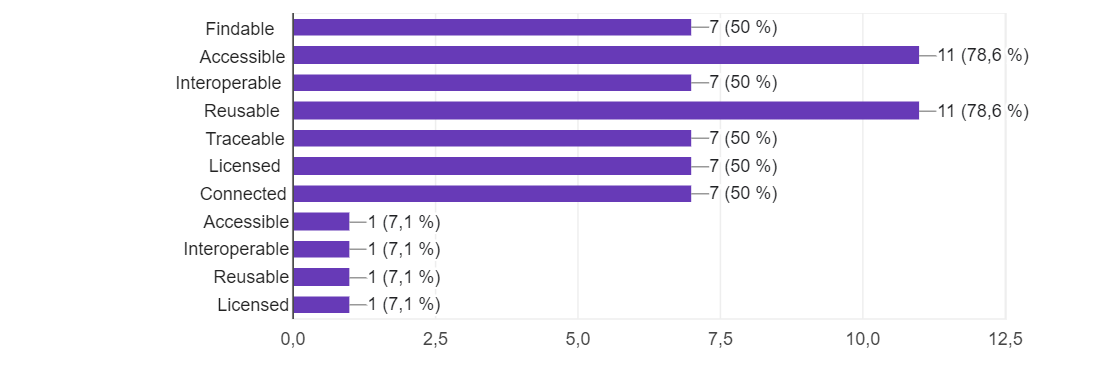}
  \caption{How compliant is your publication of software and data policy with FAIR-TLC? }
  \label{fig:question10}
\end{figure}

The answers indicate the increasing level of implementation of the reproducibility policies, considering that most of the articles are accessible and reusable, but still low in the other attributes. This could be explained because the use of open repositories limits the journals to offer the other attributes satisfactorily. Figure~\ref{fig:question10} provides the results.

\textbf{Question 11. Reproducibility validation method}

\begin{figure}
 \centering
  \includegraphics[width=0.8\linewidth]{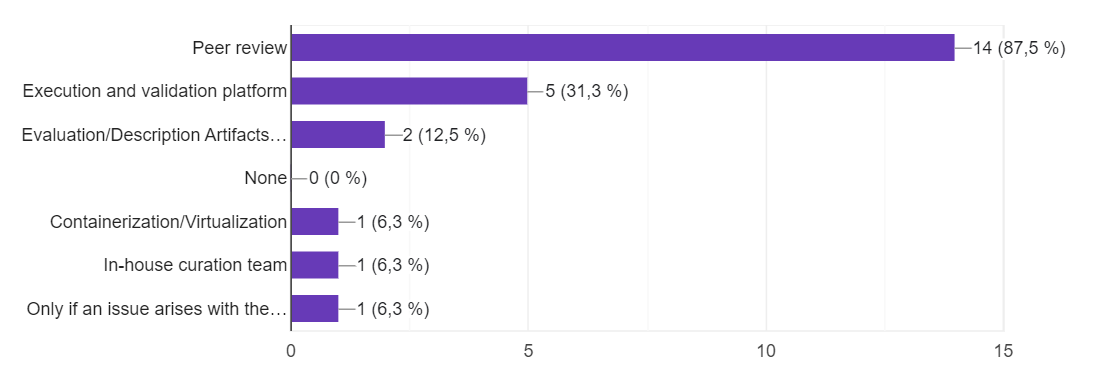}
  \caption{Reproducibility validation method.}
  \label{fig:question11}
\end{figure}

The results (Figure~\ref{fig:question11}) show that the traditional peer review model for article validation and acceptance is maintained, compared to other more automated forms of reproducibility validation. Therefore, validating the legitimacy of an article rests on one or two experts as well as their own available testing resources.

\textbf{Question 12. If you request to share the source code. What platforms or repositories do you recommend for sharing code? If others, you can write those you recommend}

\begin{figure}
 \centering
  \includegraphics[width=0.8\linewidth]{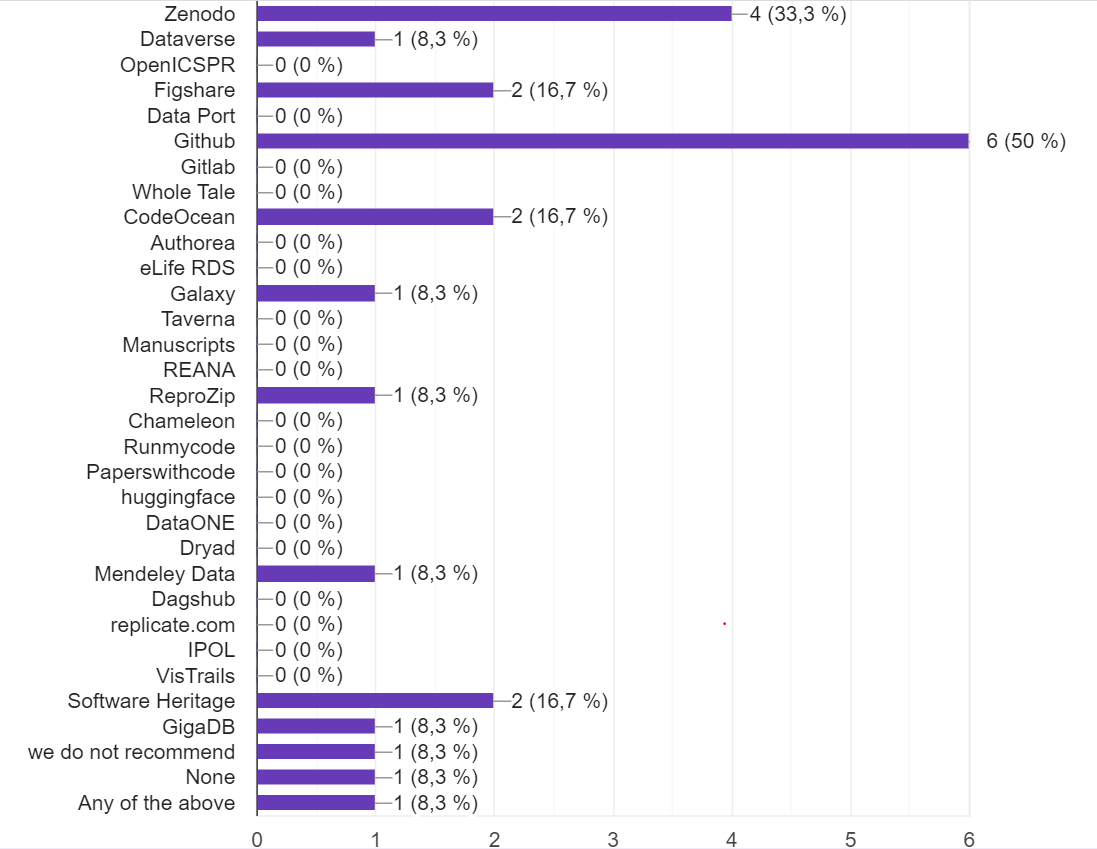}
  \caption{If you request to share source code, what platforms or repositories do you recommend for sharing code?}
  \label{fig:question12}
\end{figure}

The results (Figure~\ref{fig:question12}) describe show that Github if the preferred specialized platform, although more for developers than for publishing research results. Zenodo, on the other hand, allows the citation of code and data through its identifiers, but remains a simple non peer-reviewed repository. There is therefore still a significant lack of automation in the policies of code and data for reproducibility purposes, to validate the legitimacy and quality of the articles.

\textbf{Question 13. In the case you request reproducible research artifacts, which format is preferred?}

\begin{figure}
 \centering
  \includegraphics[width=0.8\linewidth]{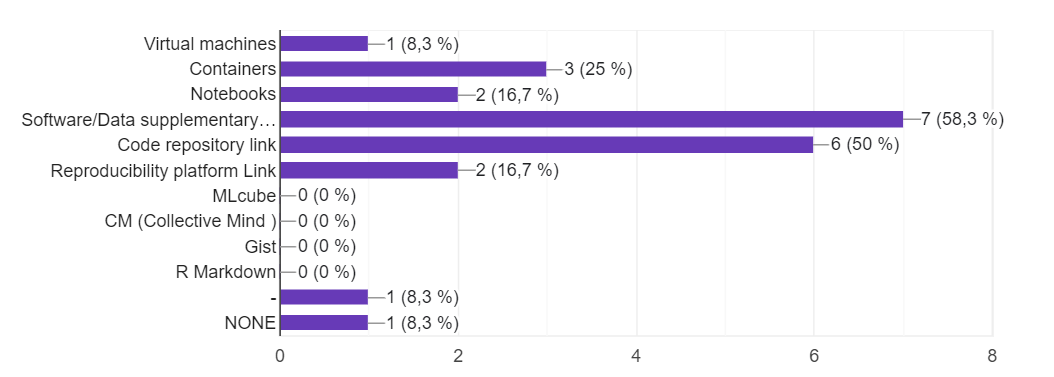}
  \caption{In the case you request reproducible research artifacts, which format is preferred? }
  \label{fig:question13}
\end{figure}

The majority of the journals indicated that indeed they request the software and data artifacts, but as supplementary material (58.3\%), with a large majority (50\%) that request a link to the source code. Figure~\ref{fig:question13} shows the results.

\textbf{Question 14. Which do you think would be the best way to reward the authors to submit reproducible articles?}

\begin{figure}
 \centering
  \includegraphics[width=0.8\linewidth]{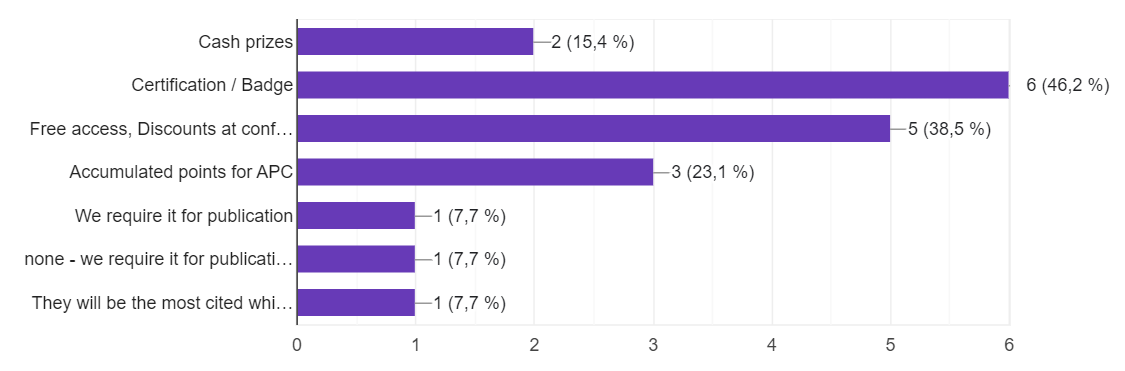}
  \caption{Which do you think would be the best way to reward the authors to submit reproducible articles? }
  \label{fig:question14}
\end{figure}

Most of the answers (Figure~\ref{fig:question14}) indicate that for most of the journals (46.2\%) the preferred way to reward authors is to grant reproducibility badges, followed by a 38.5\% of the answers which claim that the best way would be to offer free access to the journal or discounts.

\textbf{Question 15. Which do you think would be the best way to reward the reviewers of reproducible articles?}

\begin{figure}
 \centering
  \includegraphics[width=0.8\linewidth]{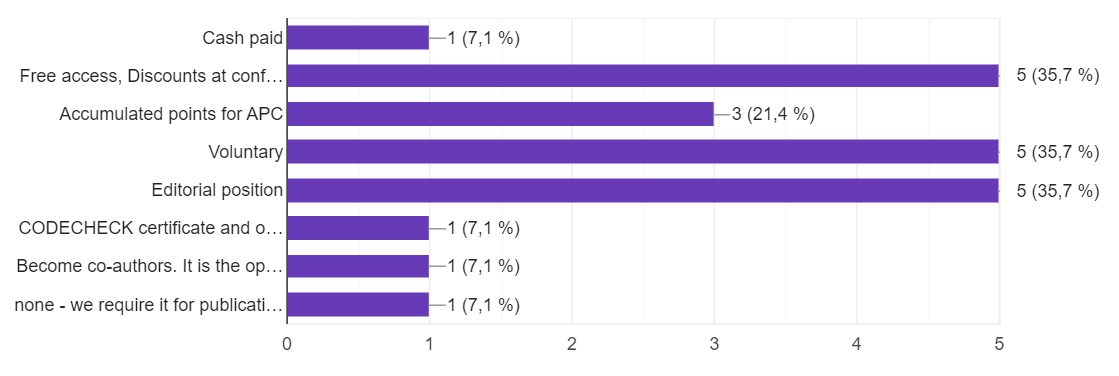}
  \caption{Which do you think would be the best way to reward the Reviewers of reproducible articles? }
  \label{fig:question15}
\end{figure}

The answers to this question (Figure~\ref{fig:question15}) do not show a strong preference for any of the options, being the most preferred to offer free access to the journal or discounts, offering the reviewers being part of the editorial board of the journals, or even considering that it should be a voluntary task.

\textbf{Question 16. Would you like to share briefly your view on the impact of implementing a reproducibility policy in the journal? For example, in terms of APC or other costs, the quality, citation impact,  credibility of articles, or any other topic you would like to address.}

This was an open questions to obtain from direct insights from the publishers. In the following we present their answers:

\begin{itemize}
    \item \textit{Observed average 15\% increase in citations after one year, 33\% after two years.}

    \item \textit{It's a great selling point and definitely has increased our visibility and content reuse.}

    \item \textit{We strongly encourage the authors to use Zenodo/Github as the repository for the code used in their experiments. We feel that both the reproducibility of the work and the ability of other to build upon it has a positive impact on science as a whole.}

    \item \textit{Credibility of the results is the main impact. However, software evaluation adds time and complexity to our reviewing procedure.}

    \item \textit{It has a deep impact on the credibility and citation of any article. I consider that our reproducibility initiative should be encouraged by most of journals. However, the most important issue is to encourage the authors to adopt a reproducibility-centered research methodology from the very beginning of their research.}

    \item \textit{As an author, I always submit my articles with a reproducibility appendix, and I always design my experiments to be fully automatic and reproducible from scratch, including the automatic generation of all data tables and figures in the articles. On the other hand, as editor, I am encouraging this same approach among our authors, and most authors who have published reproducible papers in our reproducibility section are adopting this approach. However, it demands extra effort and a lot of discipline, and for this reason, not all authors are probed to do it. For this reason, our reproducibility initiative rewards the authors with a second article (reproducible paper).}

    \textit{Finally, there is a long walk to achieve most of the authors adopt a reproducibility-centered approach. It is absolutely necessary that PhD supervisors adopt and encourage this approach among their students. A very good way of achieving it is teaching all MSc and PhD students to design and implement reproducibility protocols and submitting reproducibility appendixes with all their papers.}

    \item \textit{We do not have APC. In most cases, as much of the generated information has normally already been paid by public institutions.}

    \item \textit{Higher Impact Factor since the policy has been in place}

    \item \textit{Unfortunately, we do not implement a reproducibility policy in the BSR journal, but I would like to know more about it. On the other hand, I am not fully for the transparent publication of the data and software since it may violate privacy and licensing.}

    \item \textit{APC for original software publications is less than half of the APC for research papers. Original Software Publications are currently the strategic focus of our journal.}

    \item \textit{Surprisingly, authors are fine with our repository and requirements. We have not had any pushback. It does add editorial work and slows down the final acceptance. See the Github IJOC platform for how it works. https://github.com/INFORMSJoC".}

    \item \textit{This is very new to us - we have only introduced it this year - to start on Sept 1st. We expect it to improve the reputation of the Journal and thus its attractiveness to researchers.}
\end{itemize}

From this answers and insights, we shall in the following discuss on the technological possibilities available to address the reproducibility problem, and about the shared responsibility between authors, publishers and technology providers, including a brief gap analysis.

\section{Reproducibility technological Discussion}
\label{sec:discussion}
As pointed out by several of the works we have reviewed here, reproducibility comes with great benefits for both authors and journals. Let us briefly summarize them.

\begin{itemize}
    \item \textbf{Greater credibility and recognition}~\cite{ghimpau_incentives_2020}: reproducibility increases the credibility of the research and, therefore, can increase the recognition of the authors and the influence of their work in the scientific community.
    \item \textbf{Research results are accurate and reliable} to the extent that reproducibility is guaranteed. This is fundamental in the Scientific Method.
    \item \textbf{Increased visibility and impact}~\cite{boulbes_survey_2018}: the attention of the scientific community is attracted when articles are reproducible, thus increasing the visibility and the impact of the research.
    \item \textbf{Facilitate collaboration and reuse}~\cite{alnoamany_towards_2018}: results obtained from reproducible experiments allow other researchers to build on them, which promotes collaboration and reuse of findings. Eventually, it allows for faster scientific advancement.
    \item \textbf{}\textbf{Increase credibility and confidence in the results}~\cite{samuel_understanding_2021}: reproducibility allows the results obtained in an investigation to be verified and validated, increasing the confidence in its validity, following ethic directives~\cite{gupta_state_2022}, and transparency.
\end{itemize}

In the following, we shall discuss from the point of view of technological evolution based on the strategies presented as a reproducibility fundamental lever and support, how these interrelate into different challenges, problems, and solutions that have been proposed, and how they relate to these benefits. In particular, we discuss the problem of the responsibility of authors and publishers, such as their efforts towards reproducibility, the possibility of understanding reproducibility as a service, and finally, the impact of considering software as an important research artifact and the reward to researchers.

\subsection{Dilemma: virtualization solution or dependency}
\label{sec:dilemma_docker}

As described in the technological evolution subsection~\ref{sec:Tools}, many of the reproducibility tools and platforms (eg, Workflows) proposed so far are completely based on container technology; this strong trend leads us to discuss and draw attention to the benefits and drawbacks of relying exclusively on this technology.

It could be stated that Docker or, in general, lightweight virtualization, is the \textit{holy grail} of reproducibility, and, as will be seen, many solutions are based on this technology.
With the popularization of agile methodologies and as shown by the landscape and the containerization strategy, many of the reproducibility problems try to be solved with docker~\cite{moreau_containers_2023, canon_role_2020}, which leads us to question if there is an abuse of lightweight virtualization. As suggested, Docker is practical light and facilitates many processes that in the past were tedious; however, it is not a tool specifically designed for reproducibility, and it cannot be used indiscriminately to hide bad practices.

The possibility of packaging, freezing, and porting a code to any infrastructure and maintaining stable functionality over time make it attractive in the scientific world; however, as stated in~\cite{fursin_collective_2020-1}, this indiscriminate use brings great inconveniences, we will discuss in the following.

At this point, it is necessary to analyze in depth the problem of
reproducibility, repeatability, containers, and development. The problem is that two characteristics are desirable in systems, but actually,
they are antagonists. On the one hand, we want robust systems that will not break after an update. The classic example is a Python program that uses PyPI packages that, even if the user sets the versions in a virtual environment, the libraries may not
be available in a particular version of Python. In that case, many system designers opt for virtualization.

The containers ensure reproducibility given that the complete environment is fixed. However, if proper attention is not paid to the maintenance of the container, it might end up facing security problems given that if the environment is simply fixed and not updated, the libraries will stop receiving bug fixes and security updates.

Docker is certainly a useful tool that allows to fix the execution environment, but still maintenance is required. Regular automatic testing is recommended.

\subsection{The shared responsibility between authors and publishers}
\label{sec:shared_responsibility}
As presented in Sections \ref{sec:Tools} and \ref{sec:scientific_publishers}, complying with the criteria for reproducibility implies some costs, as well as the shared responsibility and pooling between authors and scientific journals. It also needs a commitment to transparency and reproducibility~\cite{haibe-kains_transparency_2020}. Despite the analysis of the reproducibility stakeholders in \cite{diaba-nuhoho_reproducibility_2021, feger_reproducibility_2022, macleod_improving_2022}, the roles and relationship between authors and publishers are still diffuse and at least questionable: indeed much of the reproducibility burden relies on the authors. 
Authors have a responsibility to provide detailed information about the methods and techniques used in their research, as well as to make public the data and codes that were used to generate the results. They must also ensure that their results are replicable and thus they can be verified by peers.

On the other hand, publishers have the responsibility to establish clear policies and guidelines~\cite{stodden_empirical_2018} for the submission of scientific articles, as well as looking for the transparency and reproducibility of the results.

Indeed, guaranteeing and legitimizing the reproducibility of scientific work in ML/AI implies assuming significant economic and time costs \cite{poldrack_costs_2019} depending on the size and complexity of the research project. These cannot be assumed only by the researcher.

\subsubsection{Reproducibility Cost}

Estimating the cost of reproducibility is not easy because it can be considered from the execution of a simple container on a personal laptop to a distributed execution of software in the cloud, with the market costs per hour of CPU, GPUs and storage depending on each provider and their business model (for example, GCP, Amazon, Azure, Oracle, and others).

Existing virtualization and containerization techniques and cloud computing infrastructure are key elements in this problem~\cite{6193081}. Therefore, the costs associated with cloud computing become relevant concerning reproducibility.

It is essential to highlight these associated costs~\cite{article1} and the implications for the scientific parties that have a role in the reproducibility of the scientific work. One can describe the main technological costs for the reproducibility of computational projects/experiments as follows:
$$C_R = C_{HD}+C_{HC}+C_{RC},$$
where $C_R$ is the total reproducibility cost,  $C_{HD}$ the cost of hosting data, $C_{HC}$ the cost of hosting code, and $C_{RC}$ the cost of running code.

%

The problem of increasingly complex research projects is not specific to computer science but common to other disciplines, especially when they combine different fields. Let's mention briefly the case of bioinformatics as an example. 
The researchers need to have not only knowledge of Biology but also the skills to operate the software and the data formats of the research artifacts, as well as the running environment. Typically, specialized expertise is required in Python, R~\cite{gandrud_reproducible_2020}, diverse operating systems, and database management and complex platforms such as, for example, Galaxy~\cite{afgan2018galaxy}.

IaC, virtualization and containerization, and cloud computing approach help address this divisional responsibility in a simplified manner. They allow to track the steps followed by the author so other researchers can repeat and reproduce the experiment in the same environment. However, this still requires a high level of computing skills, which should not necessarily be assumed only by the authors. In Section~\ref{sec:RaaS}, we discuss the Reproducibility as a Service (RaaS) strategy, which could be applied to manage this shared responsibility.

\section{Reproducibility efforts of authors and publishers}
\label{sec:authors_publisher_efforts}

\subsection{Effort of authors}
\label{sec:authors_efforts}

In the case of the authors, it should be understood that for several of the reasons discussed, in most cases, a scientific article cannot be reproduced in its entirety (100\%). The authors generally choose to reproduce only parts of the algorithms, demos, or data, which is essential to support the conclusions.
 
Therefore, it is the authors' effort to seek 100\% reproducibility of their work or to fully clarify the reasons that prevented reaching this objective, complying with the policies and requirements of the journals/conferences.

\subsubsection{Articles Submission Reproducibility Guide for Authors}
From our analysis of the shared responsibility between authors and journals and the most recent technological advances in computing, we shall discuss the efforts required by each of these two actors.

It is still very difficult for journals and authors to close the gap in a mutual effort, and it is even harder when the authors must comply with article submission guidelines between different journals.
An article composed of theoretical and computational parts can only be reproduced in a certain percentage and certain components that only the author is responsible for defining and specifying with the greatest of details and following a standard guide that avoids reprocessing between publishers.

MICRO2023 is a recent experience towards unified EA (artifact evaluation) guides and procedures\footnote{\url{https://ctuning.org/ae/micro2023.html}}, which allow speed up the AE process. A conference where artifacts can be complex and time-consuming to evaluate and 25\% of the submitted artifacts were awarded the artifact reusable badge. In this context, practices were developed such as Reviewers performed an initial 'smoke-test' (for example, installing the artifact, or resolving access/environment/setup issues) and also reviewed the key claims of the paper and the artifact. Likewise, two surveys were carried out consulting authors and evaluators to seek feedback on the AE process. Important insights are derived from this survey, especially in enabling authors and reviewers to faster iterate on artifacts efficiently, seamlessly, in reasonable time. For example Reviewers provided suggest that requesting authors to prepare a subset of simulations (and/or representative checkpoints) would be a good practice. Results "\textit{will appear in the ACM/IEEE MICRO 2023 conference front-matter}"\footnote{\url{https://www.linkedin.com/pulse/micro-2023-artifact-evaluation-report-56th-ieeeacm-symposium-fursin-bsgwe/}} and support a trend towards improvements to the process and clearer and standardized instructions preferable to most subjective assessment of other experiences.

Therefore, in addition to standardizing the different evaluation and description guides of Artifacts (see Table~\ref{demo-table1}), we propose to incorporate a mandatory and standardized unified guide between journals where the author contributes the \textit{effort to comply and assess} the level of reproducibility of their scientific article (see Table~\ref{tab:reproducibility}).

\begin{figure*}
    \centering
    \includegraphics[width=0.90\textwidth]{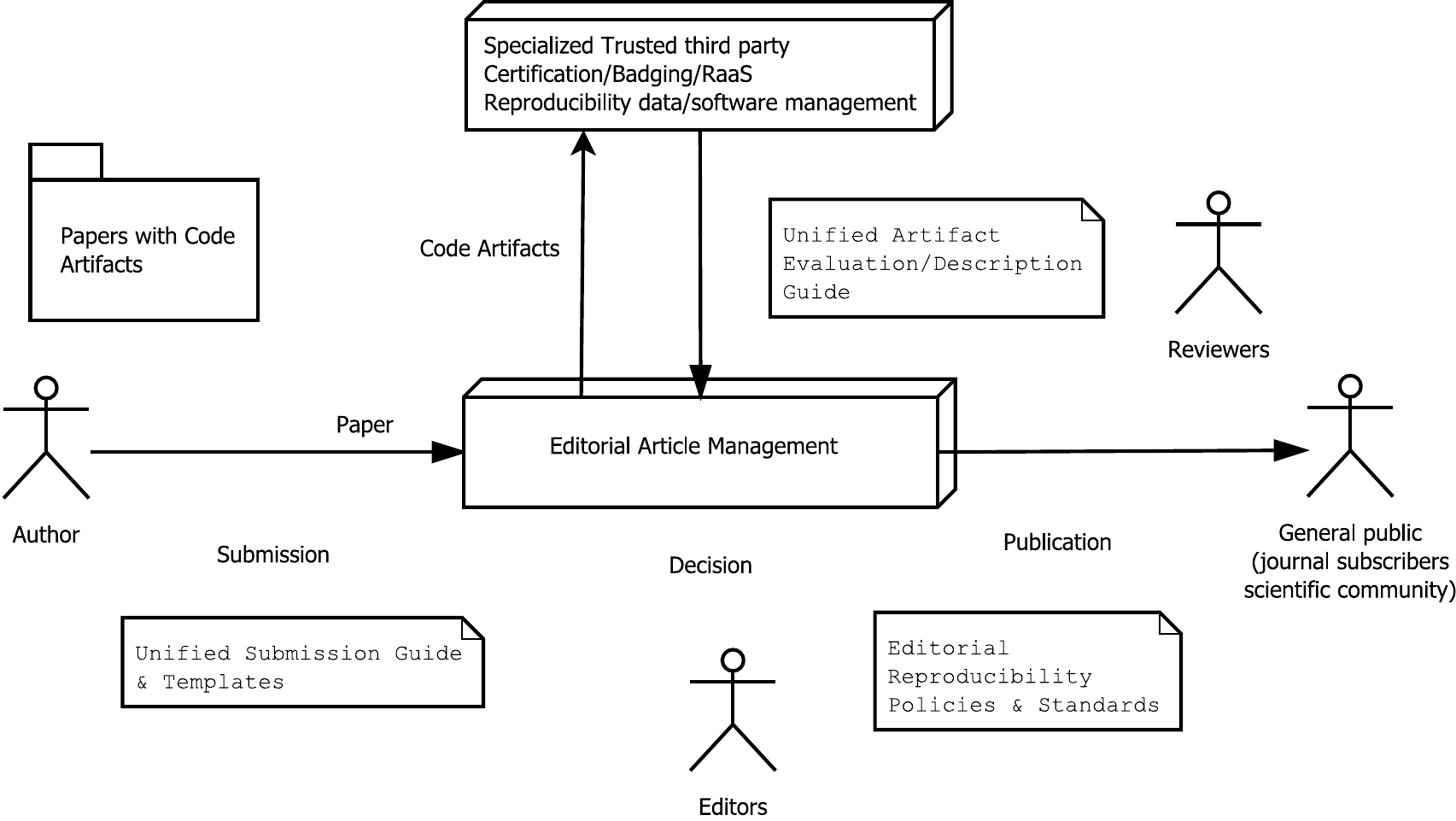}

    \caption{Proposal for a future editorial process on which the article and code are published as a whole, and third-parties certify reproducibility.}
    \label{fig:process2}
\end{figure*}

\begin{table}[tbp]
    \centering
    \begin{tabular}{p{6cm}|p{6cm}}
        \toprule
        \textbf{ITEM} & \textbf{OPTIONS} \\
        \midrule
        Article based on Software/Data? & yes/no \\
         \midrule
        Programming languages used & (e.g., Python, C++) \\
         \midrule
        Contains instructions for reproducibility & (e.g., complete, verified) \\
         \midrule
        Badges, Certified third-party Reproducibility Evaluators & (e.g., ACM badge, Ctuning) \\
         \midrule
        Infrastructure Reproducibility Required/Trusted third-party RaaS Operator & (e.g., Docker containers, MLflow, CodeOcean, Chameleon) \\
         \midrule
        Repository & (e.g., Zenodo, Software Heritage) \\
         \midrule
        Unique persistent citable identifiers of Software/Data Artifacts & (DOI, SWID, BlockchainID) \\
         \midrule
        Percentage of reproducibility of the Article & (\%) \\
         \midrule
        Reproducible components & (e.g. DEMO,virtual infrastructure, figures, tables, Backend, Frontend, Microservices, Lambda functions) \\
        \midrule
        Component Reproducibility degree & (R1,R2,R3,R4) \\
         \midrule
        Non-reproducible components (Why) & (e.g., proprietary software, sensitive data, distributed project) \\
        \bottomrule
    \end{tabular}
    \caption{Proposal of a \textit{reproducibility checklist} guide for authors. We propose to incorporate a mandatory and standardized unified guide between journals where the author contributes the effort to comply and assess the level of reproducibility of the scientific article.}
    \label{tab:reproducibility}

\end{table}

\subsection{Efforts of publishers}
\label{sec:journal_efforts}
In the case of journals, given the wide typology of submitted articles and the reproducibility costs, it is economically unfeasible that they have their own reproducibility infrastructure, which explains the current tendency to rely on third trusted parties. In the following, we shall discuss how the Reproducibility as a Service (RaaS) methodology could help discharge authors from the burden that implies running code and maintaining a complex reproducibility infrastructure, as well as the opportunity of considering a software more valued research artifacts and thus properly rewarding authors. Finally, we provide a brief gap analysis from the results of our survey in Section~\ref{sec:survey_results}.

\subsubsection{Reproducibility as a Service}
\label{sec:RaaS}
As pointed out in Section~\ref{sec:shared_responsibility}, reaching reproducibility might require for some projects a large technological investment, which should not be assumed only by the authors but shared with publishers and offered by specialized third parties. Here, we will focus on a particular strategy, Reproducibility as a Service (RaaS)~\cite{wonsil_reproducibility_2021}, which might be helpful to this purpose.

As introduced in Section~\ref{sec:repro_priciples}, RaaS is an approach to address non-reproducibility in scientific research by providing access to tools and resources that researchers and industrial actors to replicate experiments and data science projects. Also, to facilitate, manage, or overcome many of the limitations and barriers that we have identified in our review of the literature.

According to Brundage and co-authors~\cite{brundage_toward_2020}, one could label as RaaS any 3rd-party service made of tools that allow the reproducibility of scientific work.
Their proposal is to use the existing cloud computing tools to offer a service that fills the gap between two major requirements to achieve reproducibility.
On one hand, the actions taken by researchers who want to facilitate reproducibility. They provide a detailed procedure that allows to obtain the result artifacts, as well as the exact execution environment.
On the other hand, the actions taken by publishers or the industry validate reproducibility.



Moreover, there are trusted third parties that deal with big data projects and confidentiality issues of sensitive datasets. They aim to reduce the need for the strong computing skills typically required to work in complex AI/ML data science projects. Cloud Native-based RaaS~\cite{wonsil_reproducibility_2021}  adds an additional standardized layer with simplified interfaces for IaC, virtualization, and cloud computing (see Figure~\ref{fig:serv1}). As examples of these tools, we can cite Invenio~\footnote{\url{http://invenio-software.org}}, Eprints~\footnote{\url{http://www.eprints.org}}, or DSpace~\footnote{\url{http://www.dspace.org}}, among others.

Although not particularly adapted to complex workflow systems and user interactivity, a good example of the relationship between a journal and a third party that offers reproducibility services in the cloud is the partnership between IEEE and Code Ocean\footnote{\url{https://codeocean.com/signup/ieee}}. The code from IEEE articles can be browsed, discovered (assigned a DOI), run, modified, and eventually built the researcher's work on the cloud without any complex setup.

\subsubsection{Software as a valuable research artifact and reward to authors}
As presented in Section~\ref{sec:side_difficulties}, traditionally, the published article has been considered the most important and rewarded~\cite{parsons_history_2019} research artifact, leaving aside Software production. Universities, research centers, and evaluation committees usually consider the number of articles published in high-impact factor journals and the number of citations and sometimes as the major criterion to hire a researcher for increasing the salary and career evolution, among other incentives. Consequently, researchers typically do not invest large resources in the reproducibility of the results, the quality of the produced software, or even the possibility of publishing the software itself.

Many research projects are based on software contributed by others, including 
 libraries, applications, or complete frameworks, and in many cases, there is no explicit recognition of the authors of the third-party software. The recent incident about the vulnerability in the log4j library~\cite{hiesgen2022race} is a good example of a widely-spread software used by hundreds of companies, not necessarily acknowledging the library's authors.

The lack of incentives for researchers and software developers to produce quality and reproducible software has a clear negative impact ~\cite{ke_network-based_2023} on the development of Open Science. Fortunately, the criteria to evaluate researchers are evolving in parallel in the right direction. For example, CNRS (the Center for National Scientific Research) in France announced in 2022 changes in their evaluation policy to include SW projects as an equally valid element to evaluate scientific production. The reproducibility tools provided by journals and conferences are fundamental for traceability, thus allowing the proposal of new metrics specific to software.

Different works~\cite{parsons_history_2019, smith_software_2016} have focused on analyzing the citation of scientific software and data publishers~\cite{cousijn_data_2018} as a natural need to implement FAIR and paper-with-code strategies. To this purpose, FORCE11 (The Future of Research Communications and e-Scholarship)~\cite{smith_software_2016} provides guidelines for the citation of software and data.

Software needs to be properly cited and preserved. These two requirements are certainly not easy to fulfill, given their dynamic and changing nature. Indeed, millions of software repositories are constantly being updated at every instant in Github and other repositories.

The Software Heritage project, supported by UNESCO, is one important step forward in both citation and perpetual preservation of software via proper identifiers, such as the SWHID~\cite{di_cosmo_2044_2022}. Zenodo also provides a Digital Object Identifier (DOI) and Chameleon a QR-code to reference the code. It is also a great source of information to determine the provenance of software contributions.



Regarding using badges as an incentive for authors reproducibility, it must be observed that they really impact the researcher's reputation in the same way as the popularized and mature badge system awarded in e-learning by important companies, academies, to certify technical skills\cite{Stefaniak} published on reputable platforms such as Credly\footnote{\url{https://info.credly.com/}} and easily shared on Linkedin\footnote{\url{https://www.linkedin.com/}}, which allow the candidate to reinforce their CV and demonstrate to the employers in a competitive labor market. 

As pointed out by Dozmorov at al.~\cite{dozmorov_github_2018}, Github is, at the moment, the most complete database to measure the impact of software. Interestingly, they concluded that the number of \textit{forks} as a measure of software impact is not correlated with the number of citations associated with a scientific paper. This finding, at first counter-intuitive, shows that citation indices (such as the h-index and others) do not fully explain the true impact of the scientific work and the associated software. The consequence is, therefore, that producing quality software is neither properly promoted nor taken into account for the career advancement of researchers.

There is, therefore, a need for metrics that are specific to software, beyond indirect measures such as the number of \textit{forks} or \textit{stars} in public repositories. Strategies such as RaaS and more adapted metrics such as the Scientific Impact Factor (SIF)~\cite{ATM15375} could be of great help rather than the H-index or impact factor of the journal (FIJ) discussed in Section~\ref{sec:side_difficulties}.

Finally, our recommendation is depicted in Figure~\ref{fig:process2}, where there are incentives for all the actors, including 3rd-parties that implement permanent and long-term reproducibility infrastructures that support the publishing business.

\subsection{Dilemma: reproducibility sharing policies}
\label{sec:dilemma_policies}
At this point an important clarification must be made, \textit{journal reproducibility policies} should not be confused with traditional open access and open science initiatives. It could even be considered as an open topic that requires standardization.
In~\cite{stodden_empirical_2018} Stodden makes a first approximation from the analysis of Data and Code Policy Adoption by Journals, and then in ~\cite{stodden_toward_2013}~\cite{noauthor_digital_nodate} analysis of journal policy implementation and effectiveness for computational
reproducibility, however a clear concept of "reproducibility policies" is not consolidated. This leads us to consider that journals face an important dilemma in defining their internal policy of just limiting themselves to a code and data sharing policy or going further in defining veritable and strict automation tools and reproducibility evaluation article reproducibility policy.

\subsection{Brief gap analysis}
\label{sec:gap_analysis}
We provide a small gap analysis of the level of implementation of reproducibility policies that we observed from our survey. We intended to bring together all the elements of analysis. We include technological aspects, as well as the efforts required by both authors and publishers to help close or, at least, reduce the reproducibility gap. Aspects such as the standardization and implementation of reproducibility policies, adaptation of business models, and association with specialized third parties are considered. These recommendations come from analyzing the answers in our survey (Sec.~\ref{sec:survey_results}).

Table~\ref{tabla:elementos} shows the \textit{journal policy evaluation gap in identified key aspects}, indicating the survey question that helps evaluate the percentage of implementation of the reproducibility policies. With this table, each journal is evaluated in terms of its reproducibility policies and the effort it must make in the key aspects identified.

It can observed from the answers that there is a low percentage of implementation of reproducibility policies, as well as the low use of technological tools for automation, validation and sustainability of reproducibility in the long term (longevity of reproducibility).

This is explained because there is still no consensus and standardization on what should be a good reproducibility policy for journals, as well as the lack of a developed and mature market of trusted specialized RaaS services.

To determine the gap, we used the following qualitative ranking:
\begin{itemize}
    \item \textbf{High}: when there is a complete lack of accomplishment or implementation of the criterion
    \item \textbf{Intermediate}: when there is the presence of an initiative with immature development of the criterion
    \item \textbf{Low}: when there is a complete and functional implementation of the criterion
\end{itemize}

Unfortunately there is a large gap in the implementation of some aspects, mainly those related to automation, establishing reproducibility policies, management of repositories, and the use of persistent identifiers for the research artifacts, including software. Other aspects such as orienting the business model towards reproducibility itself or the use of FAIR data seem to be more developed.

Despite the observed gap, there is an opportunity to reduce the reproducibility gap with the common efforts of authors, publishers, and technological providers. See tables~\ref{tab:reproducibility}, ~\ref{tabla:elementos} and~\ref{demo-table} for more details.

\begin{table}[t]
\centering
\begin{tabular}{l|c|c}
\toprule
\textbf{Journals Reproducibility Features} &Survey questions& Gap level \\
\midrule
Automatic Validation and Execution Tool &   11, 12, 13 & High \\
\midrule
Author Incentives & 14 &Intermediate \\
\midrule
Reviewer Incentives & 15 &Intermediate \\
\midrule
Reproducibility Policy &3, 4, 5, 6, 16 &High \\
\midrule
Managed Repository &9  &High \\
\midrule
Article/Data/Software Persistent Unique Identifier & 9 &High \\
\midrule
Business Model Oriented to Reproducibility & 7, 8 &Intermediate \\
\midrule
FAIR-TLC & 10 &Intermediate \\
\bottomrule
\end{tabular}

\caption{\label{tabla:elementos}The gap in the implementation of the journal policies policies, along with the related survey's questions.}
\end{table}



\begin{table}
\begin{center}
\begin{tabular}{ 
p{1cm}|p{5cm}|p{2cm}|p{5cm} }
  \toprule

  \multicolumn{4}{c}{Summary of strategies for reproducibility} \\
 \midrule
 Type&Strategy& Papers &Examples  \\
 \midrule
 \midrule
  \multirow{3}{*}{(1)Sof} & Open Source Software, Open science, repositories, FAIR&\cite{parland-von_essen_supporting_2018,  raff_siren_2022, macleod_improving_2022,  9354557, wang_how_2023, haim_how_2023,  gonzalez-barahona_revisiting_2023,  barba_defining_2022,stodden_beyond_2020}&Github, Gitlab, bitbucket, Zenodo, softwareheritage, dataverse, Huggingface       \\
 
 \cmidrule{2-4}
  &CSharing/Documentation tools&\cite{pimentel_large-scale_2019, pimentel_understanding_2021, glavic_reproducemegit_2021, wang_restoring_2020}&  Reprozip, Notebooks, CRAN, Rmarkdown,  \\
 \cmidrule{2-4}
 &Open data formats, Baselines, SOTA Benchmarks& \cite{khritankov_mldev_2021, kazerooni_brain_2023, Fursin2014CollectiveMT}&JSON, XML, MLperf, Dataperf, Kaggel, Brats,CM, MLcube, MLdev  \\
 \midrule
 
 \multirow{3}{*}{(2)Env} &Container/ virtualization/ Cloud &\cite{6193081, canon_role_2020, moreau_containers_2023, bedo_unifying_2020}&  Docker, Vmware, singularity, AWS, GCP, AZURE, ORACLE, BioNix/ Guix \\
 \cmidrule{2-4}
  &Architectures&\cite{fritzsch_adopting_2023, jonas_cloud_2019} &monolithic/ microservice/ serverless/cloud/hybrid \\
 \cmidrule{2-4}
 &IaC-Infrastructure as a Code&\cite{bowman_improving_2023, bibliograph7, bibliograph1, bibliograph2, krzhizhanovskaya_reproducibility_2020} & Terraform, pulumi, kubernetes
CloudFormation, Ansible, puppet \\
 \midrule
 
\multirow{3}{*}{(3)Sys} & Cientific Workflows and MLOps tools BPML CWL languages&~\cite{gift_practical_2021, demchenko_experimental_2023, cohen-boulakia_scientific_2017, rosendo_kheops_2023, gundersen_machine_2022, bahaidarah_toward_2021, ghoshal_experiences_2020, kluge_watchdog_2020, article_Korkhov} &  Taverna, Galaxy, 
VisTrails, Nextflow, Neptune, Weight, Comet, Omniboard, Mlflow, TensorBoard,
Polyaxon, ClearML, Valohai, Pachyderm, Kubeflow, Verta.ai, SageMaker, DVC, kheOps, RE3, Hyperflow, watchdog ,SHIWA \\
 \cmidrule{2-4}
 &Metadata and Provenance (Traceability Lineage Logging Monitoring)&\cite{ludascher_provstore_2015, silva_junior_provenance-and_2021, wonsil_integrated_2023, samuel_end--end_2022, samuel_collaborative_2022, zirpins_towards_2022, kawamoto_ai_2020, wittek_blockchain-based_2021, glavic_machine_2021}  &MERIT, ROVPY, PROVNEO4J, PROV-DB, CONNECTOR, NOWORKFLOW, GIT2PROV,Provbook, blockchain, SWHID, DOI  \\

 \cmidrule{2-4}
 
 &RaaS-Reproducibility as a Service &\cite{wonsil_reproducibility_2021, demchenko_experimental_2023, foster_toward_2020}& Whole Tale, chameleon, CodeOcean, IPOL \\
 \midrule

\multirow{3}{*}{(4)Met} &AE/AD Peer  Code Reviews&  \cite{noauthor_supporting_2021, fostiropoulos_reproducibility_2023, malik_artifact_2020, pineau_improving_2020, athanassoulis_artifacts_2022, plale_reproducibility_2021, kerautret_reproducibility_2021}   &reviewcommons, ArVix, Peer Community In (PCI), SIGPLAN, Ctuning, NeuroIPS, Badging     \\
 \cmidrule{2-4}

& Publications with code    & \cite{de_sterck_enhancing_2023, bonsignorio_new_2017, salsabil_study_2022, trisovic_advancing_2020}& Some Journals(nature), Conferences(ACM,IEEE), runmycode\\
 \cmidrule{2-4}
&Policies, Good Practices, General frameworks, Recommendations, Methodologies Life Cycle Management & \cite{article_Korkhov,article2, milewicz_towards_2023, community_illustrations_2022, community_illustrations_2022, franch_model-driven_2022, mauerer_beyond_2022, lin_building_2022, akhlaghi_toward_2021, merz_editorial_2020, turkyilmaz-van_der_velden_reproducibility_2020, hofman_expanding_2020, samuel_understanding_2021, nichols_better_2021}& NASEM, DevSecOps, AIOps, MLops, CRISP-DM, KDD, SEMMA, Turing way, Teaching Culture Reproducibility, Journal Reproducibility Policies \\

  \bottomrule

\end{tabular}

\caption{\label{demo-table}Strategies for Reproducibility.}
\end{center}
\end{table}


\section{Conclusion}
\label{sec:conclusion}
We have presented a PRISMA-based systematic review of the existing literature on techniques and platforms which can be used in computer science and machine learning projects, putting the focus on reproducible research. In order to clarify what \textit{reproducibility} is, we have also reviewed the different definitions in the literature, which after the NASEM report have been \textit{de facto} standardized.

The main difficulties reported by researchers when trying to reproduce the work of their peers have been enumerated. We analyzed also the problem of how to measure objectively the reproducibility, especially in the case where the experiment not necessarily gives the same results each time it is repeated, but statistically equivalent.

From our discussion we have identified what we consider are the most important reproducibility strategies, such as the use of open source repositories or the use of FAIR data, or following methodologies which have been proven to be relevant to achieve reproducibility.

In this work, we have intended to address the problem of the credibility crisis specifically in computer science including ML/AI projects, from diverse reproducibility stakeholder points of view. We establish insights from the best practices, frameworks, methodologies, and technologies available at the moment.

In computer science, the variety of languages, new developments, platforms, frameworks, hardware, and architectures on which the code of scientific articles can be run are vast. In some cases, it requires third-party proprietary software and data, which is why means a significant challenge for publishers. Implementing reproducibility policies supported by RaaS or similar methodologies can certainly help reduce the reproducibility gap on publications.

We conclude that the high cost of guaranteeing the reproducibility of a software project is not properly rewarded at this moment to the reproducibility stakeholders.
Considering the costs in own infrastructure that would be required, one still needs to take into consideration business models that encourage investment by third parties in infrastructure and thus guarantee longevity and perennity in the reproducibility of scientific publications.

It is, therefore, necessary to share the efforts among the different actors. A mutually-beneficial relationship must be established between authors, reviewers, and publishers to balance the benefits and costs that authors and scientific publishers eventually assume. There are not standardized Description/Evaluation standards in this regard and, in many cases, authors reported that it is very tedious to meet the requirements of each different publisher. Therefore, some authors are discouraged from improving the quality of their papers in the so-called \textit{publish or perish} race. Moreover, some publishers (excluding those under the diamond model) are in a position of power over the authors who are charged significant article processing charges.

It is convenient to define a new metric equivalent of the Impact Factor, which could be used specifically for software. This could help properly reward the effort of software developers by acknowledging them clearly as co-authors of the scientific work and measuring the real impact of their contributions in reproducible computer science projects.

Our survey shows a growing concern about the reproducibility and how their policies can get into it. This reflected in the comments provided by the journals and the initiatives they want to push forward. Therefore, clearly defined and long-term plans are required to achieve a sustainable reproducibility model where each stakeholder obtains a benefit.
For the moment, the traditional peer review evaluation methodologies are preferred. The responsibility of the validation is mainly on the expertise of the reviewers chosen by the editors and the few functional tests to the artifacts that they can do with their limited testing infrastructure.
It also indicates that, in computer science journals indexed in SCOPUS, there is a low level of formal implementation of reproducibility policies supported by their own reproducibility platforms.

We conclude that it is imperative to bring together coordinated efforts to agree on standardized guides for authors for the submission of articles, unified reproducibility policies and artifact evaluation criteria from editors, supported by the reproducibility strategies and technological evolution discussed in this article. Consequently, there is a promising future with opportunities and potential to reduce the reproducibility gap identified with the joint effort of all actors involved to ensure reliability and trustworthiness in the knowledge conveyed by computer science-based publications.

\section*{Acknowledgements}
The authors would like to thank the financial support to the SESAME's OVD-SaaS project from Région Île de France and BPI France. Also, to the Ministry of Science, Technology and Innovation of Colombia (Minciencias), within the framework of Call 885 of 2020  for the financing of the Doctorates of Excellence Program.

\bibliographystyle{unsrt}
\bibliography{bibliography}

\begin{thebibliography}{100}

\bibitem{ivie_reproducibility_2018}
Peter Ivie and Douglas Thain.
\newblock Reproducibility in {Scientific} {Computing}.
\newblock {\em ACM Computing Surveys}, 51(3):63:1--63:36, July 2018.

\bibitem{hughes_history_2001}
T.P. Hughes.
\newblock History of {Technology}.
\newblock In {\em International {Encyclopedia} of the {Social} \& {Behavioral}
  {Sciences}}, pages 6852--6857. Elsevier, 2001.

\bibitem{hu_evolution_2020}
Y.~Hu, A.~Tareen, Y-J. Sheu, W.~T. Ireland, C.~Speck, H.~Li, L.~Joshua-Tor,
  J.~B. Kinney, and B.~Stillman.
\newblock Evolution of {DNA} replication origin specification and gene
  silencing mechanisms.
\newblock {\em Nature Communications}, 11(1):5175, October 2020.

\bibitem{plaven-sigray_readability_2017}
Pontus Plavén-Sigray, Granville~James Matheson, Björn~Christian Schiffler,
  and William~Hedley Thompson.
\newblock The readability of scientific texts is decreasing over time.
\newblock {\em eLife}, 6:e27725, September 2017.

\bibitem{gundersen_reproducibility_2020}
Odd~Erik Gundersen.
\newblock The {Reproducibility} {Crisis} {Is} {Real}.
\newblock {\em AI Magazine}, 41(3):103--106, September 2020.

\bibitem{hutson_artificial_2018}
Matthew Hutson.
\newblock Artificial intelligence faces reproducibility crisis.
\newblock {\em Science}, 359(6377):725--726, February 2018.

\bibitem{dodds_future_2019}
Francis Dodds.
\newblock The future of academic publishing: {Revolution} or evolution
  revisited.
\newblock {\em Learned Publishing}, 32(4):345--354, October 2019.

\bibitem{baillieul_reflections_2018}
John Baillieul, Gerry Grenier, and Gianluca Setti.
\newblock Reflections on the future of research curation and research
  reproducibility [point of view].
\newblock 106(5):779--783.

\bibitem{ahmed_future_2023}
Abubakari Ahmed, Aceil Al-Khatib, Yap Boum, Humberto Debat, Alonso
  Gurmendi~Dunkelberg, Lisa~Janicke Hinchliffe, Frith Jarrad, Adam Mastroianni,
  Patrick Mineault, Charlotte~R. Pennington, and J.~Andrew Pruszynski.
\newblock The future of academic publishing.
\newblock {\em Nature Human Behaviour}, 7(7):1021--1026, July 2023.

\bibitem{kitchenham_meta-analysis_2020}
Barbara Kitchenham, Lech Madeyski, and Pearl Brereton.
\newblock Meta-analysis for families of experiments in software engineering: a
  systematic review and reproducibility and validity assessment.
\newblock {\em Empirical Software Engineering}, 25(1):353--401, January 2020.

\bibitem{anchundia_resources_2020}
Carlos~E. Anchundia and Efrain~R. Fonseca~C.
\newblock Resources for reproducibility of experiments in empirical software
  engineering: Topics derived from a secondary study.
\newblock 8:8992--9004.

\bibitem{stoddart_is_2016}
Charlotte Stoddart.
\newblock Is there a reproducibility crisis in science?
\newblock {\em Nature}, pages d41586--019--00067--3, May 2016.

\bibitem{kapoor_leakage_2022}
Sayash Kapoor and Arvind Narayanan.
\newblock Leakage and the {Reproducibility} {Crisis} in {ML}-based {Science}.
\newblock 2022.

\bibitem{lucic_reproducibility_2022}
Ana Lucic, Maurits Bleeker, Sami Jullien, Samarth Bhargav, and Maarten
  De~Rijke.
\newblock Reproducibility as a {Mechanism} for {Teaching} {Fairness},
  {Accountability}, {Confidentiality}, and {Transparency} in {Artificial}
  {Intelligence}.
\newblock {\em Proceedings of the AAAI Conference on Artificial Intelligence},
  36(11):12792--12800, June 2022.

\bibitem{ahmed_measuring_2022}
Hana Ahmed, Roselyne Tchoua, and Jay Lofstead.
\newblock Measuring {Reproduciblity} of {Machine} {Learning} {Methods} for
  {Medical} {Diagnosis}.
\newblock In {\em 2022 {Fourth} {International} {Conference} on
  {Transdisciplinary} {AI} ({TransAI})}, pages 9--16, Laguna Hills, CA, USA,
  September 2022. IEEE.

\bibitem{moradi_reproducible_2021}
Amin Moradi and Alexandru Uta.
\newblock Reproducible {Model} {Sharing} for {AI} {Practitioners}.
\newblock In {\em Proceedings of the {Fifth} {Workshop} on {Distributed}
  {Infrastructures} for {Deep} {Learning} ({DIDL}) 2021}, pages 1--6, Virtual
  Event Canada, December 2021. ACM.

\bibitem{haibe-kains_transparency_2020}
Benjamin Haibe-Kains, George~Alexandru Adam, Ahmed Hosny, Farnoosh Khodakarami,
  {Massive Analysis Quality Control (MAQC) Society Board of Directors}, Thakkar
  Shraddha, Rebecca Kusko, Susanna-Assunta Sansone, Weida Tong, Russ~D.
  Wolfinger, Christopher~E. Mason, Wendell Jones, Joaquin Dopazo, Cesare
  Furlanello, Levi Waldron, Bo~Wang, Chris McIntosh, Anna Goldenberg, Anshul
  Kundaje, Casey~S. Greene, Tamara Broderick, Michael~M. Hoffman, Jeffrey~T.
  Leek, Keegan Korthauer, Wolfgang Huber, Alvis Brazma, Joelle Pineau, Robert
  Tibshirani, Trevor Hastie, John P.~A. Ioannidis, John Quackenbush, and Hugo
  J. W.~L. Aerts.
\newblock Transparency and reproducibility in artificial intelligence.
\newblock {\em Nature}, 586(7829):E14--E16, October 2020.

\bibitem{mckinney_reply_2020}
Scott~Mayer McKinney, Alan Karthikesalingam, Daniel Tse, Christopher~J. Kelly,
  Yun Liu, Greg~S. Corrado, and Shravya Shetty.
\newblock Reply to: {Transparency} and reproducibility in artificial
  intelligence.
\newblock {\em Nature}, 586(7829):E17--E18, October 2020.

\bibitem{kitamura_reproducible_2020}
Felipe~C. Kitamura, Ian Pan, and Timothy~L. Kline.
\newblock Reproducible {Artificial} {Intelligence} {Research} {Requires} {Open}
  {Communication} of {Complete} {Source} {Code}.
\newblock {\em Radiology: Artificial Intelligence}, 2(4):e200060, July 2020.

\bibitem{gibney_this_2020}
Elizabeth Gibney.
\newblock This {AI} researcher is trying to ward off a reproducibility crisis.
\newblock {\em Nature}, 577(7788):14--14, January 2020.

\bibitem{ghanta_interpretability_2018}
Sindhu Ghanta, Sriram Subramanian, Swaminathan Sundararaman, Lior Khermosh,
  Vinay Sridhar, Dulcardo Arteaga, Qianmei Luo, Dhananjoy Das, and Nisha
  Talagala.
\newblock Interpretability and {Reproducability} in {Production} {Machine}
  {Learning} {Applications}.
\newblock In {\em 2018 17th {IEEE} {International} {Conference} on {Machine}
  {Learning} and {Applications} ({ICMLA})}, pages 658--664, Orlando, FL,
  December 2018. IEEE.

\bibitem{aviyente_explainability_2022}
Selin Aviyente and Abdullah Karaaslanli.
\newblock Explainability in {Graph} {Data} {Science}: {Interpretability},
  replicability, and reproducibility of community detection.
\newblock {\em IEEE Signal Processing Magazine}, 39(4):25--39, July 2022.

\bibitem{kerautret_reproducibility_2021}
Daniel Lopresti and George Nagy.
\newblock Reproducibility: {Evaluating} the {Evaluations}.
\newblock In Bertrand Kerautret, Miguel Colom, Adrien Krähenbühl, Daniel
  Lopresti, Pascal Monasse, and Hugues Talbot, editors, {\em Reproducible
  {Research} in {Pattern} {Recognition}}, volume 12636, pages 12--23. Springer
  International Publishing, Cham, 2021.

\bibitem{raff_research_2020}
Edward Raff.
\newblock Research {Reproducibility} as a {Survival} {Analysis}.
\newblock {\em arXiv:2012.09932 [cs, stat]}, December 2020.
\newblock arXiv: 2012.09932.

\bibitem{diaba-nuhoho_reproducibility_2021}
Patrick Diaba-Nuhoho and Michael Amponsah-Offeh.
\newblock Reproducibility and research integrity: the role of scientists and
  institutions.
\newblock 14(1):451.

\bibitem{NIPS2015_86df7dcf}
D.~Sculley, Gary Holt, Daniel Golovin, Eugene Davydov, Todd Phillips, Dietmar
  Ebner, Vinay Chaudhary, Michael Young, Jean-Fran\c{c}ois Crespo, and Dan
  Dennison.
\newblock Hidden technical debt in machine learning systems.
\newblock In C.~Cortes, N.~Lawrence, D.~Lee, M.~Sugiyama, and R.~Garnett,
  editors, {\em Advances in Neural Information Processing Systems}, volume~28.
  Curran Associates, Inc., 2015.

\bibitem{kitzes_practice_2018}
Justin Kitzes, Daniel Turek, and Fatma Deniz, editors.
\newblock {\em The practice of reproducible research: case studies and lessons
  from the data-intensive sciences}.
\newblock University of California Press, Oakland, California, 2018.

\bibitem{baker_workshop_2019}
Nathan Baker, Frank Alexander, Timo Bremer, Aric Hagberg, Yannis Kevrekidis,
  Habib Najm, Manish Parashar, Abani Patra, James Sethian, Stefan Wild, Karen
  Willcox, and Steven Lee.
\newblock Workshop {Report} on {Basic} {Research} {Needs} for {Scientific}
  {Machine} {Learning}: {Core} {Technologies} for {Artificial} {Intelligence}.
\newblock Technical Report 1478744, February 2019.

\bibitem{parashar_research_2022}
Manish Parashar, Michael~A. Heroux, and Victoria Stodden.
\newblock Research {Reproducibility}.
\newblock {\em Computer}, 55(8):16--18, August 2022.

\bibitem{article_Thompson}
Paul Thompson and Andrew Burnett.
\newblock Reproducible research.
\newblock {\em CORE Issues in Professional and Research Ethics}, 1, 01 2012.

\bibitem{raghupathi_reproducibility_2022}
Wullianallur Raghupathi, Viju Raghupathi, and Jie Ren.
\newblock Reproducibility in {Computing} {Research}: {An} {Empirical} {Study}.
\newblock {\em IEEE Access}, 10:29207--29223, 2022.

\bibitem{gundersen_fundamental_2021}
Odd~Erik Gundersen.
\newblock The {Fundamental} {Principles} of {Reproducibility}.
\newblock {\em Philosophical Transactions of the Royal Society A: Mathematical,
  Physical and Engineering Sciences}, 379(2197):20200210, May 2021.
\newblock arXiv:2011.10098 [cs].

\bibitem{feger_reproducibility_2022}
Sebastian~Stefan Feger and Paweł~W. Woźniak.
\newblock Reproducibility: {A} {Researcher}-{Centered} {Definition}.
\newblock {\em Multimodal Technologies and Interaction}, 6(2):17, February
  2022.

\bibitem{macleod_improving_2022}
Malcolm Macleod and {the University of Edinburgh Research Strategy Group}.
\newblock Improving the reproducibility and integrity of research: what can
  different stakeholders contribute?
\newblock {\em BMC Research Notes}, 15(1):146, April 2022.

\bibitem{committee_on_reproducibility_and_replicability_in_science_reproducibility_2019}
{Committee on Reproducibility and Replicability in Science}, {Board on
  Behavioral, Cognitive, and Sensory Sciences}, {Committee on National
  Statistics}, {Division of Behavioral and Social Sciences and Education},
  {Nuclear and Radiation Studies Board}, {Division on Earth and Life Studies},
  {Board on Mathematical Sciences and Analytics}, {Committee on Applied and
  Theoretical Statistics}, {Division on Engineering and Physical Sciences},
  {Board on Research Data and Information}, {Committee on Science, Engineering,
  Medicine, and Public Policy}, {Policy and Global Affairs}, and {National
  Academies of Sciences, Engineering, and Medicine}.
\newblock {\em Reproducibility and {Replicability} in {Science}}.
\newblock National Academies Press, Washington, D.C., September 2019.

\bibitem{plesser_reproducibility_2018}
Hans~E. Plesser.
\newblock Reproducibility vs. {Replicability}: {A} {Brief} {History} of a
  {Confused} {Terminology}.
\newblock {\em Frontiers in Neuroinformatics}, 11:76, January 2018.

\bibitem{essawy_taxonomy_2020}
Bakinam~T. Essawy, Jonathan~L. Goodall, Daniel Voce, Mohamed~M. Morsy,
  Jeffrey~M. Sadler, Young~Don Choi, David~G. Tarboton, and Tanu Malik.
\newblock A taxonomy for reproducible and replicable research in environmental
  modelling.
\newblock {\em Environmental Modelling \& Software}, 134:104753, December 2020.

\bibitem{heroux_toward_2018}
Michael~A. Heroux, Lorena Barba, Manish Parashar, Victoria Stodden, and Michela
  Taufer.
\newblock Toward a {Compatible} {Reproducibility} {Taxonomy} for
  {Computational} and {Computing} {Sciences}.
\newblock Technical Report SAND2018-11186, 1481626, October 2018.

\bibitem{jose_reproducibility_2020}
Jimmy Lin and Qian Zhang.
\newblock Reproducibility is a {Process}, {Not} an {Achievement}: {The}
  {Replicability} of {IR} {Reproducibility} {Experiments}.
\newblock In Joemon~M. Jose, Emine Yilmaz, João Magalhães, Pablo Castells,
  Nicola Ferro, Mário~J. Silva, and Flávio Martins, editors, {\em Advances in
  {Information} {Retrieval}}, volume 12036, pages 43--49. Springer
  International Publishing, Cham, 2020.

\bibitem{gundersen_state_2018}
Odd~Erik Gundersen and Sigbjørn Kjensmo.
\newblock State of the {Art}: {Reproducibility} in {Artificial} {Intelligence}.
\newblock {\em Proceedings of the AAAI Conference on Artificial Intelligence},
  32(1), April 2018.

\bibitem{impagliazzo_reproducibility_2022}
Russell Impagliazzo, Rex Lei, Toniann Pitassi, and Jessica Sorrell.
\newblock Reproducibility in learning.
\newblock In {\em Proceedings of the 54th {Annual} {ACM} {SIGACT} {Symposium}
  on {Theory} of {Computing}}, pages 818--831, Rome Italy, June 2022. ACM.

\bibitem{akhlaghi_toward_2021}
Mohammad Akhlaghi, Raul Infante-Sainz, Boudewijn~F. Roukema, Mohammadreza
  Khellat, David Valls-Gabaud, and Roberto Baena-Galle.
\newblock Toward {Long}-{Term} and {Archivable} {Reproducibility}.
\newblock {\em Computing in Science \& Engineering}, 23(3):82--91, May 2021.

\bibitem{baker_why_2016}
Monya Baker.
\newblock Why scientists must share their research code.

\bibitem{fan_escaping_2020}
Gang Fan, Chengpeng Wang, Rongxin Wu, Xiao Xiao, Qingkai Shi, and Charles
  Zhang.
\newblock Escaping dependency hell: finding build dependency errors with the
  unified dependency graph.
\newblock In {\em Proceedings of the 29th {ACM} {SIGSOFT} {International}
  {Symposium} on {Software} {Testing} and {Analysis}}, {ISSTA} 2020, pages
  463--474, New York, NY, USA, July 2020. Association for Computing Machinery.

\bibitem{mack_how_2018}
Chris~A. Mack.
\newblock {\em How to write a good scientific paper}.
\newblock SPIE Press, Bellingham, Washington, 2018.
\newblock OCLC: on1019885580.

\bibitem{bailey_reproducibility_2020}
David~H Bailey.
\newblock Reproducibility and variable precision computing.
\newblock {\em The International Journal of High Performance Computing
  Applications}, 34(5):483--490, September 2020.

\bibitem{jalal_apostal_improving_2020}
Sara~Faraji Jalal~Apostal, David Apostal, and Ronald Marsh.
\newblock Improving {Numerical} {Reproducibility} of {Scientific} {Software} in
  {Parallel} {Systems}.
\newblock In {\em 2020 {IEEE} {International} {Conference} on {Electro}
  {Information} {Technology} ({EIT})}, pages 066--074, Chicago, IL, USA, July
  2020. IEEE.

\bibitem{jezequel_first_2014}
Fabienne Jézéquel, Philippe Langlois, and Nathalie Revol.
\newblock First steps towards more numerical reproducibility.
\newblock {\em ESAIM: Proceedings and Surveys}, 45:229, September 2014.

\bibitem{dolstra_nixos_2010}
Eelco Dolstra, Andres Löh, and Nicolas Pierron.
\newblock {NixOS}: {A} purely functional {Linux} distribution.
\newblock {\em Journal of Functional Programming}, 20(5-6):577--615, November
  2010.

\bibitem{vallet_toward_2022}
Nicolas Vallet, David Michonneau, and Simon Tournier.
\newblock Toward practical transparent verifiable and long-term reproducible
  research using {Guix}.
\newblock {\em Scientific Data}, 9(1):597, October 2022.

\bibitem{liu_promoting_2022}
Jing Liu, Jacob Carlson, Josh Pasek, Brian Puchala, Arvind Rao, and H.~V.
  Jagadish.
\newblock Promoting and {Enabling} {Reproducible} {Data} {Science} {Through} a
  {Reproducibility} {Challenge}.
\newblock {\em Harvard Data Science Review}, July 2022.

\bibitem{ghimpau_incentives_2020}
Valentina Ghimpau.
\newblock Incentives, rewards, and recognition - what really motivates a
  researcher?
\newblock In {\em Judging Research}. {MDPI}.

\bibitem{article5}
Hugh Desmond.
\newblock Incentivizing replication is insufficient to safeguard default trust.
\newblock {\em Philosophy of Science}, 88, 03 2020.

\bibitem{aksnes_citations_2019}
Dag~W. Aksnes, Liv Langfeldt, and Paul Wouters.
\newblock Citations, {Citation} {Indicators}, and {Research} {Quality}: {An}
  {Overview} of {Basic} {Concepts} and {Theories}.
\newblock {\em SAGE Open}, 9(1):215824401982957, January 2019.

\bibitem{ATM15375}
Giuseppe Lippi and Camilla Mattiuzzi.
\newblock Scientist impact factor (sif): a new metric for improving
  scientists’ evaluation?
\newblock {\em Annals of Translational Medicine}, 5(15), 2017.

\bibitem{raff_does_2022}
Edward Raff.
\newblock Does the {Market} of {Citations} {Reward} {Reproducible} {Work}?
\newblock {\em arXiv:2204.03829 [cs]}, April 2022.
\newblock arXiv: 2204.03829.

\bibitem{frery_badging_2020}
Alejandro~C. Frery, Luis Gomez, and Antonio~C. Medeiros.
\newblock A {Badging} {System} for {Reproducibility} and {Replicability} in
  {Remote} {Sensing} {Research}.
\newblock {\em IEEE Journal of Selected Topics in Applied Earth Observations
  and Remote Sensing}, 13:4988--4995, 2020.

\bibitem{mauerer_beyond_2022}
Wolfgang Mauerer, Stefan Klessinger, and Stefanie Scherzinger.
\newblock Beyond the badge: reproducibility engineering as a lifetime skill.
\newblock In {\em Proceedings of the 4th {International} {Workshop} on
  {Software} {Engineering} {Education} for the {Next} {Generation}}, pages
  1--4, Pittsburgh Pennsylvania, May 2022. ACM.

\bibitem{edwards_academic_2017}
Marc~A. Edwards and Siddhartha Roy.
\newblock Academic {Research} in the 21st {Century}: {Maintaining} {Scientific}
  {Integrity} in a {Climate} of {Perverse} {Incentives} and {Hypercompetition}.
\newblock {\em Environmental Engineering Science}, 34(1):51--61, January 2017.

\bibitem{cukier_checklists_2020}
Samantha Cukier, Lucas Helal, Danielle~B. Rice, Justina Pupkaite, Nadera
  Ahmadzai, Mitchell Wilson, Becky Skidmore, Manoj~M. Lalu, and David Moher.
\newblock Checklists to detect potential predatory biomedical journals: a
  systematic review.
\newblock {\em BMC Medicine}, 18(1):104, December 2020.

\bibitem{benureau_re-run_2018}
Fabien C.~Y. Benureau and Nicolas~P. Rougier.
\newblock Re-run, repeat, reproduce, reuse, replicate: Transforming code into
  scientific contributions.
\newblock 11:69.

\bibitem{rosenblatt_epistemic_2023}
Lucas Rosenblatt, Bernease Herman, Anastasia Holovenko, Wonkwon Lee, Joshua
  Loftus, Elizabeth McKinnie, Taras Rumezhak, Andrii Stadnik, Bill Howe, and
  Julia Stoyanovich.
\newblock Epistemic {Parity}: {Reproducibility} as an {Evaluation} {Metric} for
  {Differential} {Privacy}.
\newblock {\em Proceedings of the VLDB Endowment}, 16(11):3178--3191, July
  2023.

\bibitem{raff_step_2019}
Edward Raff.
\newblock A {Step} {Toward} {Quantifying} {Independently} {Reproducible}
  {Machine} {Learning} {Research}.
\newblock 2019.

\bibitem{nordling_literature_2022}
Torbjörn Nordling and Tomas~Melo Peralta.
\newblock A literature review of methods for assessment of reproducibility in
  science.
\newblock preprint, In Review, November 2022.

\bibitem{Collberg2014MeasuringRI}
Christian~S. Collberg.
\newblock Measuring reproducibility in computer systems research.
\newblock 2014.

\bibitem{Schelter2015}
Sebastian Schelter, Felix Biessmann, Tim Januschowski, David Salinas, Stephan
  Seufert, and Gyuri Szarvas.
\newblock On challenges in machine learning model management.
\newblock {\em IEEE Data Engineering Bulletin}, 2015.

\bibitem{freire_computational_2012}
Juliana Freire, Philippe Bonnet, and Dennis Shasha.
\newblock Computational reproducibility: state-of-the-art, challenges, and
  database research opportunities.
\newblock In {\em Proceedings of the 2012 {ACM} {SIGMOD} {International}
  {Conference} on {Management} of {Data}}, pages 593--596, Scottsdale Arizona
  USA, May 2012. ACM.

\bibitem{saltz_current_2022}
Jeffrey~S. Saltz and Iva Krasteva.
\newblock Current approaches for executing big data science projects—a
  systematic literature review.
\newblock {\em PeerJ Computer Science}, 8:e862, February 2022.

\bibitem{bibliograph5}
Jiating~Chen Gonzalo~Rivero.
\newblock Best coding practices to ensure reproducibility.
\newblock 2020.

\bibitem{6171147}
Randall~J. LeVeque, Ian~M. Mitchell, and Victoria Stodden.
\newblock Reproducible research for scientific computing: Tools and strategies
  for changing the culture.
\newblock {\em Computing in Science \& Engineering}, 14(4):13--17, 2012.

\bibitem{ray_review_2022}
Partha~Pratim Ray.
\newblock A review on {TinyML}: {State}-of-the-art and prospects.
\newblock {\em Journal of King Saud University - Computer and Information
  Sciences}, 34(4):1595--1623, April 2022.

\bibitem{pouchard_computational_2019}
Line Pouchard, Sterling Baldwin, Todd Elsethagen, Shantenu Jha, Bibi Raju, Eric
  Stephan, Li~Tang, and Kerstin~Kleese Van~Dam.
\newblock Computational reproducibility of scientific workflows at extreme
  scales.
\newblock {\em The International Journal of High Performance Computing
  Applications}, 33(5):763--776, September 2019.

\bibitem{barba_defining_2022}
Lorena~A. Barba.
\newblock Defining the {Role} of {Open} {Source} {Software} in {Research}
  {Reproducibility}.
\newblock {\em Computer}, 55(8):40--48, August 2022.

\bibitem{9354557}
Ryan~P. Abernathey, Tom Augspurger, Anderson Banihirwe, Charles~C.
  Blackmon-Luca, Timothy~J. Crone, Chelle~L. Gentemann, Joseph~J. Hamman, Naomi
  Henderson, Chiara Lepore, Theo~A. McCaie, Niall~H. Robinson, and Richard~P.
  Signell.
\newblock Cloud-native repositories for big scientific data.
\newblock {\em Computing in Science \& Engineering}, 23(2):26--35, 2021.

\bibitem{gonzalez-barahona_revisiting_2023}
Jesus~M. Gonzalez-Barahona and Gregorio Robles.
\newblock Revisiting the reproducibility of empirical software engineering
  studies based on data retrieved from development repositories.
\newblock {\em Information and Software Technology}, 164:107318, December 2023.

\bibitem{wang_how_2023}
Aaron Haim, Stacy~T. Shaw, and Neil~T. Heffernan.
\newblock How to {Open} {Science}: {Promoting} {Principles} and
  {Reproducibility} {Practices} {Within} the {Artificial} {Intelligence} in
  {Education} {Community}.
\newblock In Ning Wang, Genaro Rebolledo-Mendez, Vania Dimitrova, Noboru
  Matsuda, and Olga~C. Santos, editors, {\em Artificial {Intelligence} in
  {Education}. {Posters} and {Late} {Breaking} {Results}, {Workshops} and
  {Tutorials}, {Industry} and {Innovation} {Tracks}, {Practitioners},
  {Doctoral} {Consortium} and {Blue} {Sky}}, volume 1831, pages 74--78.
  Springer Nature Switzerland, Cham, 2023.

\bibitem{haim_how_2023}
Aaron Haim, Stacy Shaw, and Neil Heffernan.
\newblock How to {Open} {Science}: {A} {Principle} and {Reproducibility}
  {Review} of the {Learning} {Analytics} and {Knowledge} {Conference}.
\newblock In {\em {LAK23}: 13th {International} {Learning} {Analytics} and
  {Knowledge} {Conference}}, pages 156--164, Arlington TX USA, March 2023. ACM.

\bibitem{stodden_beyond_2020}
Victoria Stodden.
\newblock Beyond {Open} {Data}: {A} {Model} for {Linking} {Digital} {Artifacts}
  to {Enable} {Reproducibility} of {Scientific} {Claims}.
\newblock In {\em Proceedings of the 3rd {International} {Workshop} on
  {Practical} {Reproducible} {Evaluation} of {Computer} {Systems}}, pages
  9--14, Stockholm Sweden, June 2020. ACM.

\bibitem{stodden_empirical_2018}
Victoria Stodden, Jennifer Seiler, and Zhaokun Ma.
\newblock An empirical analysis of journal policy effectiveness for
  computational reproducibility.
\newblock {\em Proceedings of the National Academy of Sciences},
  115(11):2584--2589, March 2018.

\bibitem{parland-von_essen_supporting_2018}
Jessica Parland-von Essen, Katja Fält, Zubair Maalick, Miika Alonen, and
  Eduardo Gonzalez.
\newblock Supporting {FAIR} data: categorization of research data as a tool in
  data management.
\newblock {\em Informaatiotutkimus}, 37(4), December 2018.

\bibitem{albertoni_reproducibility_2023}
Riccardo Albertoni, Sara Colantonio, Piotr Skrzypczyński, and Jerzy
  Stefanowski.
\newblock Reproducibility of {Machine} {Learning}: {Terminology},
  {Recommendations} and {Open} {Issues}, February 2023.
\newblock arXiv:2302.12691 [cs].

\bibitem{article_Vanschoren}
Joaquin Vanschoren, Mikio Braun, and Cheng~Soon Ong.
\newblock Open science in machine learning.
\newblock {\em Implementing Reproducible Research}, 02 2014.

\bibitem{baracaldo_towards_2022}
Nathalie Baracaldo, Ali Anwar, Mark Purcell, Ambrish Rawat, Mathieu Sinn,
  Bashar Altakrouri, Dian Balta, Mahdi Sellami, Peter Kuhn, Ulrich Schopp, and
  Matthias Buchinger.
\newblock Towards an {Accountable} and {Reproducible} {Federated} {Learning}:
  {A} {FactSheets} {Approach}.
\newblock 2022.

\bibitem{zirpins_towards_2022}
José~A. Peregrina, Guadalupe Ortiz, and Christian Zirpins.
\newblock Towards a {Metadata} {Management} {System} for {Provenance},
  {Reproducibility} and {Accountability} in {Federated} {Machine} {Learning}.
\newblock In Christian Zirpins, Guadalupe Ortiz, Zoltan Nochta, Oliver
  Waldhorst, Jacopo Soldani, Massimo Villari, and Damian Tamburri, editors,
  {\em Advances in {Service}-{Oriented} and {Cloud} {Computing}}, volume 1617,
  pages 5--18. Springer Nature Switzerland, Cham, 2022.

\bibitem{vitek_repeatability_2011}
Jan Vitek and Tomas Kalibera.
\newblock Repeatability, reproducibility, and rigor in systems research.
\newblock In {\em Proceedings of the ninth {ACM} international conference on
  {Embedded} software}, pages 33--38, Taipei Taiwan, October 2011. ACM.

\bibitem{Fursin2014CollectiveMT}
Grigori Fursin, Renato Miceli, Anton Lokhmotov, Michael Gerndt, Marc Baboulin,
  Allen~D. Malony, Zbigniew Chamski, Diego Novillo, and Davide~Del Vento.
\newblock Collective mind: Towards practical and collaborative auto-tuning.
\newblock {\em Sci. Program.}, 22:309--329, 2014.

\bibitem{fursin_invited_2018}
Grigori Fursin.
\newblock Invited {Talk} {Abstract}: {Introducing} {ReQuEST}: {An} {Open}
  {Platform} for {Reproducible} and {Quality}-{Efficient} {Systems}-{ML}
  {Tournaments}.
\newblock In {\em 2018 1st {Workshop} on {Energy} {Efficient} {Machine}
  {Learning} and {Cognitive} {Computing} for {Embedded} {Applications}
  ({EMC2})}, pages 3--3, Williamsburg, VA, March 2018. IEEE.

\bibitem{kazerooni_brain_2023}
Anahita~Fathi Kazerooni, Nastaran Khalili, Xinyang Liu, Debanjan Haldar, Zhifan
  Jiang, Syed~Muhammed Anwar, Jake Albrecht, Maruf Adewole, Udunna Anazodo,
  Hannah Anderson, Sina Bagheri, Ujjwal Baid, Timothy Bergquist, Austin~J.
  Borja, Evan Calabrese, Verena Chung, Gian-Marco Conte, Farouk Dako, James
  Eddy, Ivan Ezhov, Ariana Familiar, Keyvan Farahani, Shuvanjan Haldar,
  Juan~Eugenio Iglesias, Anastasia Janas, Elaine Johansen, Blaise~V Jones,
  Florian Kofler, Dominic LaBella, Hollie~Anne Lai, Koen Van~Leemput,
  Hongwei~Bran Li, Nazanin Maleki, Aaron~S McAllister, Zeke Meier, Bjoern
  Menze, Ahmed~W Moawad, Khanak~K Nandolia, Julija Pavaine, Marie Piraud, Tina
  Poussaint, Sanjay~P Prabhu, Zachary Reitman, Andres Rodriguez, Jeffrey~D
  Rudie, Ibraheem~Salman Shaikh, Lubdha~M. Shah, Nakul Sheth, Russel~Taki
  Shinohara, Wenxin Tu, Karthik Viswanathan, Chunhao Wang, Jeffrey~B Ware,
  Benedikt Wiestler, Walter Wiggins, Anna Zapaishchykova, Mariam Aboian, Miriam
  Bornhorst, Peter de~Blank, Michelle Deutsch, Maryam Fouladi, Lindsey Hoffman,
  Benjamin Kann, Margot Lazow, Leonie Mikael, Ali Nabavizadeh, Roger Packer,
  Adam Resnick, Brian Rood, Arastoo Vossough, Spyridon Bakas, and Marius~George
  Linguraru.
\newblock The {Brain} {Tumor} {Segmentation} ({BraTS}) {Challenge} 2023:
  {Focus} on {Pediatrics} ({CBTN}-{CONNECT}-{DIPGR}-{ASNR}-{MICCAI}
  {BraTS}-{PEDs}).
\newblock 2023.

\bibitem{fritzsch_adopting_2023}
Jonas Fritzsch, Justus Bogner, Markus Haug, Ana Cristina~Franco da~Silva,
  Carolin Rubner, Matthias Saft, Horst Sauer, and Stefan Wagner.
\newblock Adopting {Microservices} and {DevOps} in the {Cyber}-{Physical}
  {Systems} {Domain}: {A} {Rapid} {Review} and {Case} {Study}.
\newblock {\em Software: Practice and Experience}, 53(3):790--810, March 2023.
\newblock arXiv:2210.06858 [cs].

\bibitem{jonas_cloud_2019}
Eric Jonas, Johann Schleier-Smith, Vikram Sreekanti, Chia-Che Tsai, Anurag
  Khandelwal, Qifan Pu, Vaishaal Shankar, Joao Carreira, Karl Krauth, Neeraja
  Yadwadkar, Joseph~E. Gonzalez, Raluca~Ada Popa, Ion Stoica, and David~A.
  Patterson.
\newblock Cloud {Programming} {Simplified}: {A} {Berkeley} {View} on
  {Serverless} {Computing}.
\newblock 2019.

\bibitem{pimentel_large-scale_2019}
João~Felipe Pimentel, Leonardo Murta, Vanessa Braganholo, and Juliana Freire.
\newblock A {Large}-{Scale} {Study} {About} {Quality} and {Reproducibility} of
  {Jupyter} {Notebooks}.
\newblock In {\em 2019 {IEEE}/{ACM} 16th {International} {Conference} on
  {Mining} {Software} {Repositories} ({MSR})}, pages 507--517, May 2019.
\newblock ISSN: 2574-3864.

\bibitem{samuel2023computational}
Sheeba Samuel and Daniel Mietchen.
\newblock Computational reproducibility of jupyter notebooks from biomedical
  publications, 2023.

\bibitem{pimentel_understanding_2021}
João~Felipe Pimentel, Leonardo Murta, Vanessa Braganholo, and Juliana Freire.
\newblock Understanding and improving the quality and reproducibility of
  {Jupyter} notebooks.
\newblock {\em Empirical Software Engineering}, 26(4):65, July 2021.

\bibitem{glavic_reproducemegit_2021}
Sheeba Samuel and Birgitta König-Ries.
\newblock {ReproduceMeGit}: {A} {Visualization} {Tool} for {Analyzing}
  {Reproducibility} of {Jupyter} {Notebooks}.
\newblock In Boris Glavic, Vanessa Braganholo, and David Koop, editors, {\em
  Provenance and {Annotation} of {Data} and {Processes}}, volume 12839, pages
  201--206. Springer International Publishing, Cham, 2021.

\bibitem{wang_restoring_2020}
Jiawei Wang, Tzu-yang Kuo, Li~Li, and Andreas Zeller.
\newblock Restoring reproducibility of {Jupyter} notebooks.
\newblock In {\em Proceedings of the {ACM}/{IEEE} 42nd {International}
  {Conference} on {Software} {Engineering}: {Companion} {Proceedings}}, pages
  288--289, Seoul South Korea, June 2020. ACM.

\bibitem{wolke_reproducible_2016}
Andreas Wolke, Martin Bichler, Fernando Chirigati, and Victoria Steeves.
\newblock Reproducible experiments on dynamic resource allocation in cloud data
  centers.
\newblock {\em Information Systems}, 59:98--101, July 2016.

\bibitem{congo_building_2015}
Faical Congo.
\newblock Building a {Cloud} {Service} for {Reproducible} {Simulation}
  {Management}.
\newblock pages 187--193, Austin, Texas, 2015.

\bibitem{6193081}
Bill Howe.
\newblock Virtual appliances, cloud computing, and reproducible research.
\newblock {\em Computing in Science \& Engineering}, 14(4):36--41, 2012.

\bibitem{bahaidarah_toward_2021}
Layan Bahaidarah, Ethan Hung, Andreas F. De~Melo Oliveira, Jyotsna Penumaka,
  Lukas Rosario, and Ana Trisovic.
\newblock Toward {Reusable} {Science} with {Readable} {Code} and
  {Reproducibility}.
\newblock {\em arXiv:2109.10387 [cs]}, September 2021.
\newblock arXiv: 2109.10387.

\bibitem{canon_role_2020}
R.~Shane Canon.
\newblock The {Role} of {Containers} in {Reproducibility}.
\newblock In {\em 2020 2nd {International} {Workshop} on {Containers} and {New}
  {Orchestration} {Paradigms} for {Isolated} {Environments} in {HPC}
  ({CANOPIE}-{HPC})}, pages 19--25, Atlanta, GA, USA, November 2020. IEEE.

\bibitem{krzhizhanovskaya_reproducibility_2020}
Michał Orzechowski, Bartosz Baliś, Renata~G. Słota, and Jacek Kitowski.
\newblock Reproducibility of {Computational} {Experiments} on
  {Kubernetes}-{Managed} {Container} {Clouds} with {HyperFlow}.
\newblock In Valeria~V. Krzhizhanovskaya, Gábor Závodszky, Michael~H. Lees,
  Jack~J. Dongarra, Peter M.~A. Sloot, Sérgio Brissos, and João Teixeira,
  editors, {\em Computational {Science} – {ICCS} 2020}, volume 12137, pages
  220--233. Springer International Publishing, Cham, 2020.

\bibitem{vasyukov_using_2018}
Alexey Vasyukov and Igor Petrov.
\newblock Using {Computing} {Containers} and {Continuous} {Integration} to
  {Improve} {Numerical} {Research} {Reproducibility}.
\newblock {\em International Journal of Computer (IJC)}, 30(1):27--33, July
  2018.

\bibitem{forde_reproducible_2018}
Jessica Forde, Tim Head, Chris Holdgraf, Yuvi Panda, Gladys Nalvarete, Benjamin
  Ragan-Kelley, and Erik Sundell.
\newblock Reproducible {Research} {Environments} with {Repo2Docker}.
\newblock June 2018.

\bibitem{sugimura_building_2018}
Peter Sugimura and Florian Hartl.
\newblock Building a {Reproducible} {Machine} {Learning} {Pipeline}.
\newblock {\em arXiv:1810.04570 [cs, stat]}, October 2018.
\newblock arXiv: 1810.04570.

\bibitem{steidl_pipeline_2023}
Monika Steidl, Michael Felderer, and Rudolf Ramler.
\newblock The pipeline for the continuous development of artificial
  intelligence models—{Current} state of research and practice.
\newblock {\em Journal of Systems and Software}, 199:111615, May 2023.

\bibitem{kluge_watchdog_2020}
Michael Kluge, Marie-Sophie Friedl, Amrei~L Menzel, and Caroline~C Friedel.
\newblock Watchdog 2.0: {New} developments for reusability, reproducibility,
  and workflow execution.
\newblock {\em GigaScience}, 9(6):giaa068, June 2020.

\bibitem{glavic_machine_2021}
Sheeba Samuel, Frank Löffler, and Birgitta König-Ries.
\newblock Machine {Learning} {Pipelines}: {Provenance}, {Reproducibility} and
  {FAIR} {Data} {Principles}.
\newblock In Boris Glavic, Vanessa Braganholo, and David Koop, editors, {\em
  Provenance and {Annotation} of {Data} and {Processes}}, volume 12839, pages
  226--230. Springer International Publishing, Cham, 2021.

\bibitem{da_silva_community_2021}
Rafael~Ferreira Da~Silva, Henri Casanova, Kyle Chard, Ilkay Altintas, Rosa~M
  Badia, Bartosz Balis, Taina Coleman, Frederik Coppens, Frank Di~Natale,
  Bjoern Enders, Thomas Fahringer, Rosa Filgueira, Grigori Fursin, Daniel
  Garijo, Carole Goble, Dorran Howell, Shantenu Jha, Daniel~S. Katz, Daniel
  Laney, Ulf Leser, Maciej Malawski, Kshitij Mehta, Loic Pottier, Jonathan
  Ozik, J.~Luc Peterson, Lavanya Ramakrishnan, Stian Soiland-Reyes, Douglas
  Thain, and Matthew Wolf.
\newblock A {Community} {Roadmap} for {Scientific} {Workflows} {Research} and
  {Development}.
\newblock In {\em 2021 {IEEE} {Workshop} on {Workflows} in {Support} of
  {Large}-{Scale} {Science} ({WORKS})}, pages 81--90, St. Louis, MO, USA,
  November 2021. IEEE.

\bibitem{franch_model-driven_2022}
Fran Melchor, Roberto Rodriguez-Echeverria, Jose~M. Conejero, and Juan Prieto,
  Alvaro amd~Gutierrez.
\newblock A {Model}-{Driven} {Approach} for {Systematic} {Reproducibility} and
  {Replicability} of {Data} {Science} {Projects}.
\newblock In Xavier Franch, Geert Poels, Frederik Gailly, and Monique Snoeck,
  editors, {\em Advanced {Information} {Systems} {Engineering}}, volume 13295,
  pages 147--163. Springer International Publishing, Cham, 2022.

\bibitem{francoise_marcelle_2021}
Jules Françoise, Baptiste Caramiaux, and Téo Sanchez.
\newblock Marcelle: {Composing} {Interactive} {Machine} {Learning} {Workflows}
  and {Interfaces}.
\newblock In {\em The 34th {Annual} {ACM} {Symposium} on {User} {Interface}
  {Software} and {Technology}}, pages 39--53, Virtual Event USA, October 2021.
  ACM.

\bibitem{prabhu_reproducible_2020}
Anirudh Prabhu and Peter Fox.
\newblock Reproducible {Workflow}.
\newblock 2020.

\bibitem{ghoshal_experiences_2020}
Devarshi Ghoshal, Drew Paine, Gilberto Pastorello, Abdelrahman Elbashandy, Dan
  Gunter, Oluwamayowa Amusat, and Lavanya Ramakrishnan.
\newblock Experiences with {Reproducibility}: {Case} {Studies} from
  {Scientific} {Workflows}.
\newblock In {\em Proceedings of the 4th {International} {Workshop} on
  {Practical} {Reproducible} {Evaluation} of {Computer} {Systems}}, pages 3--8,
  Virtual Event Sweden, June 2020. ACM.

\bibitem{rosendo_kheops_2023}
Daniel Rosendo, Kate Keahey, Alexandru Costan, Matthieu Simonin, Patrick
  Valduriez, and Gabriel Antoniu.
\newblock {KheOps}: {Cost}-effective {Repeatability}, {Reproducibility}, and
  {Replicability} of {Edge}-to-{Cloud} {Experiments}.
\newblock In {\em Proceedings of the 2023 {ACM} {Conference} on
  {Reproducibility} and {Replicability}}, pages 62--73, Santa Cruz CA USA, June
  2023. ACM.

\bibitem{cohen-boulakia_scientific_2017}
Sarah Cohen-Boulakia, Khalid Belhajjame, Olivier Collin, Jérôme Chopard,
  Christine Froidevaux, Alban Gaignard, Konrad Hinsen, Pierre Larmande, Yvan~Le
  Bras, Frédéric Lemoine, Fabien Mareuil, Hervé Ménager, Christophe Pradal,
  and Christophe Blanchet.
\newblock Scientific workflows for computational reproducibility in the life
  sciences: {Status}, challenges and opportunities.
\newblock {\em Future Generation Computer Systems}, 75:284--298, October 2017.

\bibitem{plale_reproducibility_2021}
Beth~A. Plale, Tanu Malik, and Line~C. Pouchard.
\newblock Reproducibility {Practice} in {High}-{Performance} {Computing}:
  {Community} {Survey} {Results}.
\newblock {\em Computing in Science \& Engineering}, 23(5):55--60, September
  2021.

\bibitem{santana-perez_towards_2015}
Idafen Santana-Perez and María~S. Pérez-Hernández.
\newblock Towards {Reproducibility} in {Scientific} {Workflows}: {An}
  {Infrastructure}-{Based} {Approach}.
\newblock {\em Scientific Programming}, 2015:e243180, February 2015.

\bibitem{8258038}
Eric Breck, Shanqing Cai, Eric Nielsen, Michael Salib, and D.~Sculley.
\newblock The ml test score: A rubric for ml production readiness and technical
  debt reduction.
\newblock In {\em 2017 IEEE International Conference on Big Data (Big Data)},
  pages 1123--1132, 2017.

\bibitem{gift_practical_2021}
Noah Gift and Alfredo Deza.
\newblock {\em Practical {MLOps}: operationalizing machine learning models}.
\newblock O'Reilly Media Inc, Sebastopol, CA, first edition edition, 2021.
\newblock OCLC: on1249501065.

\bibitem{8804457}
Saleema Amershi, Andrew Begel, Christian Bird, Robert DeLine, Harald Gall, Ece
  Kamar, Nachiappan Nagappan, Besmira Nushi, and Thomas Zimmermann.
\newblock Software engineering for machine learning: A case study.
\newblock In {\em 2019 IEEE/ACM 41st International Conference on Software
  Engineering: Software Engineering in Practice (ICSE-SEIP)}, pages 291--300,
  2019.

\bibitem{gundersen_machine_2022}
Odd~Erik Gundersen, Saeid Shamsaliei, and Richard~Juul Isdahl.
\newblock Do machine learning platforms provide out-of-the-box reproducibility?
\newblock {\em Future Generation Computer Systems}, 126:34--47, January 2022.

\bibitem{schlegel_management_2022}
Marius Schlegel and Kai-Uwe Sattler.
\newblock Management of {Machine} {Learning} {Lifecycle} {Artifacts}: {A}
  {Survey}.
\newblock 2022.

\bibitem{bibliograph3}
Preprint.
\newblock Ml reproducibility systems: Status and research agenda.
\newblock 2021.

\bibitem{fursin_collective_2020}
Grigori Fursin.
\newblock Collective {Knowledge}: organizing research projects as a database of
  reusable components and portable workflows with common {APIs}.
\newblock 2020.

\bibitem{adams_yawl_2020}
Michael Adams, Andreas~V. Hense, and Arthur~H.M. ter Hofstede.
\newblock {YAWL}: {An} open source {Business} {Process} {Management} {System}
  from science for science.
\newblock {\em SoftwareX}, 12:100576, July 2020.

\bibitem{article_Korkhov}
Vladimir Korkhov, Dagmar Krefting, Johan Montagnat, Tram Truong-Huu, Tamas
  Kukla, Gabor Terstyanszky, David Manset, Matthan Caan, and Silvia
  Olabarriaga.
\newblock Shiwa workflow interoperability solutions for neuroimaging data
  analysis.
\newblock {\em Studies in health technology and informatics}, 175:109--10, 09
  2012.

\bibitem{demchenko_experimental_2023}
Yuri Demchenko, Sebastian Gallenmuller, Serge Fdida, Panayiotis Andreou, Cedric
  Crettaz, and Mathias Kirkeng.
\newblock Experimental {Research} {Reproducibility} and {Experiment} {Workflow}
  {Management}.
\newblock In {\em 2023 15th {International} {Conference} on {COMmunication}
  {Systems} \& {NETworkS} ({COMSNETS})}, pages 835--840, Bangalore, India,
  January 2023. IEEE.

\bibitem{bowman_improving_2023}
Richard~W. Bowman.
\newblock Improving instrument reproducibility with open source hardware.
\newblock 3(1):27.

\bibitem{bibliograph7}
https://octoverse.github.com/.
\newblock State of the octoverse 2022.
\newblock 2023.

\bibitem{bibliograph1}
Pedro~Mestre Daniel Adorno~Gomes and Carlos Serodio.
\newblock Infrastructure-as-code for scientific computing environments.
\newblock CENTRIC 2019.

\bibitem{bibliograph2}
pulumi.com.
\newblock Delivering cloud native infrastructure as code - pulumi.
\newblock 2019.

\bibitem{silva_junior_provenance-and_2021}
Daniel Silva~Junior, Esther Pacitti, Aline Paes, and Daniel De~Oliveira.
\newblock Provenance-and machine learning-based recommendation of parameter
  values in scientific workflows.
\newblock {\em PeerJ Computer Science}, 7:e606, July 2021.

\bibitem{ludascher_provstore_2015}
Trung~Dong Huynh and Luc Moreau.
\newblock {ProvStore}: {A} {Public} {Provenance} {Repository}.
\newblock In Bertram Ludäscher and Beth Plale, editors, {\em Provenance and
  {Annotation} of {Data} and {Processes}}, volume 8628, pages 275--277.
  Springer International Publishing, Cham, 2015.

\bibitem{inproceedings}
Fernando Chirigati, Dennis Shasha, and Juliana Freire.
\newblock Reprozip: Using provenance to support computational reproducibility.
\newblock 01 2013.

\bibitem{wonsil_integrated_2023}
Joseph Wonsil, Jack Sullivan, Margo Seltzer, and Adam Pocock.
\newblock Integrated {Reproducibility} with {Self}-describing {Machine}
  {Learning} {Models}.
\newblock In {\em Proceedings of the 2023 {ACM} {Conference} on
  {Reproducibility} and {Replicability}}, pages 1--14, Santa Cruz CA USA, June
  2023. ACM.

\bibitem{samuel_collaborative_2022}
Sheeba Samuel and Birgitta König-Ries.
\newblock A collaborative semantic-based provenance management platform for
  reproducibility.
\newblock {\em PeerJ Computer Science}, 8:e921, March 2022.

\bibitem{samuel_end--end_2022}
Sheeba Samuel and Birgitta König-Ries.
\newblock End-to-{End} provenance representation for the understandability and
  reproducibility of scientific experiments using a semantic approach.
\newblock {\em Journal of Biomedical Semantics}, 13(1):1, December 2022.

\bibitem{wittek_blockchain-based_2021}
Kevin Wittek, Neslihan Wittek, James Lawton, Iryna Dohndorf, Alexander Weinert,
  and Andrei Ionita.
\newblock A {Blockchain}-{Based} {Approach} to {Provenance} and
  {Reproducibility} in {Research} {Workflows}.
\newblock In {\em 2021 {IEEE} {International} {Conference} on {Blockchain} and
  {Cryptocurrency} ({ICBC})}, pages 1--6, Sydney, Australia, May 2021. IEEE.

\bibitem{kawamoto_ai_2020}
Yasutaka Kawamoto and Akihiro Kobayashi.
\newblock {AI} pedigree verification platform using blockchain.
\newblock In {\em 2020 2nd {Conference} on {Blockchain} {Research} \&
  {Applications} for {Innovative} {Networks} and {Services} ({BRAINS})}, pages
  204--205, Paris, France, September 2020. IEEE.

\bibitem{wonsil_reproducibility_2021}
Joseph Wonsil.
\newblock Reproducibility as a service.

\bibitem{crick_reproducibility_2015}
Tom Crick, Benjamin~A. Hall, and Samin Ishtiaq.
\newblock Reproducibility as a {Technical} {Specification}.
\newblock 2015.

\bibitem{colom_ipol_2015}
Miguel Colom, Bertrand Kerautret, Nicolas Limare, Pascal Monasse, and
  Jean-Michel Morel.
\newblock {IPOL}: a new journal for fully reproducible research; analysis of
  four years development.
\newblock July 2015.

\bibitem{milewicz_towards_2023}
Reed Milewicz and Miranda Mundt.
\newblock Towards {Evidence}-{Based} {Software} {Quality} {Practices} for
  {Reproducibility}: {Preliminary} {Results} and {Research} {Directions}.
\newblock In {\em Proceedings of the 2023 {ACM} {Conference} on
  {Reproducibility} and {Replicability}}, pages 85--88, Santa Cruz CA USA, June
  2023. ACM.

\bibitem{stodden_best_2014}
Victoria Stodden and Sheila Miguez.
\newblock Best practices for computational science: Software infrastructure and
  environments for reproducible and extensible research.
\newblock 2(1):e21.

\bibitem{community_illustrations_2022}
The Turing~Way Community and {Scriberia}.
\newblock Illustrations from {The} {Turing} {Way}: {Shared} under {CC}-{BY} 4.0
  for reuse, May 2022.

\bibitem{alnoamany_towards_2018}
Yasmin AlNoamany and John~A. Borghi.
\newblock Towards computational reproducibility: researcher perspectives on the
  use and sharing of software.
\newblock {\em PeerJ Computer Science}, 4:e163, September 2018.

\bibitem{martinez_survey_2021}
Inigo Martinez, Elisabeth Viles, and Igor~G Olaizola.
\newblock A survey study of success factors in data science projects.
\newblock In {\em 2021 {IEEE} {International} {Conference} on {Big} {Data}
  ({Big} {Data})}, pages 2313--2318, Orlando, FL, USA, December 2021. IEEE.

\bibitem{article2}
Christoph Schröer, Felix Kruse, and Jorge Marx~Gómez.
\newblock A systematic literature review on applying crisp-dm process model.
\newblock {\em Procedia Computer Science}, 181:526--534, 01 2021.

\bibitem{turkyilmaz-van_der_velden_reproducibility_2020}
Yasemin Turkyilmaz-van~der Velden, Nicolas Dintzner, and Marta Teperek.
\newblock Reproducibility {Starts} from {You} {Today}.
\newblock {\em Patterns}, 1(6):100099, September 2020.

\bibitem{merz_editorial_2020}
Kenneth~M. Merz, Rommie Amaro, Zoe Cournia, Matthias Rarey, Thereza Soares,
  Alexander Tropsha, Habibah~A. Wahab, and Renxiao Wang.
\newblock Editorial: Method and data sharing and reproducibility of scientific
  results.
\newblock 60(12):5868--5869.

\bibitem{samuel_understanding_2021}
Sheeba Samuel and Birgitta König-Ries.
\newblock Understanding experiments and research practices for reproducibility:
  an exploratory study.
\newblock {\em PeerJ}, 9:e11140, April 2021.

\bibitem{nichols_better_2021}
James~D. Nichols, Madan~K. Oli, William.~L. Kendall, and G.~Scott Boomer.
\newblock A better approach for dealing with reproducibility and replicability
  in science.
\newblock 118(7):e2100769118.

\bibitem{Heesen2017-HEECAT}
Remco Heesen.
\newblock Communism and the incentive to share in science.
\newblock {\em Philosophy of Science}, 84(4):698--716, 2017.

\bibitem{bonsignorio_new_2017}
Fabio Bonsignorio.
\newblock A {New} {Kind} of {Article} for {Reproducible} {Research} in
  {Intelligent} {Robotics} [{From} the {Field}].
\newblock {\em IEEE Robotics \& Automation Magazine}, 24(3):178--182, September
  2017.

\bibitem{de_sterck_enhancing_2023}
Hans De~Sterck, Chi-Wang Shu, and Rémi Abgrall.
\newblock Enhancing {Reproducibility} of {Research} {Papers} in {SISC}, {JSC}
  and {JCP}.
\newblock {\em Journal of Scientific Computing}, 95(3):77,
  s10915--023--02193--7, June 2023.

\bibitem{trisovic_advancing_2020}
Ana Trisovic, Philip Durbin, Tania Schlatter, Gustavo Durand, Sonia Barbosa,
  Danny Brooke, and Mercè Crosas.
\newblock Advancing {Computational} {Reproducibility} in the {Dataverse} {Data}
  {Repository} {Platform}.
\newblock In {\em Proceedings of the 3rd {International} {Workshop} on
  {Practical} {Reproducible} {Evaluation} of {Computer} {Systems}}, pages
  15--20, Stockholm Sweden, June 2020. ACM.

\bibitem{pineau_improving_2020}
Joelle Pineau, Philippe Vincent-Lamarre, Koustuv Sinha, Vincent Larivière,
  Alina Beygelzimer, Florence d'Alché Buc, Emily Fox, and Hugo Larochelle.
\newblock Improving {Reproducibility} in {Machine} {Learning} {Research} ({A}
  {Report} from the {NeurIPS} 2019 {Reproducibility} {Program}).
\newblock {\em arXiv:2003.12206 [cs, stat]}, December 2020.
\newblock arXiv: 2003.12206.

\bibitem{foster_toward_2020}
Kyle Chard, Niall Gaffney, Mihael Hategan, Kacper Kowalik, Bertram Ludäscher,
  Timothy {McPhillips}, Jarek Nabrzyski, Victoria Stodden, Ian Taylor, Thomas
  Thelen, Matthew~J. Turk, and Craig Willis.
\newblock Toward enabling reproducibility for data-intensive research using the
  whole tale platform.
\newblock In Ian Foster, Gerhard~R. Joubert, Luděk Kučera, Wolfgang~E. Nagel,
  and Frans Peters, editors, {\em Advances in Parallel Computing}. {IOS} Press.

\bibitem{brinckman_computing_2019}
Adam Brinckman, Kyle Chard, Niall Gaffney, Mihael Hategan, Matthew~B. Jones,
  Kacper Kowalik, Sivakumar Kulasekaran, Bertram Ludäscher, Bryce~D. Mecum,
  Jarek Nabrzyski, Victoria Stodden, Ian~J. Taylor, Matthew~J. Turk, and
  Kandace Turner.
\newblock Computing environments for reproducibility: {Capturing} the
  “{Whole} {Tale}”.
\newblock {\em Future Generation Computer Systems}, 94:854--867, May 2019.

\bibitem{chard_implementing_2019}
Kyle Chard, Niall Gaffney, Matthew~B. Jones, Kacper Kowalik, Bertram
  Ludäscher, Jarek Nabrzyski, Victoria Stodden, Ian Taylor, Matthew~J. Turk,
  and Craig Willis.
\newblock Implementing {Computational} {Reproducibility} in the {Whole} {Tale}
  {Environment}.
\newblock In {\em Proceedings of the 2nd {International} {Workshop} on
  {Practical} {Reproducible} {Evaluation} of {Computer} {Systems}}, pages
  17--22, Phoenix AZ USA, June 2019. ACM.

\bibitem{yildiz_reproducedpapersorg_2021}
Burak Yildiz, Hayley Hung, Jesse~H. Krijthe, Cynthia C.~S. Liem, Marco Loog,
  Gosia Migut, Frans Oliehoek, Annibale Panichella, Przemyslaw Pawelczak,
  Stjepan Picek, Mathijs de~Weerdt, and Jan van Gemert.
\newblock {ReproducedPapers}.org: {Openly} teaching and structuring machine
  learning reproducibility.
\newblock {\em arXiv:2012.01172 [cs]}, 12636:3--11, 2021.
\newblock arXiv: 2012.01172.

\bibitem{kerautret_rescience_2019}
Nicolas~P. Rougier and Konrad Hinsen.
\newblock {ReScience} {C}: {A} {Journal} for {Reproducible} {Replications} in
  {Computational} {Science}.
\newblock In Bertrand Kerautret, Miguel Colom, Daniel Lopresti, Pascal Monasse,
  and Hugues Talbot, editors, {\em Reproducible {Research} in {Pattern}
  {Recognition}}, volume 11455, pages 150--156. Springer International
  Publishing, Cham, 2019.

\bibitem{inbook}
Kate Keahey, Pierre Riteau, Dan Stanzione, Tim Cockerill, Joe Mambretti, Paul
  Rad, and Paul Ruth.
\newblock {\em Chameleon: A Scalable Production Testbed for Computer Science
  Research}, pages 123--148.
\newblock 05 2019.

\bibitem{salsabil_study_2022}
Lamia Salsabil, Jian Wu, Muntabir~Hasan Choudhury, William~A. Ingram, Edward~A.
  Fox, Sarah~M. Rajtmajer, and C.~Lee Giles.
\newblock A {Study} of {Computational} {Reproducibility} using {URLs} {Linking}
  to {Open} {Access} {Datasets} and {Software}.
\newblock In {\em Companion {Proceedings} of the {Web} {Conference} 2022},
  pages 784--788, Virtual Event, Lyon France, April 2022. ACM.

\bibitem{corcho_documenting_2022}
Al~Idrissou, Veruska Zamborlini, and Tobias Kuhn.
\newblock Documenting the {Creation}, {Manipulation} and {Evaluation} of
  {Links} for {Reuse} and {Reproducibility}.
\newblock In Oscar Corcho, Laura Hollink, Oliver Kutz, Nicolas Troquard, and
  Fajar~J. Ekaputra, editors, {\em Knowledge {Engineering} and {Knowledge}
  {Management}}, volume 13514, pages 81--96. Springer International Publishing,
  Cham, 2022.

\bibitem{raff_siren_2022}
Edward Raff and Andrew~L. Farris.
\newblock A {Siren} {Song} of {Open} {Source} {Reproducibility}.
\newblock {\em arXiv:2204.04372 [cs]}, April 2022.
\newblock arXiv: 2204.04372.

\bibitem{gomes_why_2022}
Dylan G.~E. Gomes, Patrice Pottier, Robert Crystal-Ornelas, Emma~J. Hudgins,
  Vivienne Foroughirad, Luna~L. Sánchez-Reyes, Rachel Turba, Paula~Andrea
  Martinez, David Moreau, Michael~G. Bertram, Cooper~A. Smout, and Kaitlyn~M.
  Gaynor.
\newblock Why don't we share data and code? {Perceived} barriers and benefits
  to public archiving practices.
\newblock {\em Proceedings of the Royal Society B: Biological Sciences},
  289(1987):20221113, November 2022.

\bibitem{noauthor_supporting_2021}
Supporting computational reproducibility through code review.
\newblock {\em Nature Human Behaviour}, 5(8):965--966, August 2021.

\bibitem{athanassoulis_artifacts_2022}
Manos Athanassoulis, Peter Triantafillou, Raja Appuswamy, Rajesh Bordawekar,
  Badrish Chandramouli, Xuntao Cheng, Ioana Manolescu, Yannis Papakonstantinou,
  and Nesime Tatbul.
\newblock Artifacts {Availability} \& {Reproducibility} ({VLDB} 2021 {Round}
  {Table}).
\newblock {\em ACM SIGMOD Record}, 51(2):74--77, July 2022.

\bibitem{radha_verifiable_2021}
Swapna~Krishnakumar Radha, Ian Taylor, Jarek Nabrzyski, and Iain Barclay.
\newblock Verifiable {Badging} {System} for scientific data reproducibility.
\newblock {\em Blockchain: Research and Applications}, 2(2):100015, June 2021.

\bibitem{fursin_enabling_2020}
Grigori Fursin.
\newblock Enabling reproducible {ML} and {Systems} research: the good, the bad,
  and the ugly.
\newblock August 2020.

\bibitem{fostiropoulos_reproducibility_2023}
Iordanis Fostiropoulos, Bowman Brown, and Laurent Itti.
\newblock Reproducibility {Requires} {Consolidated} {Artifacts}.
\newblock In {\em 2023 {IEEE}/{ACM} 2nd {International} {Conference} on {AI}
  {Engineering} – {Software} {Engineering} for {AI} ({CAIN})}, pages
  100--101, Melbourne, Australia, May 2023. IEEE.

\bibitem{malik_artifact_2020}
Tanu Malik.
\newblock Artifact {Description}/{Artifact} {Evaluation}: {A} {Reproducibility}
  {Bane} or a {Boon}.
\newblock In {\em Proceedings of the 4th {International} {Workshop} on
  {Practical} {Reproducible} {Evaluation} of {Computer} {Systems}}, pages 1--1,
  Virtual Event Sweden, June 2020. ACM.

\bibitem{perignon_certify_2019}
Christophe Pérignon, Kamel Gadouche, Christophe Hurlin, Roxane Silberman, and
  Eric Debonnel.
\newblock Certify reproducibility with confidential data.
\newblock {\em Science}, 365(6449):127--128, July 2019.

\bibitem{cousijn_data_2018}
Helena Cousijn, Amye Kenall, Emma Ganley, Melissa Harrison, David Kernohan,
  Thomas Lemberger, Fiona Murphy, Patrick Polischuk, Simone Taylor, Maryann
  Martone, and Tim Clark.
\newblock A data citation roadmap for scientific publishers.
\newblock {\em Scientific Data}, 5(1):180259, November 2018.

\bibitem{seibold_computational_2021}
Heidi Seibold, Severin Czerny, Siona Decke, Roman Dieterle, Thomas Eder,
  Steffen Fohr, Nico Hahn, Rabea Hartmann, Christoph Heindl, Philipp Kopper,
  Dario Lepke, Verena Loidl, Maximilian Mandl, Sarah Musiol, Jessica Peter,
  Alexander Piehler, Elio Rojas, Stefanie Schmid, Hannah Schmidt, Melissa
  Schmoll, Lennart Schneider, Xiao-Yin To, Viet Tran, Antje Völker, Moritz
  Wagner, Joshua Wagner, Maria Waize, Hannah Wecker, Rui Yang, Simone Zellner,
  and Malte Nalenz.
\newblock A computational reproducibility study of {PLOS} {ONE} articles
  featuring longitudinal data analyses.
\newblock 16(6):e0251194.

\bibitem{karathanasis_reproducibility_2022}
Nestoras Karathanasis, Daniel Hwang, Vibol Heng, Rimal Abhimannyu, Phillip
  Slogoff-Sevilla, Gina Buchel, Victoria Frisbie, Peiyao Li, Dafni Kryoneriti,
  and Isidore Rigoutsos.
\newblock Reproducibility efforts as a teaching tool: {A} pilot study.
\newblock {\em PLOS Computational Biology}, 18(11):e1010615, November 2022.

\bibitem{hofman_expanding_2020}
Jake~M. Hofman, Daniel~G. Goldstein, Siddhartha Sen, and Forough
  Poursabzi-Sandegh.
\newblock Expanding the {Scope} of {Reproducibility} {Research} {Through}
  {Data} {Analysis} {Replications}.
\newblock In {\em Companion {Proceedings} of the {Web} {Conference} 2020},
  pages 567--571, Taipei Taiwan, April 2020. ACM.

\bibitem{fund_we_2023}
Fraida Fund.
\newblock We {Need} {More} {Reproducibility} {Content} {Across} the {Computer}
  {Science} {Curriculum}.
\newblock In {\em Proceedings of the 2023 {ACM} {Conference} on
  {Reproducibility} and {Replicability}}, pages 97--101, Santa Cruz CA USA,
  June 2023. ACM.

\bibitem{lin_building_2022}
Jimmy Lin.
\newblock Building a {Culture} of {Reproducibility} in {Academic} {Research}.
\newblock 2022.

\bibitem{parsons_history_2019}
Mark~A. Parsons, Ruth~E. Duerr, and Matthew~B. Jones.
\newblock The {History} and {Future} of {Data} {Citation} in {Practice}.
\newblock {\em Data Science Journal}, 18:52, November 2019.

\bibitem{dozmorov_github_2018}
Mikhail~G. Dozmorov.
\newblock {GitHub} {Statistics} as a {Measure} of the {Impact} of
  {Open}-{Source} {Bioinformatics} {Software}.
\newblock {\em Frontiers in Bioengineering and Biotechnology}, 6:198, December
  2018.

\bibitem{stodden_toward_2013}
Victoria Stodden, Peixuan Guo, and Zhaokun Ma.
\newblock Toward {Reproducible} {Computational} {Research}: {An} {Empirical}
  {Analysis} of {Data} and {Code} {Policy} {Adoption} by {Journals}.
\newblock {\em PLoS ONE}, 8(6):e67111, June 2013.

\bibitem{lewis_policy_nodate}
Thu-Mai Lewis.
\newblock {\em From policy to practice: {How} journal-based data policies
  encourage scientists' adoption of reproducible research practices}.
\newblock PhD thesis, The University of North Carolina at Chapel Hill
  University Libraries.

\bibitem{noauthor_digital_nodate}
Zhaokun Stodden~Victoria, Guo~Peixuan.
\newblock How journals are adopting open data and code policies.

\bibitem{bosman_oa_2021}
Jeroen Bosman, Jan~Erik Frantsvåg, Bianca Kramer, Pierre-Carl Langlais, and
  Vanessa Proudman.
\newblock {OA} {Diamond} {Journals} {Study}. {Part} 1: {Findings}.
\newblock Technical report, Zenodo, March 2021.

\bibitem{vasilevsky_reproducible_2017}
Nicole~A. Vasilevsky, Jessica Minnier, Melissa~A. Haendel, and Robin~E.
  Champieux.
\newblock Reproducible and reusable research: are journal data sharing policies
  meeting the mark?
\newblock {\em PeerJ}, 5:e3208, April 2017.

\bibitem{konkol_publishing_2020}
Markus Konkol, Daniel Nüst, and Laura Goulier.
\newblock Publishing computational research - a review of infrastructures for
  reproducible and transparent scholarly communication.
\newblock {\em Research Integrity and Peer Review}, 5(1):10, December 2020.

\bibitem{willis_trust_2020}
Craig Willis and Victoria Stodden.
\newblock Trust but {Verify}: {How} to {Leverage} {Policies}, {Workflows}, and
  {Infrastructure} to {Ensure} {Computational} {Reproducibility} in
  {Publication}.
\newblock {\em Harvard Data Science Review}, 2(4), December 2020.

\bibitem{boulbes_survey_2018}
Delphine~R. Boulbes, Tracy Costello, Keith Baggerly, Fan Fan, Rui Wang, Rajat
  Bhattacharya, Xiangcang Ye, and Lee~M. Ellis.
\newblock A {Survey} on {Data} {Reproducibility} and the {Effect} of
  {Publication} {Process} on the {Ethical} {Reporting} of {Laboratory}
  {Research}.
\newblock {\em Clinical Cancer Research}, 24(14):3447--3455, July 2018.

\bibitem{gupta_state_2022}
Abhishek Gupta, Connor Wright, Marianna~Bergamaschi Ganapini, Masa Sweidan, and
  Renjie Butalid.
\newblock State of {AI} {Ethics} {Report} ({Volume} 6, {February} 2022),
  February 2022.
\newblock arXiv:2202.07435 [cs].

\bibitem{moreau_containers_2023}
David Moreau, Kristina Wiebels, and Carl Boettiger.
\newblock Containers for computational reproducibility.
\newblock {\em Nature Reviews Methods Primers}, 3(1):1--16, July 2023.

\bibitem{fursin_collective_2020-1}
Grigori Fursin.
\newblock The {Collective} {Knowledge} project: making {ML} models more
  portable and reproducible with open {APIs}, reusable best practices and
  {MLOps}.
\newblock 2020.

\bibitem{poldrack_costs_2019}
Russell~A. Poldrack.
\newblock The {Costs} of {Reproducibility}.
\newblock {\em Neuron}, 101(1):11--14, January 2019.

\bibitem{article1}
Armbrust.
\newblock Above the clouds: A berkeley view of cloud computing.
\newblock 01 2009.

\bibitem{gandrud_reproducible_2020}
Christopher Gandrud.
\newblock {\em Reproducible research with {R} and {RStudio}}.
\newblock The {R} series. CRC Press, Boca Raton, FL, third edition edition,
  2020.

\bibitem{afgan2018galaxy}
Enis Afgan, Dannon Baker, B{\'e}r{\'e}nice Batut, Marius Van Den~Beek, Dave
  Bouvier, Martin {\v{C}}ech, John Chilton, Dave Clements, Nate Coraor,
  Bj{\"o}rn~A Gr{\"u}ning, et~al.
\newblock The galaxy platform for accessible, reproducible and collaborative
  biomedical analyses: 2018 update.
\newblock {\em Nucleic acids research}, 46(W1):W537--W544, 2018.

\bibitem{brundage_toward_2020}
Miles Brundage, Shahar Avin, Jasmine Wang, Haydn Belfield, Gretchen Krueger,
  Gillian Hadfield, Heidy Khlaaf, Jingying Yang, Helen Toner, Ruth Fong, Tegan
  Maharaj, Pang~Wei Koh, Sara Hooker, Jade Leung, Andrew Trask, Emma Bluemke,
  Jonathan Lebensold, Cullen O'Keefe, Mark Koren, Théo Ryffel, JB~Rubinovitz,
  Tamay Besiroglu, Federica Carugati, Jack Clark, Peter Eckersley, Sarah
  de~Haas, Maritza Johnson, Ben Laurie, Alex Ingerman, Igor Krawczuk, Amanda
  Askell, Rosario Cammarota, Andrew Lohn, David Krueger, Charlotte Stix, Peter
  Henderson, Logan Graham, Carina Prunkl, Bianca Martin, Elizabeth Seger, Noa
  Zilberman, Seán hÉigeartaigh, Frens Kroeger, Girish Sastry, Rebecca Kagan,
  Adrian Weller, Brian Tse, Elizabeth Barnes, Allan Dafoe, Paul Scharre, Ariel
  Herbert-Voss, Martijn Rasser, Shagun Sodhani, Carrick Flynn, Thomas~Krendl
  Gilbert, Lisa Dyer, Saif Khan, Yoshua Bengio, and Markus Anderljung.
\newblock Toward {Trustworthy} {AI} {Development}: {Mechanisms} for
  {Supporting} {Verifiable} {Claims}.
\newblock 2020.

\bibitem{hiesgen2022race}
Raphael Hiesgen, Marcin Nawrocki, Thomas~C Schmidt, and Matthias W{\"a}hlisch.
\newblock The race to the vulnerable: Measuring the log4j shell incident.
\newblock {\em arXiv preprint arXiv:2205.02544}, 2022.

\bibitem{ke_network-based_2023}
Qing Ke, Alexander~J. Gates, and Albert-László Barabási.
\newblock A network-based normalized impact measure reveals successful periods
  of scientific discovery across disciplines.
\newblock {\em Proceedings of the National Academy of Sciences},
  120(48):e2309378120, November 2023.

\bibitem{smith_software_2016}
Arfon~M. Smith, Daniel~S. Katz, Kyle~E. Niemeyer, and {FORCE11 Software
  Citation Working Group}.
\newblock Software citation principles.
\newblock {\em PeerJ Computer Science}, 2:e86, September 2016.

\bibitem{di_cosmo_2044_2022}
Roberto Di~Cosmo, Morane Gruenpeter, and Stefano Zacchiroli.
\newblock 204.4 {Identifiers} for {Digital} {Objects}: {The} case of software
  source code preservation.
\newblock August 2022.

\bibitem{Stefaniak}
Jill Stefaniak and Kimberly Carey.
\newblock Instilling purpose and value in the implementation of digital badges
  in higher education.
\newblock {\em International Journal of Educational Technology in Higher
  Education}, 16, 12 2019.

\bibitem{khritankov_mldev_2021}
Anton Khritankov, Nikita Pershin, Nikita Ukhov, and Artem Ukhov.
\newblock {MLDev}: {Data} {Science} {Experiment} {Automation} and
  {Reproducibility} {Software}.
\newblock {\em arXiv:2107.12322 [cs]}, July 2021.
\newblock arXiv: 2107.12322.

\bibitem{bedo_unifying_2020}
Justin Bedő, Leon Di Stefano, and Anthony~T Papenfuss.
\newblock Unifying package managers, workflow engines, and containers:
  Computational reproducibility with {BioNix}.
\newblock 9(11):giaa121.

\bibitem{salvador_olivan_reproducibilidad_2023}
José~Antonio Salvador~Oliván, Gonzalo Marco~Cuenca, and Rosario
  Arquero~Avilés.
\newblock reproducibilidad de las estrategias de búsqueda en revisiones
  sistemáticas publicadas en revistas españolas de {Biblioteconomía} y
  {Documentación}.
\newblock {\em Ibersid: revista de sistemas de información y documentación},
  17(1):129--137, June 2023.

\bibitem{page_prisma_2021}
Matthew~J Page, Joanne~E McKenzie, Patrick~M Bossuyt, Isabelle Boutron, Tammy~C
  Hoffmann, Cynthia~D Mulrow, Larissa Shamseer, Jennifer~M Tetzlaff, Elie~A
  Akl, Sue~E Brennan, Roger Chou, Julie Glanville, Jeremy~M Grimshaw, Asbjørn
  Hróbjartsson, Manoj~M Lalu, Tianjing Li, Elizabeth~W Loder, Evan
  Mayo-Wilson, Steve McDonald, Luke~A McGuinness, Lesley~A Stewart, James
  Thomas, Andrea~C Tricco, Vivian~A Welch, Penny Whiting, and David Moher.
\newblock The {PRISMA} 2020 statement: an updated guideline for reporting
  systematic reviews.
\newblock {\em BMJ}, page n71, March 2021.

\end{thebibliography}

\appendix

\section{Research Methods}
\subsection{PRISMA Literature Review}

\label{sec:literature_review}
To perform our review two strategies~\cite{salvador_olivan_reproducibilidad_2023} were adopted. On one hand, active search for articles highly cited reports, and follow-up of citation threads on computational reproducibility. On the other hand, we applied the PRISMA 2020~\cite{page_prisma_2021} methodology with the SCOPUS \& WoS databases between 2020 to the present day (Table~\ref{demo-table01}). Eventually, we reduced our analysis to 100 representative works.

\begin{table}[H]
  \label{tab:freq}
  \begin{tabular}{ p{2cm} p{2.5cm} p{7cm} }
    \toprule
 \multicolumn{3}{c}{PRISMA} \\
  \midrule
 Step& Num items & Condition  \\
 \midrule
 \midrule
  Identification   &   SCOPUS (413)  &   TITLE ( reproducibility ) AND PUBYEAR $>$ 2019 AND PUBYEAR $<$ 2024 AND ( LIMIT-TO ( SUBJAREA , "COMP" ) )  \\
 \midrule
   Identification   &   WoS (371)  &   TI=(reproducibility) and 2023 or 2022 or 2021 or 2020 (Publication Years) and Multidisciplinary Sciences or Computer Science Interdisciplinary Applications or Computer Science Theory Methods or Computer Science Information Systems (Web of Science Categories)  \\
 \midrule

 Deleting repeated &  (144) & Repeated articles\\ 
 \midrule
Screening &  (80) &   Main theme reproducibility     \\
 \midrule
 Included &    (60) &    
Systematically classified   \\
 \bottomrule

\end{tabular}

\caption{The PRISMA screening was used in this review to select recent relevant articles related to reproducibility in scientific research.}
\label{demo-table01}
\end{table}

\subsection{Journals Survey}
\label{sec:survey_methodology}

To avoid bias in the research and not only address journals that are known to apply reproducibility policies, e.g. (IEEE with CodeOcean \footnote{\url{https://innovate.ieee.org/ieee-code-ocean/}}, ACM with reproducibility Badges \footnote{\url{https://www.acm.org/publications/policies/artifact-review-badging}}, Journal Nature\footnote{\url{https://www.nature.com/nature-portfolio/editorial-policies/reporting-standards}} ), a request for participation was sent to several journals specialized in computer science.

Information was voluntarily requested through a 16-question form from a list of 500 journals specialized in computer science, scopus indexed from Q1 to Q4 in the period from July 1, 2023 to September 1, 2023. Although the question was very precise, they were left open to comments.

\subsection{Classification of reproducibility criteria.}
Table~\ref{demo-table1} shows the classification of reproducibility criteria for Artifacts Description (AD) and Artifacts Evaluation (AE).

\begin{longtable}{| p{5cm} | p{8cm} | p{2cm} |}
 \toprule
 \textbf{Criterion} & \textbf{Description} & \textbf{Type}  \\ \hline
 \midrule
 \midrule
 
Results& Documented result and analysis  &  Experiment \\
Analysis&  Supported claims & Experiment   \\
Justification& Justified method, metrics, datasets  & Experiment   \\
Workflow& Summarized experiment execution and configurations  & Experiment   \\
Workflow execution& Tracked execution with configuration  &Experiment    \\
Hardware& Specified hardware &  Experiment  \\
Software& Documented software dependencies.  &  Experiment  \\
Citation Export& Reference automatically generated& Experiment   \\
Code repository& Shared code in repository  & Experiment   \\
Code metadata&  Code metadata included &  Experiment  \\
Code license& Code license included  &  Experiment  \\
Code citeable&Code (DOI) or (PURL) assigned.   & Experiment  \\
Hypothesis&  Documented hypothesis &  Method  \\
Prediction& Documented predictions  &  Method  \\
Setup& Documented parameters and conditions, statistical significance of results.  &  Method  \\
Problem description&  Clearly described problem &  Method  \\
Outline& Conceptually described method   & Method   \\
Pseudo code &  Documented pseudo code &  Method \\
\midrule
Data repository & Data shared in accessible repository &  Data \\
Data metadata& Metadata included in the datasets  & Data  \\
Data license& Licensed data &  Data \\
Data citeable&  DOI or P-URL of data assigned &  Data \\
  \midrule
  \midrule
NEUROips Checklist \cite{pineau_improving_2020}&   &   \\
 \midrule
 \midrule
Model and algorithms& Clarified mathematical models, algorithms, settings - assumptions explained - algorithm complexity analyzed & Experiment  \\
Theoretical claim&  Clarified claim statements - fully proven claims &  Method \\
Datasets& Relevant statistics, details of train/validation/test split, explanation of excluded data and preprocessing, link to downloadable version of environment and dataset, description of quality Control methods  & Data  \\
Code&  Specified dependencies, Training code, evaluation code, README file with results table, pre trained models & Experiment  \\
Rxperimental result& Selection method, range, specification of best hyperparameters, exact number of training and evaluation runs, clear definition of metrics and results statistics, description of results with trend and central variation, average energy cost, results runtime.  &  Method \\
 \midrule
 \midrule
SIGPLAN&   &   \\
\midrule
\midrule
Clearly stated claims& Explicit claims, appropriately scoped, recognize limitations   &  Method \\
Suitable comparison& Compare with the appropriate baseline, comparison is fair  &   Method\\
Principled benchmark choice& Appropriate and fair use of non-standard suit, using applications instead of kernels   &  Method \\
Adequate data analysis& Sufficient number of trials, appropriate summary of statistics, data distribution reported   & Method  \\
Relevant Metrics&  Direct and appropriate proxy metrics, successful in measuring all effects  &  Method \\
Appropriate and clear Experimental Design& Enough information to repeat, reasonable platform, Consider all key design parameters, open workload generator, evaluated in test set  & Method  \\
Appropriate presentation of results&  Clear summary of results, appropriate truncate axes, ratios plotted correctly, appropriate level of precision  & Method  \\
\midrule
\midrule
Ctuning&   &   \\
 \midrule
 \midrule
Abstract& Clearly stated the problem, solution and supporting results?& Method  \\
Algorithm& Is it a new algorithm?  & Experiment  \\
Program& Are any benchmarks used?   &  Method \\
Compilation& Does it require a specific compiler?   & Experiment  \\
Transformations& Does it require a program transformation tool? &  Experiment\\
Binary& Are binaries included?  & Experiment  \\
Model&  Are specific models used?   &  Experiment \\
Data set& Are specific data sets used ?  &  Experiment \\
Run-time environment& Are there any OS-specific artifact? & Experiment  \\
Hardware& Is specific hardware required?   & Experiment  \\
Run-time state& Is the state sensitive to run-time?  & Experiment  \\
Execution& Is the software running under specific conditions?  & Experiment  \\
Metrics& How are the metrics evaluated?  & Experiment  \\
Output& What is the output?  & Experiment  \\
Experiments& How to prepare and reproduce results?& Method \\
Disk space & How much disk space is required?  &  Experiment \\
Workflow & How much time is needed to prepare the workflow?  & Experiment  \\
Time evaluation&  How much time is needed to complete the experiments? &  Experiment \\ 
Publicly available&   &  Data \\
Code licenses&  Is the software under any licenses? & Data  \\
Workflow frameworks& Are workflow framework used for automation?   & Method  \\ 
Archived& Is the software archived and make it public?  & Data   \\
Description&   &   \\
Access& Does the system describe how reviewers will access the research artifacts?& Data  \\
Hardware dependencies& Does the system describe any specific hardware and specific features?  &  Experiment \\
Software dependencies& Does the system describe OS and software packages required to evaluate your artifacts?  & Experiment  \\
Data sets& Are there any third-party data sets used in your packages?& Data  \\
Installation& Does the system describe the setup procedures?  & Method  \\
Experiment workflow& Does the system describe how the workflow is implemented and executed?&  Experiment\\
Evaluation and expected result& Does the system describe how to reproduce the key results from the paper?& Method  \\
Experiment customization& Does the system provide special instructions to customize and tune the experiments?& Method  \\
Notes&   &   \\
\midrule
\midrule
NISO&   &   \\
 \midrule
 \midrule
Artifact Available& A DOI or URL link to the repository along with a unique identifier for the object is provided  &  Data \\
Artifacts Evaluated-Functional& the artifacts are documented, consistent, complete, exercisable, and include appropriate evidence of verification and validation.  & Method  \\
Artifacts Evaluate-Reusable& Artifacts associated with the paper are of a quality, documented and well-structured that significantly exceeds minimal functionality to the extent that reuse and re-purposing is facilitated.  & Method  \\
Open Research Objects (ORO) & A DOI or URL link to the repository along with a unique identifier for the object is provided for Functional + placed on a publicly accessible archival repository.  &  Method \\
Research Object Reviewed (ROR) & The results of the article have been obtained independently in a study by a team or reviewer other than the authors, without the use of artifacts provided by the authors.  & Method  \\
Results Replicated (RER) & ROR + ORO + without the use of artifacts provided by the authors, the main results of the article were obtained independently in a subsequent evaluation by a reviewer or team other than the authors,  & Method  \\
Results Reproduced (ROR-R)&  ROR + ORO + the main results of the article have been obtained in a subsequent evaluation carried out by a reviewer or team other than the authors, using, in part, artifacts provided by the author. & Method  \\
 \midrule
 \midrule
FAIR-TLC\cite{parland-von_essen_supporting_2018}& &   \\
\midrule
\midrule
Findable &Data must have rich and accurate metadata that allows for easy discovery and identification  &  Data\\
Accessible& Data must be available and accessible for free or through clear and well-defined mechanisms& Data  \\
Interoperable& Data must be structured and organized in a coherent way so that it can be combined, integrated, and used in conjunction with other data& Data  \\
Reusable & The data must be made available under a clearly specified open license that allows its use and reuse by other users & Data \\
Traceable &  The data must be accompanied by information that makes it possible to trace its origin and the way in which it has been modified or processed. This includes provenance  information &  Data\\
Licensed &  The data must have a license or a legal agreement that specifies the terms and conditions for its use and reuse &  Data\\
Connected &  The data must be linked to other related data sets and relevant resources, such as scientific publications, source codes, and documentation, among others & Data \\
\bottomrule
\caption{\label{demo-table1}Classification of reproducibility criteria for Artifacts Description (AD) and Artifacts Evaluation (AE).}

\end{longtable}

\end{document}